\documentclass[usenatbib]{mn2e}
\usepackage{apjfonts,amsfonts,amsmath,amssymb,bm,ctable,verbatim,color}

\definecolor{clorange}{rgb}{0.85,0.45,0.0}

\newcommand{\GIZMO}{{\small GIZMO}}
\newcommand{\gizmourl}{\href{http://www.tapir.caltech.edu/~phopkins/Site/GIZMO.html}{\url{http://www.tapir.caltech.edu/~phopkins/Site/GIZMO.html}}}

\newcommand\plotone[2]{\centering \leavevmode \includegraphics[width={#2\columnwidth}]{#1}}

\newcommand{\plottwo}[2]{\centering \leavevmode \includegraphics[width={0.5\textwidth}]{#1}\includegraphics[width={0.5\textwidth}]{#2}}
\newcommand{\acknowledgments}[1]{\begin{small}\section*{Acknowledgments}\end{small}{\noindent #1}\vspace{5pt}}

\title[Lagrangian Vlasov Methods]{An Efficient Method for Directly Integrating the Vlasov Equation in Collisionless and Weakly-Collisional Simulations}

\author[Hopkins]{
\parbox[t]{\textwidth}{Philip F.~Hopkins$^1$} \vspace*{4pt} \\
$^1$ TAPIR, Mailcode 350-17, California Institute of Technology, Pasadena, CA 91125, USA. E-mail:phopkins@caltech.edu 
}

\date{}
\begin{document}
\maketitle

\begin{abstract}
We describe and test a new numerical method to solve the Vlasov/Boltzmann equation in systems with arbitrary levels of collisionality. The method combines ideas from recent Lagrangian hydrodynamics, radiation, and plasma schemes to evolve the distribution function (DF) directly in $6+1$ dimensional phase-space in a manner which is Lagrangian in both position and velocity. The ``particles'' of traditional Monte-Carlo/Particle-In-Cell/N-body (MC/PIC/NB) methods become mesh-generating points defining a volume partition, each cell of which carries a set of momentum-Lagrangian basis functions representing a local, arbitrarily anisotropic momentum DF. The DF is then integrated between cells and bases via a higher-order reconstruction and Godunov-type method with a series of one-sided Riemann problems, evolving a 10 or 13-moment representation of each basis to ensure manifest mass/momentum/energy conservation. The method allows one to represent arbitrary DFs, including multiple independent streams but also collisional or phase-mixed or thermal distributions (unlike traditional ``moments'' or waterbag/tessellation methods), and the cost is bounded to a modest multiplier of MC/PIC/NB costs (much cheaper than direct or lattice Boltzmann methods). We provide an example implementation which uses a spatially-Lagrangian volume partition, second-order reconstruction, and simple set of bases which ensure constant cell masses and couple directly to existing MC/PIC/NB machinery, made public in the code \GIZMO. We validate in a series of standard test problems, demonstrating manifest conservation, Galilean and translation and rotation invariance, and ability to represent complex DFs without spurious dissipation or instability. We discuss applications where this method can have advantages compared to traditional algorithms for simulations of dark matter, stellar, weakly-collisional neutral gas, and particulate/dust dynamics; multiphase/polydisperse flows; cosmic ray and radiation transport/hydrodynamics; and plasma physics. 
\end{abstract}


\section{Introduction}
\label{sec:intro}

The dynamics of collisionless or weakly-collisional systems are fundamental to astrophysics, atmospheric sciences, and many other domains, appearing in e.g.\ plasma physics, radiation hydrodynamics (RHD), stellar dynamics, dark matter/cosmology, cosmic ray physics, particulate/dust/grain/pebble dynamics, polydisperse and multiphase flows, atomic/molecular kinetics, and more. 
In all of these applications, the highly non-linear, chaotic, multi-physics nature of the interesting problems necessitates numerical simulations. 
And although a huge variety of methods have been developed to solve the equations-of-motion of these systems, 
for very-weakly collisional or truly collisionless systems (for which one cannot use typical ``fluid like'' or Boltzmann moments methods, as these ignore the fundamentally collisionless nature of these systems in favor of assuming quantities like the velocity/flux/intensity are single-valued at each point), 
the most widely used methods are often variants of MC/PIC/NB methods. 
Those allow one to capture truly collisionless behaviors, but at the cost of representing some enormous true number of microphysical particles or elements via an orders-of-magnitude smaller (and therefore computationally manageable) discrete $N$ ``super-particles,'' each evolving akin to a single micro-particle.
For example, in gravitational dynamics N-body methods, one solves the equations of motion for $N$ point masses \citep{hernquist:1990.barnes.nbody.review,dehnen:2011.nbody.review}, but while this might be appropriate for a system which actually has $N$ point-like masses (e.g.\ a small star cluster with $\lesssim 10^{5}$ stars) in e.g.\ cosmological simulations a typical MC/PIC/NB super-particle might represent $\sim 10^{60}-10^{90}$ individual dark matter particles. 
Clearly, in these limits MC/PIC/NB sampling can generate a number of spurious effects, including discreteness noise and far-too-rapid relaxation \citep{melott:1997.discreteness.error.nbody,binney:2004.discreteness.in.nbody,chavanis:2006.violent.relaxation.dynamics,campa:2009.stamech.of.finite.nbody}. Since the continuum (smooth) density is not well-defined in the numerical method (a MC/PIC/NB particle is, by definition, a $\delta$-function in phase space), force softening for long-range forces and often regularization or Monte-Carlo sampling prescriptions for short-range forces are necessary but {ad hoc} \citep[for discussion in gravitational dynamics, see e.g.][]{merritt:1996.optimal.softening,dehnen:2001.optimal.softening,athanassoula:2000.optimal.force.softening.collisionless.sims,zhang:2019.optimal.cosmo.sim.softenings}, and the adaptive softening schemes which can be reliably applied to continuum systems (e.g.\ gas, \citealt{price:2007.lagrangian.adaptive.softening,hopkins:gizmo,hopkins:fire2.methods}) become formally ill-defined for MC/PIC/NB super-particles \citep[see][]{iannuzzi:2011.collisionless.adaptive.softening.gadget,iannuzzi:2013.no.need.adaptive.softening.for.dm,barnes:2012.softening.is.smoothing,mansfield.avestruz:2021.halo.bias.over.softening}, and themselves enable spurious heating, equipartition, and disruption effects \citep{ludlow:2019.spurious.equipartition.heating.galaxy.sizes,ludlow:2021.disk.heating.by.dm.particles,vandenbosch:2018.oversoft.disruption.numerical.discreteness.noise,vandenbosch:2018.disruption.over.soft,garrisonkimmel:fire.subhalo.destruction,chiang:2025.numerical.convergence.critera.for.subhalo.tidal.evolution.tidal.softening.improves.greatly.need.to.resolve.in.units.of.tidal.radius,hopkins:tidal.softening}. The formal/analytic convergence properties for continuum predictions are similarly ill-defined, and are instead defined in practice via brute-force resolution tests \citep[see e.g.][]{power:2003.nfw.models.convergence,springel:2008.aquarius,diemand:2008.via.lactea.clumps.streams,stadel:2009.ghalo.multibillion.halo.zoom}, leading to long-standing questions about whether e.g.\ predictions for the central structure of collapsed objects or merged particulate piles or long-range (self-gravitating or Coulomb) force equilibrium are actually robust or numerical artifacts \citep[again, for gravity see e.g.][and references therein]{nfw:profile,deblock:2008.core.cusp.halo.low.densities.in.things,pontzen.governato:2012.feedback.cusp.core.dwarfs,pontzen.governato:2014.cdm.heats.up.review}. And it is difficult if not impossible to couple such approaches to certain types of collision or relaxation operators with non-trivial phase-space interactions (a limitation for many particle dark matter models, types of plasma collisions and interactions, coupling terms for aerosols or polydisperse media, cosmic ray losses and scattering, van der Waals forces, anisotropic radiation scattering/absorption, annihilation operators, solid boundaries, etc.).

\begin{figure*}
\centering
\includegraphics[width=0.99\textwidth]{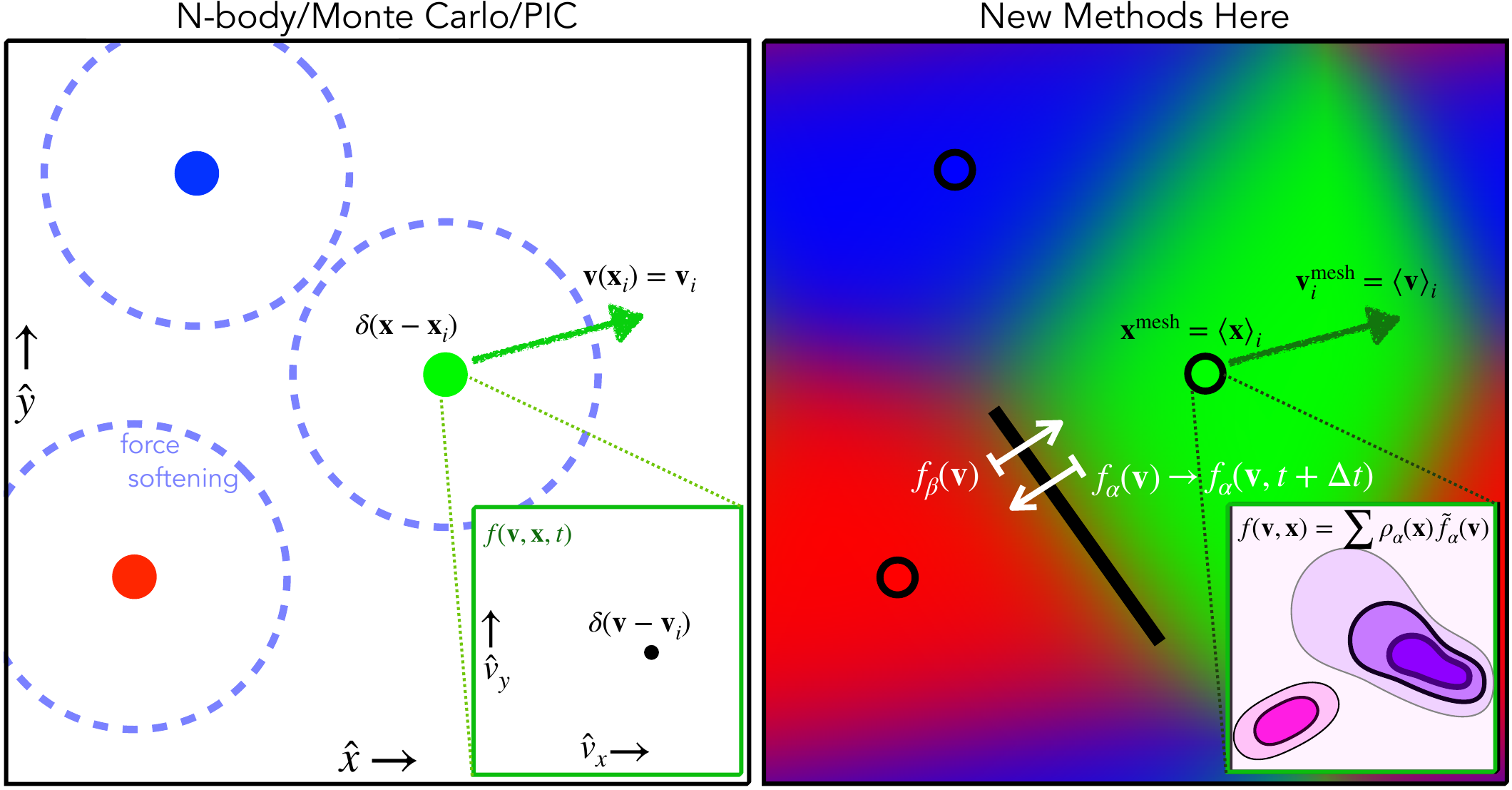}
\caption{
High-level schematic illustrating the method here (\textit{right}), as compared to standard MC/PIC/NB methods (\textit{left}).  
\textit{Left:} In MC/PIC/NB: particles $i$ sample trajectories of individual micro-particles, i.e.\ a distribution function (DF) $f({\bf x},\,{\bf v},\,t) \propto \delta({\bf x}-{\bf x}_{i},\,{\bf v}-{\bf v}_{i},\,t)$, but then get assigned a ``weight'' corresponding to many micro-particles ($\Delta m_{i} \gg m_{\rm micro}$). As such, concepts like ``force softening'' must be defined ad-hoc since the volume is not formally partitioned in any meaningful manner, and collisions or inter-particle forces must be sampled in a Monte-Carlo or coarse-grained manner. 
\textit{Right:} The volume domain obeys a partition of unity into cells/domains $\Omega_{i}$, around mesh-generating points ${\bf x}^{\rm mesh}_{i}$. For the MFM-like method we implement, these involve a smooth, overlapping decomposition with effective faces (\textit{black line}) between neighbors, with Lagrangian motion ${\bf x}^{\rm mesh}_{i} = \langle {\bf x}\rangle_{i} = m_{i}^{-1} \int_{\Omega_{i}} \rho({\bf x}) {\rm d}^{3} {\bf x}$ and ${\bf v}^{\rm mesh}_{i} \equiv {\rm d}_{t} {\bf x}^{\rm mesh}_{i} = \langle {\bf v} \rangle_{i}$. The DF is partitioned within each cell according to a set of basis functions $f({\bf v}|{\bf x},\,t) = \sum \rho_{i,\,\alpha} \tilde{f}_{i,\,\alpha}({\bf v})$ which are carried with the cells and evolve with time. The DF evolution is determined directly by the solution to the Vlasov equation, via fluxes between neighbors constructed at the face (after some higher-order reconstruction of $f\to f^{\ast}$ from each side). Thus the method is a full 6+1D representation of the DF, Lagrangian in both ${\bf x}$ and ${\bf v}$.
\label{fig:cartoon.xvdecomp}}\vspace{-0.3cm}
\end{figure*}

For continuum systems where the number of ``micro-particles'' is large, the obvious ``correct'' approach would be to directly integrate the phase-space distribution function (DF) obeying the Vlasov equation in $6+1$ (spatial and momentum + time) dimensions. But simply discretizing phase space into a uniform 6D ``grid'' (in $\Delta^{3}{\bf x}\,\Delta^{3}{\bf v}$) is prohibitively expensive for most problems. However, a variety of advances in simulation techniques for hydrodynamics, RHD, and moment as well as PIC schemes for plasma physics suggest promising routes forward \citep[e.g.][]{fox:2008.eqmom.gaussian.quadrature.moment.method,marchisio.fox:2013.qmom.book.multiphase.moments,jiang:2014.rhd.solver.local,hopkins:gizmo,miller.shumlak:2016.multispecies.13.moment.moderately.collisional.plasma,grudic:starforge.methods}.

In this paper, we propose and test a method for directly integrating the continuous, 6+1 dimensional Vlasov equation for weakly-collisional or truly collisionless systems (which interpolates smoothly into fully-collisional systems as well). The method is quasi-Lagrangian or arbitrary-Lagrangian-Eulerian as desired in both position and velocity space, providing adaptively higher resolution automatically in regions of phase space of greatest interest. It manifestly conserves mass, momentum, and energy; can stably handle arbitrary shell-crossing and phase-wrapping without increasing CPU cost; and couples trivially to existing MC/PIC/NB methods for integrating long-range or collective forces like gravity or magnetic forces. Most importantly, it introduces only a modest, fixed cost multiplier compared to typical methods already in use in these fields -- either ``particle'' methods like MC/PIC/NB \textit{or} ``moments'' methods like 13-moment/M1/CR-fluid/dust-fluid methods (both of which it extends and improves by evolving the \textit{full} DF). This makes it feasible to run full high-resolution, high-dynamic range, multi-physics simulations with this family of methods. We specifically outline and test one example implementation, which we make public in the code \GIZMO.\footnote{The public version of \GIZMO\ is available at: \gizmourl}

\section{Numerical Principle \&\ Theory}
\label{sec:theory}

Conceptually, the method is simple. Configuration space $d^{D}{\bf x}$ is partitioned among a set of mesh-generating points $a$ (positions ${\bf x}_{a}$), according to a standard partition of unity (domains $\Omega_{a}$, surfaces $\partial \Omega_{a}$). Momentum (velocity) space $d^{D}{\bf p}$ ($d^{D}{\bf v}$) is reconstructed according to a set of independent basis functions, each of which evolves via fluxes between cells+bases and external forces independently according to a generalized anisotropic $N$-moment scheme. 

\subsection{Theoretical Basis: Equations Solved}

\subsubsection{Continuum Vlasov-Poisson Equation}
\label{sec:cbe.continuum}

Consider a system of $N$ point-like test particles at spatial coordinate/positions ${\bf x}$ and individual rest masses $\mu$, with velocity/momentum coordinate ${\bf v}$/${\bf p}$ defined on a $D$-dimensional domain, evolving in time $t$ subject to some force ${\bf F}={\bf F}({\bf x},\,{\bf v},\,t,\,...)$. When $N$ is large, the ensemble is described by a continuous phase-space DF $f = f({\bf x},\,{\bf v},\, ...)$, whose evolution is generically determined by the Vlasov equation:
\begin{align}
\label{eqn:vlasov.general} \frac{\partial f}{\partial t} + {{\bf v}} \cdot \nabla f + {\bf F} \cdot \frac{\partial f}{\partial {\bf p}} = \left( \frac{\partial f}{\partial t} \right)_{\rm collisions}\ , 
\end{align}
where ${\bf F}$ collects all external forces, including e.g.\ gravity ${\bf F} = -\mu \nabla \Phi$ determined by the coupled Poisson equation ($\nabla^{2} \Phi = 4\pi\,G\,\rho = 4\pi\,G\,\int f\,d^{D} {\bf v}$, in terms of the mass density $\rho$). We will consider the specific example of gravity here but any external force could be applied (and we discuss other cases below). 
We will consider non-relativistic systems, although the generalization of our methods to relativistic cases is, in principle, straightforward. Here $f$ is the differential mass $dM$ per interval in coordinate space ${\bf x}$ and velocity space ${\bf v} = {\bf p} / \mu$:
\begin{align}
f \equiv \frac{d M}{d^{D}{\bf x}\,d^{D}\,{\bf v}}\ .
\end{align}
The right-hand side of Eq.~\ref{eqn:vlasov.general} represents the collision operator. This can be operator-split from the remaining equation, and solved separately, giving an update of the form 
\begin{align}
\Delta f_{\rm collisions} &= \left( \frac{\partial f}{\partial t} \right)_{\rm collisions} \Delta t 	
\end{align}
(note this can be arbitrarily high-order, we simply use the $\partial_{t} f \Delta t$ notation as symbolic of the operator-split update). 
For collisionless systems it vanishes. With this and the non-relativistic approximation we obtain: 
\begin{align}
\frac{\partial f}{\partial t} = -{\bf v} \cdot \nabla f - {\bf a}_{\rm ext} \cdot \frac{\partial f}{\partial {\bf v}} \ , 
\end{align}
where ${\bf a}_{\rm ext} = {\bf F}/\mu$ is the bulk external acceleration.

\subsubsection{External-Force Operator Split}
\label{sec:cbe.aext.split}

If ${\bf a}_{\rm ext}$ is independent of the velocity ${\bf v}$ (as it is for gravity), this term amounts to pure advection in ${\bf v}$ space. It is trivial to operator-split this as well, giving an update of the form
\begin{align}
f({\bf x},\,{\bf v}+\Delta {\bf v}[{\bf x}]) &= f({\bf x},\,{\bf v}) \\ 
\nonumber \Delta {\bf v}[{\bf x}] &\equiv {\bf a}_{\rm ext}({\bf x})\,\Delta t\ , 
\end{align}
which can be solved via usual numerical methods.

\subsubsection{Finite-Volume Discretization}
\label{sec:cbe.fv}

This leaves the phase-space advection equation:
\begin{align}
\frac{\partial f}{\partial t} = -{\bf v} \cdot \nabla f  \ . 
\end{align}
Since ${\bf v}$ is an independent variable here (not a function of position), this can be integrated over volume in the usual finite-volume fashion, giving the rate-of-change of the velocity DF $G \equiv dM/d^{D}{\bf v} = \int f\,d^{D}{\bf x}$ within a differential volume $\Omega$ around the point ${\bf x}$: 
\begin{align}
\frac{\partial }{\partial t}\,\left(  \frac{d M}{d^{D}{\bf v}}  \right)_{\Omega} &= \int_{\Omega} \frac{\partial f}{\partial t}\,d^{D}{\bf x} = -{\bf v} \cdot \int_{\Omega} \nabla f\,d^{D}{\bf x} \\
\nonumber & = -{\bf v}\cdot \int_{\partial \Omega} f\,d^{D-1}\,{\bf A} \ .
\end{align}
In the last step, we follow the usual finite-volume Godunov step of using Gauss's law to replace the integral over the $D$-dimensional volume $\Omega$ with the integral over the $D-1$-dimensional surface $\partial \Omega$. 
Around a mesh-generating point $a$ (the analog of an MC/PIC/NB ``particle''), in standard fashion, this is then discretized at a desired order by replacing the integral with a discrete sum over the geometric faces ${\bf A}_{ab}$ between $a$ and neighboring mesh-generating points $b$:
\begin{align}
\label{eqn:cbe.discrete} \frac{\partial }{\partial t}\,\left(  \frac{d m_{a}}{d^{D}{\bf v}} \right) \approx -{\bf v}\cdot \sum_{b} f^{\ast}_{ab}({\bf x}^{\ast}_{ab},\,{\bf v})\,{\bf A}_{ab}
= - \sum_{b} {\bf v}\cdot f^{\ast}_{ab}({\bf x}^{\ast}_{ab},\,{\bf v})\, {\bf A}_{ab} \ , 
\end{align}
where $f_{ab}^{\ast}({\bf x}_{ab}^{\ast},\,{\bf v})$ denotes the ``interface value'' of $f$ (with the desired quadrature rule for the order of the solver). 

\begin{figure*}
\plottwo{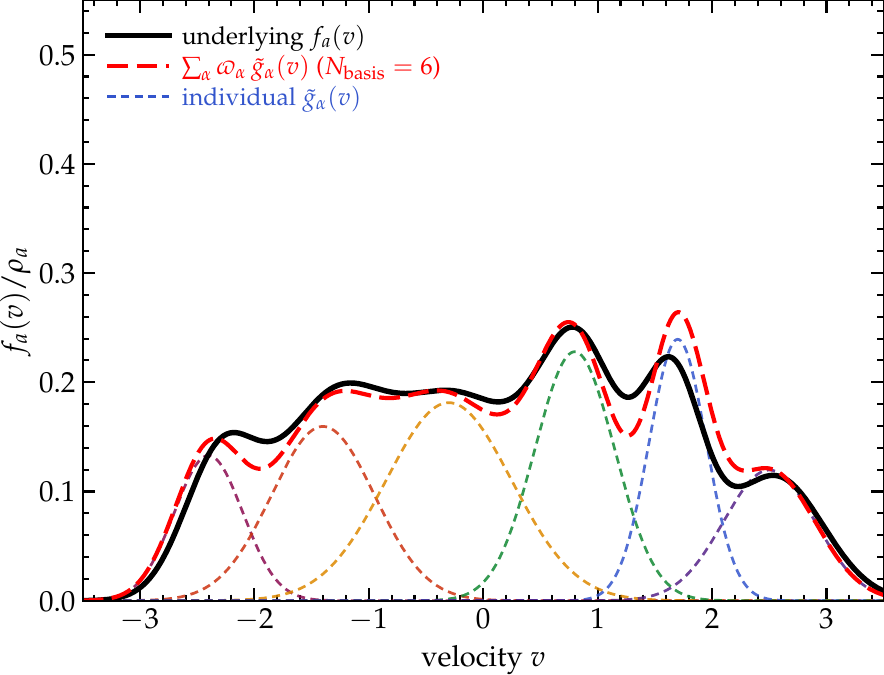}{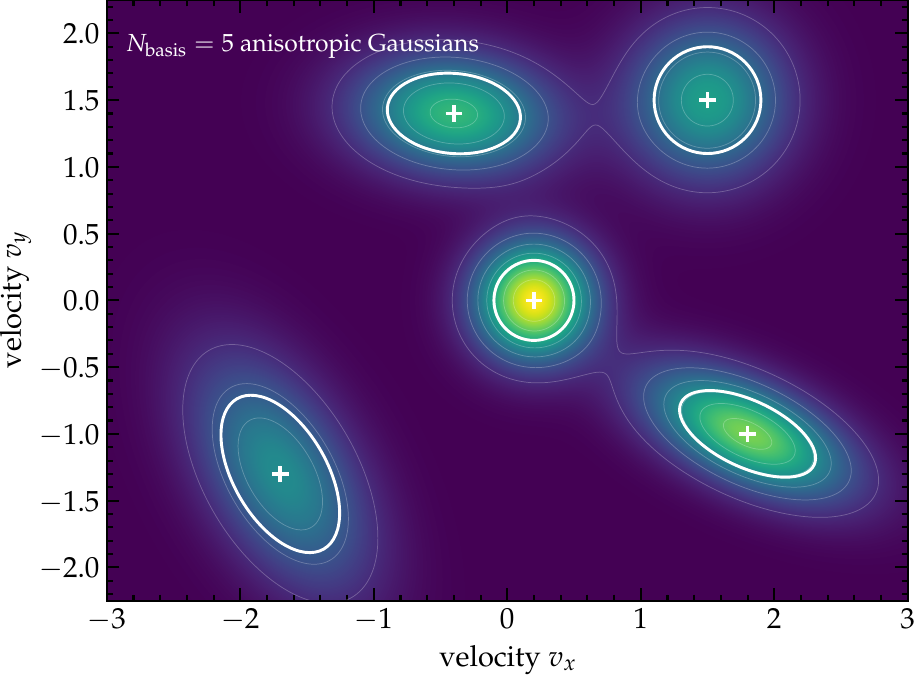}\vspace{-0.5cm}
\caption{
Schematic of the per-cell basis-function representation of the velocity DF $f_{a}({\bf v})$.
\textit{Left:} a stylized 1D distribution (heavy black) with multiple unresolved peaks, reconstructed (red dashed) as a sum of $N_{\rm basis}=6$ Gaussian components (thin dashed coloured lines), each with its own weight $\varpi_{\alpha}$, mean $\langle v_{\alpha}\rangle$, and dispersion $\sigma_{\alpha}$, all Lagrangian-evolved with the local flow.
\textit{Right:} a stylized 2D distribution $f_{a}(v_{x},v_{y})$ (filled contours of the mixture) built from $N_{\rm basis}=5$ anisotropic Gaussian basis components (here $5$ independent  streams), with $1\sigma$ ellipses (white) marking the orientation and aspect of each individual $\tilde{g}_{\alpha}$, and centres (white $+$) at $\langle{\bf v}_{\alpha}\rangle$. The same parameters carry over directly to $D=3$, where each basis carries $(m,\,{\bf p},\,{\bf T})$ with ${\bf T}$ a symmetric $3\times 3$ tensor encoding an anisotropic local dispersion. The fact that $\langle{\bf v}_{\alpha}\rangle$ and ${\bf S}_{\alpha}$ are themselves evolved Lagrangian variables -- not fixed on a global $(v_{x},v_{y},v_{z})$ grid -- is what makes this representation many orders of magnitude cheaper than a uniform-velocity-grid Vlasov scheme at comparable effective phase-space resolution.
\label{fig:cartoon.vdf}}\vspace{-0.3cm}
\end{figure*}

\subsection{Velocity-Space Partition: Basis Function Expansion}

To discretize the Vlasov equation, we must partition the independent variables into some discrete representation. First consider velocity space. 

\subsubsection{Basis Function Definitions}

Let $f_{a}({\bf v})\equiv V_{a}^{-1} \int_{\Omega} f({\bf x}\in \Omega_{a} ,\,{\bf v})\,d^{D} {\bf x}$ be the phase-space DF averaged in the domain $\Omega = \Omega_{a}$ surrounding the mesh-generating point $a$, and represent this with the basis functions $\tilde{g}_{\alpha}$, 
\begin{align}
f_{a}({\bf v}) &\equiv \rho_{a}\,\sum_{\alpha}\, \varpi_{\alpha}\,\tilde{g}_{\alpha}({\bf v})  \ , 
\end{align}
where $\rho_{a}$ is the mean mass density\footnote{By definition, 
$\rho_{a} \equiv m_{a} / V_{a}$ where $V_{a} = \int_{\Omega} d^{D}{\bf x}$ is the volume of the domain belonging to the mesh-generating point $a$, $m_{a} = \int_{\Omega} \rho({\bf x})\,d^{D}{\bf x} = \int_{\Omega}\,f\,d^{D}{\bf x}\,d^{D}{\bf v}$ is the total mass of $f$ contained in the domain, and $f_{a}({\bf v}) = \langle d m_{a} / d^{D} {\bf x} \,d^{D}{\bf v} \rangle \equiv V_{a}^{-1} \int_{\Omega} f({\bf x},\,{\bf v})\,d^{D}{\bf x}$ is the velocity DF averaged over the domain.} within $\Omega$, $\varpi_{\alpha}$ are dimensionless weights, and the basis functions are normalized to unity:
\begin{align}
	1 &= \int \tilde{g}_{\alpha}({\bf v})\,d^{D}{\bf v} \ .
\end{align}
By definition $\rho_{a} = \int f_{a}\,d^{D}{\bf v}$, so the weights must sum to unity, $\sum \varpi_{\alpha}=1$. We also define the first and second moments of each basis:
\begin{align}
\langle {\bf v}_{\alpha} \rangle &= \int {\bf v}\,\tilde{g}_{\alpha}({\bf v})\,d^{D}{\bf v} \ , \\ 
{\bf S}_{\alpha} &= \left( \int {\bf v} \otimes {\bf v} \,\tilde{g}_{\alpha}({\bf v})\,d^{D}{\bf v} \right) - \left( \langle {\bf v}_{\alpha} \rangle \otimes \langle {\bf v}_{\alpha} \rangle \right) \ .
\end{align}

Any velocity distribution can be represented with a sufficiently large set of basis functions; of course, for computational reasons we will use a finite set. This means our ``velocity resolution'' is limited by the number of basis functions, times the number of ``particles'' (finite-element mesh-generating points) used. Fig.~\ref{fig:cartoon.vdf} illustrates the kind of representation this enables \textit{at a single position in space} ${\bf x}_{a}$ (i.e.\ within a single spatial resolution element), with a complex multi-peaked $f_{a}(v)$ in 1D being captured by an $N_{\rm basis}=6$ Gaussian mixture, and an anisotropic, multi-stream 2D $f_{a}(v_{x},v_{y})$ captured by $N_{\rm basis}=5$ Gaussian components with non-trivial covariance. We discuss the meaning of this resolution further below.

Conceptually, we stress that the basis functions are \textit{not} required to be either (a) orthogonal, nor (b) complete (in the per-cell instantaneous sense). For (a), overlap is completely fine and creates no issues, since $f_{a}({\bf v})$ (the real object) is reconstructed identically. If two become strictly degenerate it creates a practical implementation challenge (and redundancy/waste), but that is trivially solved by merging/splitting as noted below. For (b), because the basis functions are adaptive \textit{in velocity space}, even if a given cell only has a few bases (so can only represent a few independent non-overlapping streams ``at a time''), all velocities are realizable in principle (it can always adapt to move $\langle {\bf v} \rangle$) if new material advects in with new velocities. They are thus automatically complete/spanning in a general sense. Of course, the \textit{accuracy} of the representation scales with the basis number, with a defined convergence order discussed below.

\subsubsection{Corresponding Conservative Variables}
\label{sec:Q.vs.U}

If we consider each basis function as an independent DF $f_{a,\,\alpha} \equiv \rho_{a}\,\varpi_{\alpha}\,\tilde{g}_{\alpha}({\bf v})$, then each has a set of conserved variables corresponding to the moments defined above, integrated over its domain in phase-space. In standard finite-volume fashion we will often explicitly work in these conserved variables: 
\begin{align}
m_{a,\,\alpha} &\equiv \int_{\Omega}\,\Delta f_{a,\,\alpha} = \varpi_{\alpha}\,m_{a} \ , \\ 
{\bf p}_{a,\,\alpha} &\equiv \int_{\Omega}\,({\bf v})\,\Delta f_{a,\,\alpha} = m_{a,\,\alpha}\,\langle {\bf v}_{\alpha} \rangle \ , \\ 
{\bf T}_{a,\,\alpha} &\equiv \int_{\Omega}\,({\bf v}\otimes{\bf v})\,\Delta f_{a,\,\alpha} = m_{a,\,\alpha}\,\left[{\bf S}_{\alpha} + \langle {\bf v}_{\alpha} \rangle\otimes \langle {\bf v}_{\alpha} \rangle \right]  \ , \\ 
\Delta f_{a,\,\alpha} &\equiv \rho_{a}\,\varpi_{\alpha}\,\tilde{g}_{\alpha}({\bf v})\,d^{D}{\bf x}\,d^{D}{\bf v} \ .
\end{align}
The total mass ($m_{a} = \sum m_{a,\,\alpha}$), momentum (${\bf p}_{a} = \sum {\bf p}_{a,\,\alpha}$), stress (${\bf T}_{a} = \sum {\bf T}_{a,\,\alpha}$), and energy ($E_{a} = {\rm Trace}({\bf T}_{a})/2 + U_{a}^{\rm gravity}$) on the domain are simply sums over the basis functions. Note that ${\bf T}$ is symmetric, so in $D=(1,\,2,\,3)$ dimensions this system has $(3,\,6,\,10)$ independent variables per basis function $\alpha$. 
We can continue to define higher-order conserved variables, e.g. ${\bf G}_{a,\,\alpha} \equiv \int_{\Omega}\,({\bf v}\otimes{\bf v} \otimes {\bf v})\,\Delta f_{a,\,\alpha}$; where to truncate depends on the basis definition.

\subsubsection{Closure Order and Choice of Bases}
\label{sec:Q.vs.U.choice}

We can of course continue to define higher-order moments. However, if our basis functions are chosen to be any functions which close at second-order (i.e.\ have higher-order moments which are entirely defined by the zeroth, first, and second moments), then these moments are sufficient to completely specify the basis functions. Since we will truncate our flux approximations at second-order in our study here, we go no further now. Second-order moments are also important because they are sufficient to ensure manifest mass, momentum, and energy conservation with a finite basis set. 

We stress that truncating at second order does \textit{not} mean we are unable to represent terms like skew or kurtosis in the DF, nor that we miss physical effects like heat flux/conduction in the weakly-collisional regime. Because we use multiple basis functions, in principle (with sufficient bases), any arbitrarily complex DF can be represented. In terms of reconstruction, introducing more complex basis functions is formally degenerate with simply adding additional (well-chosen) second-order basis functions. This is an important distinction from e.g.\ 13-moment methods in plasma physics, which we discuss further below. There, one represents the DF via a \textit{single} basis function, and generally assumes the DF is close-to-Maxwellian (often a good approximation for ionized plasmas owing to kinetic instabilities if the DF deviates too far). In that case a Gaussian basis with 3 additional moments physically corresponding to a (small/weakly non-Gaussian) skew (producing the reduced heat flux vector ${\bf q}_{i}$) is useful.

One is completely free, in our method, to choose bases which close at third or higher order, provided they can be specified via the set of evolved moments (or moments are added appropriately) -- one simply needs to calculate the correct fluxes for those moment definitions. We discuss the pros and cons of extending the methods here to higher-moments (like 13+ moment methods), versus including more basis functions, below. As one might expect, this is highly problem-dependent.

\section{Numerical Implementation \&\ Choices}
\label{sec:implementation}

Everything below outlines an implementation in detail: we attempt to specify throughout what is ``intrinsic'' to the method, versus what is simply a specific implementation choice we have adopted in our particular numerical examples.

\subsection{Volume (Configuration-Space) Partition}

The definition of the volume partition $\Omega$ and corresponding faces ${\bf A}_{ab}$ can, in principle, follow any partition method for fluid dynamics, around some set of mesh-generating points ${\bf x}_{a}$. For example, an Eulerian fixed or adaptive-mesh-refinement Cartesian grid is a simple option that lends itself well to high-order methods and idealized, homogeneous problems. But for the applications we'll study here, where there are cold streams in phase-space, Eulerian partitions involve large numerical diffusion as the streams advect across the grid and have to be constantly ``jumping across'' basis functions that may not be well-aligned. In this regime quasi-Lagrangian methods, e.g.\ moving mesh or meshless finite-volume methods are convenient.

\subsubsection{Specific Implementation: Meshless Finite Mass}
\label{sec:volume.partition.mfm}

Here we will follow the meshless finite mass (MFM) method developed in \citet{hopkins:gizmo} for \GIZMO\ (following earlier work by \citealt{lanson.vila:2008.meshfree.consistency,lanson.vila:2008.meshfree.convergence,gaburov:2012.public.moving.mesh.code}). The advantage of this is that (1) the method can define the partition $\Omega_{a}$, densities $\rho_{a}$ and faces ${\bf A}_{ab}$ from an arbitrary, moving, irregular set of generating points (naturally present if we are evolving subject to strong external or long-range forces), (2) the method is naturally second-order, and (3) it conserves element mass in each domain, which reduces MC/PIC/NB noise in the integration scheme and eliminates the need for particle merging/splitting.  

The derivations of these properties and expressions are in \citet{hopkins:gizmo}; here we simply quote them. The faces and densities are given by:
\begin{align}
\label{eqn:mfm.face.area.def} {\bf A}_{ab} &\equiv  \bar{n}_{a}^{-1}\,\bar{\bf q}_{b}({\bf x}_{a}) + \bar{n}_{b}^{-1}\,\bar{\bf q}_{a}({\bf x}_{b}) \ , \\ 
\label{eqn:mfm.face.area.def.sub1} \bar{\bf q}_{b}({\bf x}_{a}) &\equiv {\bf E}_{a}^{-1} \cdot {\bf x}_{ba}\, W({\bf x}_{ba},\,h_{a}) \ ,  \\
\label{eqn:mfm.face.area.def.sub2} {\bf E}_{a} &\equiv \sum_{c}\,({\bf x}_{ca} \otimes {\bf x}_{ca}) \,W({\bf x}_{ca},\,h_{a}) \ ,  \\
\label{eqn:n.ngb}\bar{n}_{a} &\equiv \sum_{c} W({\bf x}_{ca},\,h_{a}) = V_{a}^{-1} \ ,  \\ 
\label{eqn:h.ngb} \rho_{a} &\equiv m_{a}\,\bar{n}_{a} = m_{a}\,h_{a}^{-D} \ , 
\end{align}
where ${\bf x}_{ba} \equiv {\bf x}_{b} - {\bf x}_{a} $, ``$\cdot$'' and ``$\otimes$'' denote the inner and outer product, respectively. Eqs.~\ref{eqn:n.ngb}-\ref{eqn:h.ngb} form an implicit system for $h_{a}$ (solved iteratively as in \citealt{springel:entropy}). $W$ is a kernel function which should have various properties (symmetry, isotropy, smooth first derivatives, compact support, and Gaussian-like shape): here we adopt a standard cubic spline\footnote{We have briefly experimented with other kernels and see little difference in behavior. We have also applied a slightly different face estimator using the consistent reproducing kernel to define the face, following \citet{liu:1995.reproducing.kernel.particle.methods}, but again see little difference, as expected since the difference in consistency vs.\ the usual MFM face is at higher-order than our reconstruction in the flux computation below.} with maximum radius of compact support (analogous to the stencil for regular-grid methods) $H_{a}=2\,h_{a}$. 

\subsubsection{Motion of Mesh-Generating Points \&\ Faces}
\label{sec:volume.partition.mfm:motion}

In MFM, the mesh-generating points move strictly with the bulk center-of-mass velocity, ${\bf v}_{a} \equiv {\bf p}_{a} / m_{a}$, i.e.\  
\begin{align}
{\bf x}_{a} \rightarrow {\bf x}_{a} + \frac{{\bf p}_{a}}{m_{a}}\,\Delta t \ .
\end{align}
This is done using a standard leapfrog scheme (see \citealt{springel:gadget}). The volume partition (e.g.\ $V_{a}$, $h_{a}$) is therefore re-calculated each timestep (for more details on how to efficiently evolve $H_{a}$ to optimize this, see \citealt{hopkins:gizmo}). 

This means the faces ${\bf A}_{ab}$ move and deform with an effective face-centered velocity ${\bf v}_{F,\,ab}$ (usually $\sim v_{a} \sim v_{b}$). We therefore must re-write Eq.~\ref{eqn:cbe.discrete} in terms of the Lagrangian derivative moving with ${\bf v}_{F,\,ab}$: 
\begin{align}
\frac{d }{d t}\,\left(  \frac{d m_{a}}{d^{D}{\bf v}} \right)  &\equiv \sum_{b}  \frac{d }{d t}\,\left(  \frac{d m_{a,\,b}}{d^{D}{\bf v}} \right) \ , \\
\label{eqn:cbe.discrete.lagrangian} \frac{d }{d t}\,\left(  \frac{d m_{a,\,b}}{d^{D}{\bf v}} \right) &\equiv  -\left({\bf v} -{\bf v}_{F,\,ab}\right)\cdot f^{\ast}_{ab}({\bf x}^{\ast}_{ab},\,{\bf v})\,{\bf A}_{ab}\ .
\end{align}
Following \citet{hopkins:gizmo}, the face motions ${\bf v}_{F,\,ab}$ in MFM are \textit{defined} by the mean motion of the effective Lagrangian surface between the cells, meaning that the  net mass flux through each face vanishes identically (ensuring the mass per domain $\Omega$ is conserved to machine precision, independent of the time-integration scheme). 
As shown therein, this is second-order consistent (given the mesh-generating-point motion), and for a collisional fluid (e.g.\ hydrodynamics) with an HLLC-type solver corresponds to the face moving with the $S^{\ast}$ contact-wave. The generalization of this is to solve for the face velocity ${\bf v}_{F,\,ab}$ that ensures $d m_{a,b}/dt = 0$ integrated over all ${\bf v}$ from both sides of the face. The solution depends on how the fluxes are defined; we discuss this further below.

\subsection{Reconstruction}
\label{sec:reconstruction}

\subsubsection{General Considerations}
\label{sec:reconstruction:general}

A wide range of higher-order methods exist by which to reconstruct fluid quantities $Q$ -- e.g.\ primitive variables like $\rho,\,\langle {\bf v} \rangle, {\bf S}$ -- at the effective interaction faces or quadrature points (${\bf x}_{f,\,ab}$). Any of these can be combined with the basic method here, in principle. 
However, care is needed thinking about reconstruction here. 

The meaningful quantity one wishes to reconstruct is the DF $f^{\ast}_{ab}$ -- ${\bf v}$ is an independent variable like ${\bf x}$. If the DF $f({\bf x},\,{\bf v})$ is smooth and continuous, then the representation $f^{\ast} \approx \sum_{\rm bases} \rho^{\ast}\tilde{g}^{\ast}$ can be reconstructed via interpolation of the bases. At second-order spatial accuracy as we implement below, since our bases are fully-specified by their primitive $\rho_{\alpha}$, $\langle {\bf v} \rangle_{\alpha}$, ${\bf S}_{\alpha}$, it is sufficient to reconstruct those variables, associating each reconstructed $\rho^{\ast}_{ab,\,\alpha}\tilde{g}^{\ast}_{ab,\,\alpha}(\langle {\bf v}\rangle^{\ast}_{ab,\alpha},\,{\bf S}^{\ast}_{ab,\alpha})$ with its reconstruction base $\rho_{a,\alpha} \tilde{g}_{a,\,\alpha}(\langle {\bf v} \rangle_{a,\alpha},\,{\bf S}_{a,\alpha})$. But this means that there is more freedom in non-smooth flows. Because we solve a series of one-sided vacuum Riemann problems reflecting a collisionless flow, monotonicity is not a stability criterion in the usual sense -- an illustrative example being the case of propagation of a cold stream into true ``vacuum'' (what we call ``empty basis'' below), i.e.\ a region where such a stream does not exist. Obviously it does not make sense for the ``empty'' side (where the basis has not yet reached) any role in the gradient, since there is -- as of yet -- no basis function in that cell representing the stream before it propagates in, so its empty bases into which the stream will flow have undefined (vacuum) values of $\langle {\bf v} \rangle$. But it is critical to enforce \textit{realizability} of the reconstructed DF.

In practice, we have experimented with piecewise-linear and piecewise-constant reconstructions. Consistent with most fluid dynamics results, we find piecewise-constant is so diffusive on our test problems that it is not practically useful. Piecewise linear does much better. Higher order methods like PPM or WENO-type methods are conceptually straightforward, with the primary challenges being (1) memory overhead (since these must apply to all moments of all bases), relevant on many GPU architectures, and (2) ensuring realizability (amplified here by arbitrary velocity anisotropy).

\subsubsection{Piecewise-Linear Reconstruction at Faces}
\label{sec:reconstruction.linear}

In our default implementation, we consider a piecewise-linear reconstruction of $Q$ at the interface ${\bf A}_{ab}$ between $a$ and $b$, defined from each side for a given basis $\alpha$: 
\begin{align}
\label{eqn:Qface} Q_{ab,\,\alpha}^{\ast,\,a} &= Q_{a,\,\alpha} + \nabla Q_{a,\,\alpha} \cdot ({\bf x}_{f,\,ab} - {\bf x}_{a})\ ,
\end{align}
with the face center ${\bf x}_{f,\,ab} \equiv {\bf x}_{a} + \psi_{ab} ({\bf x}_{b} - {\bf x}_{a})$ located at the fractional distance $\psi_{ab}$ along the line connecting the centers of $a$ and $b$ \citep[to second-order accuracy given our MFM partition, with $\psi_{ab} \approx h_{b}/(h_{a}+h_{b})$, this is its effective geometric location; see][]{hopkins:gizmo}. The cell-centered gradients $\nabla Q_{a,\,\alpha}$ are computed by the same matrix-weighted least-squares (MLS) used for MFM hydrodynamics:
\begin{align}
\label{eqn:lsq} \nabla Q_{a,\,\alpha} &= \phi^{Q}_{a,\,\alpha}\, {\bf E}_{a}^{-1} \cdot \sum_{b} \left[ Q_{b,\,\alpha(b)} - Q_{a,\,\alpha} \right]\,{\bf x}_{ba}\, W({\bf x}_{ba},\,h_{a})\ ,
\end{align}
where ${\bf E}_{a}$ is the same kernel-moment matrix defined in Eq.~\ref{eqn:mfm.face.area.def.sub2} (saved for the face computation), and $\phi^{Q}_{a,\,\alpha}$ is a slope limiter.

\subsubsection{Slope-Limiting}
\label{sec:reconstruction.slopelim}

For stability, the gradients must be limited and must ensure realizability of the reconstructed DF (i.e.\ that ${\bf S}_{ab,\,\alpha}^{\ast,\,a}$ remains symmetric positive-definite [SPD] and $\rho_{ab,\,\alpha}^{\ast,\,a}$ remains positive-definite, corresponding to a positive-definite $f$). This is not trivial. 
But as usual one desires a less-aggressive limiter where not needed, as too-strong limiters can easily ensure stability but effectively reduce the method to piecewise-constant. 
For the simple implementation here, we attempt to balance these by mirroring the component-wise limiting\footnote{We have also experimented with a single $\phi^{Q}_{a,\,\alpha} \to \phi_{a,\alpha}\equiv {\rm MIN}[\phi^{Q}_{a,\,\alpha}]$ defined by the strongest component-wise limiter. Unsurprisingly this leads to much stronger numerical diffusion, akin to lower-order methods.} of the independent primitive fluid variables already standard in the MFM hydrodynamics \citep{hopkins:gizmo}, but err on the side of stronger limiters. An independent $\phi^{Q}_{a,\,\alpha}$ is applied to the reconstructed gradient of each primitive $Q \in \{\rho,\,\langle {\bf v} \rangle^{k},\,{\bf S}\}$. 

For $Q=\rho$, we adopt the following \citet{barth.jesperson:1989.upwind.schemes.for.unstructured.meshes}-style pairwise limiter, adapted from the standard global-stencil form to the per-basis moments here. For each matched neighbor pair, $\phi^{Q}_{a,\,\alpha}$ is the largest value $\in [0,1]$ such that
\begin{align}
\label{eqn:bj.limiter}
\min( Q_{a,\alpha}\ ,\ Q_{b,\alpha(b) }  ) \le Q_{ab,\alpha}^{\ast,\,a} \le \max ( Q_{a,\alpha}\ ,\ Q_{b,\alpha(b)} )
\end{align}
taking the minimum of $\phi^Q_{a,\,\alpha}$ over \textit{all} matched neighbors $b$.
This preserves monotonicity for smooth flows and, since $\rho_{a,\,\alpha} \to 0$ in the vacuum limit and $\rho_{a,\,\alpha} > 0$ elsewhere, is trivially positivity-preserving and correctly interpolates in the stable sense for cold flows into un-represented bases. 

For $Q=\langle {\bf v} \rangle$, there is no realizability concern, and we verify that our implementation remains stable in every problem we consider, with $\phi^{v_{k}}_{a,\,\alpha}=1$. However in the unlimited case a second problem can arise in discontinuities or disordered/noisy flows. If the gradients are large, the reconstructed values of $\langle {\bf v} \rangle_{a,\,\alpha_{1}},\, \langle {\bf v} \rangle_{a,\,\alpha_{2}}$ from two bases can ``cross'' between $a$ and the face $ab$, which causes our adopted neighbor bases-matching algorithm (\S~\ref{sec:pairing}) to flip assignments, which in turn further modifies the gradients and can lead to the incorrect basis ``receiving'' flux from neighbors. 
We therefore adopt a conservative limiter to prevent this: when the gradients are sufficiently large, such that the change in $\langle {\bf v} \rangle$ from the reconstruction becomes comparable to some inter-particle velocity scale $\varsigma$, we enforce smoothness/monotonicity with Eq.~\ref{eqn:bj.limiter}; otherwise we use the second-order gradient estimator.
We define $\varsigma_{k}\equiv {\rm max}(|\langle v_{a,\alpha,k}\rangle - \langle v_{a,k} \rangle|,\,\sigma_{a,\alpha,k},\,|\langle v_{b,\beta,k}\rangle - \langle v_{b,k} \rangle|,\,\sigma_{b,\beta,k})$ component-wise (so bulk velocities are normalized to cell-mean and preserve Galilean invariance), the threshold at a conservative $\epsilon_{v} \varsigma$ with $\epsilon_{v}=0.1$, and
\begin{align}
\label{eqn:bj.limiter.v}
\phi^{v_{k}}_{a,\,\alpha} = \begin{cases} 1 & {\rm if}\ |\nabla v_{k,\,a,\alpha}\cdot {\bf x}_{ba}| < \epsilon_{v}\,\varsigma_{k}  \\ \phi^{Q}_{a,\,\alpha}\ {\rm from\ Eq.~\ref{eqn:bj.limiter}} & {\rm otherwise}\ . \end{cases}
\end{align}
We show below (\S~\ref{sec:test.harmonic}) the $\langle {\bf v}\rangle$ limiter has a large effect on numerical dissipation. 

For $Q={\bf S}$, taking the gradient of each component $S^{k}$ with a monotonicity-preserving limiter can still produce unrealizable ${\bf S}_{ab,\,\alpha}^{\ast,\,a}$ (even if every cell-centered ${\bf S}$ is SPD). So we adopt the more conservative approach of defining a {single} $\phi^{S}_{a,\alpha}$, applied to all components $S^{k}$. This is defined as the largest value $\in [0,1]$ such that Eq.~\ref{eqn:bj.limiter} is satisfied for {every} component (in $a,\alpha$), {and} the reconstructed ${\bf S}_{ab,\,\alpha}^{\ast,\,a}$ remains SPD. This is solved using an iterative bisection algorithm: first taking the minimum $\phi^{S,\,{\rm guess}}_{ab,\alpha} \equiv \min [ \phi^{S,\,k}_{ab,\alpha} ]$ satisfying Eq.~\ref{eqn:bj.limiter} for all components $k$ for a given pair $ab$, then iteratively reducing this to find the largest value $\phi^{S}_{ab,\alpha} \in [0,\,\phi^{S,\,{\rm guess}}_{ab,\alpha}]$ such that the reconstructed ${\bf S}_{ab,\,\alpha}^{\ast,\,a}$ is SPD, then taking the minimum over all neighbors $\phi^{S}_{a,\alpha} \equiv \min_{b}[\phi^{S}_{ab,\alpha}]$. This enforces monotonicity and realizability. It is stronger than the minimal needed, and further improvements are definitely warranted.

\subsubsection{Matched-Basis Reconstruction Stencil}
\label{sec:reconstruction.matched}

Note that because the basis index $\alpha$ refers to a spatial-and-velocity-Lagrangian basis component, two (spatially) neighboring cells are not ensured to have exactly identical bases. The neighbor value $Q_{b,\,\alpha(b)}$ entering Eq.~\ref{eqn:lsq} must therefore use the basis index $\alpha(b)$ on side $b$ that is paired to $\alpha$ on side $a$ by the \textit{consistent} pairing operator described below (\S~\ref{sec:pairing}) and used in the actual flux computation.

\subsection{Flux Computation}
\label{sec:flux}

\subsubsection{Per-Basis Flux Integrals}
\label{sec:flux.integrals}

The flux of $f$ at a specific value of ${\bf v}$ through the face ${\bf A}_{ab}$ is given by Eq.~\ref{eqn:cbe.discrete.lagrangian}.
This is just the linear sum of the flux from each basis component $\tilde{g}_{\alpha}$: 
\begin{align}
\label{eqn:cbe.discrete.lagrangian.decomp}
\frac{d }{d t} &\,\left(  \frac{d m_{a,\,b}}{d^{D}{\bf v}} \right) \equiv  -\left({\bf v} -{\bf v}_{F,\,ab}\right) f^{\ast}_{ab}({\bf x}^{\ast}_{ab},\,{\bf v}) \cdot {\bf A}_{ab}  \\ 
\nonumber &= -\left({\bf v} -{\bf v}_{F,\,ab}\right) \left[ \sum_{\alpha} \rho^{\ast,a}_{ab,\alpha} \tilde{g}_{ab,\alpha}^{\ast,a} + \sum_{\beta} \rho^{\ast,b}_{ba,\beta} \tilde{g}_{ba,\beta}^{\ast,b}   \right] \cdot {\bf A}_{ab} 
\end{align}
where $\rho^{\ast,a}_{ab,\,\alpha} \equiv \rho^{\ast,a}_{ab} \varpi_{ab,\alpha}^{\ast,a}$ and $\tilde{g}_{ab,\alpha}^{\ast,a}({\bf v}) \equiv \tilde{g}_{a,\alpha}({\bf v}\, | \langle {\bf v} \rangle^{\ast,a}_{ab,\alpha} \, , \, {\bf S}_{ab,\alpha}^{\ast,a},\,\dots )$ are the reconstructed density and velocity DFs,  (and the sum goes over those reconstructed from the $a$ and $b$ side). 

As noted above, we want to define a strictly conservative method, so we want to consider fluxes of the conservative variables ${\bf U}V \equiv \{ m,\, {\bf p},\, {\bf T} \}$. This is given by integrating over ${\bf v}^{n} d^{D} {\bf v}$ ($m,\,{\bf p},\,{\bf T}$ corresponding to $n=0,\,1,\,2$), i.e.\ 
\begin{align}
\label{eqn:conserved.update.total} \frac{d ({\bf U} V)_{a}}{d t} &= \sum_{b} \frac{d ({\bf U} V)_{a,\,b}}{d t}  \\ 
\nonumber & =\sum_{b} \left[ \sum_{\alpha} \frac{d ({\bf U} V)_{ab,\,\alpha}}{d t} \Bigr|_{b} + \sum_{\beta} \frac{d ({\bf U} V)_{ab,\,\beta}}{d t} \Bigr|_{b} \right] \\ 
\nonumber & = - \sum_{b} \left[ \sum_{\alpha} {\bf F}^{\ast}_{a \rightarrow b,\alpha} + \sum_{\beta} {\bf F}^{\ast}_{b \rightarrow a,\beta}  \right] \cdot {\bf A}_{ab}
\end{align}
where the flux from $a$ to $b$ associated with $a$'s basis $\alpha$ is defined by 
\begin{align}
\label{eqn:flux.integral} {\bf F}^{\ast\,(n)}_{a \rightarrow b,\alpha} &= \int {\bf v}^{(n)}  \left({\bf v} -{\bf v}_{F,\,ab}\right) f^{\ast}_{ab,\,\alpha}({\bf x}^{\ast}_{ab},\,{\bf v}) \, d^{D} {\bf v} \ .
\end{align}
For properly defined fluxes, this is automatically conservative of mass, momentum, and energy on equal timesteps, because the $d({\bf U}V)_{a,\,b}/dt = -d({\bf U}V)_{b,\,a}/dt$ are antisymmetric. Like with most existing fluid-dynamics and gravity methods, momentum and energy conservation becomes integration-accurate for adaptive individual timesteps, unless one carries additional variables to represent the shorter-timestep active-side updates at all times. In either case mass is conserved to machine precision, owing to our definition of the face motion.

\subsubsection{Direct and Lax-Friedrichs Type Fluxes}

If we simply insert $f^{\ast}_{ab,\,\alpha} \equiv \rho^{\ast,a}_{ab,\alpha} \tilde{g}_{ab,\alpha}^{\ast,a}$ and evaluate Eq.~\ref{eqn:flux.integral} for the system of primitive variables: 
\begin{align}
{\bf U} &\equiv 
\begin{pmatrix}
\rho \\
\rho\,\langle {\bf v} \rangle \\
\rho\,\langle {\bf v}\otimes{\bf v} \rangle
\end{pmatrix}
= 
\frac{1}{V} \begin{pmatrix}
m \\
{\bf p} \\
{\bf T}
\end{pmatrix}, 
\end{align}
we immediately obtain
\begin{align}
\label{eqn:flux.direct}
{\bf F}_{a\rightarrow b,\alpha}^{\ast} 
&= \rho
\begin{pmatrix}
\nonumber {\bf u} \\
\langle {\bf v} \rangle \,{\bf u} + {\bf S}  \\
\langle {\bf v}\otimes{\bf v} \rangle \,{\bf u} + \langle {\bf v} \rangle \otimes {\bf S} + {\bf S}\otimes\langle {\bf v}\rangle
\end{pmatrix}^{\ast,a}_{ab,\,\alpha} \ , \\
 {\bf u} &\equiv \langle {\bf v}\rangle - {\bf v}_{F,ab} \ , \ \langle {\bf v} \otimes {\bf v} \rangle = \langle {\bf v} \rangle \otimes  \langle {\bf v} \rangle + {\bf S} \ .
\end{align}
The $()^{\ast,a}_{ab,\alpha}$ notation means these indices should be applied to all variables on the right-hand side. 
The expressions for the $m$ and ${\bf p}$ fluxes are valid for any form of the bases; for ${\bf T}$ we assume only that the basis is even in $({\bf v}-\langle {\bf v}_{\alpha} \rangle)$ consistent with our earlier representation.

However, since all we are doing here is integrating a series of simple hyperbolic equations, it is well-known that this flux will be numerically unstable, unless we add some upwind dissipation terms. We confirm this in our own implementations. One option, akin to smoothed-particle-hydrodynamics methods, would be to add some sort of ``artificial viscosity'' to the problem to cure the instability. 
For example, in 13-moment methods for plasma physics, one often combines a more diffusive HLL or GLF-like solver with an artificial diffusion term and an artificial collisionality term (to damp the DF back to quasi-thermal; \citealt{torrilhon:2010.hyperbolic.moment.equations.pearson.iv.kinetic.gas,mcdonald.torrilhon:2013.affordable.robust.maximum.entropy.moment.closures,cai.fan.li:2014.globally.hyperbolic.regularization.grad.moment.system,miller.shumlak:2016.multispecies.13.moment.moderately.collisional.plasma}). 
But this has many well-known challenges and potential pitfalls, especially for a collisionless problem like this one (where there may be no physical viscosity/damping/collisionality). 

Even a lower-order method suffers from this. Consider (purely for illustrative purposes) a reconstruction which assumes a pairwise-taper (e.g.\ $q = q_{a}(x_{b}-x)/(x_{b}-x_{a})$, along the line connecting ${\bf x}_{a}$, ${\bf x}_{b}$) for $\rho$ with constant specific properties, giving: 
$\rho^{\ast,a}_{ab,\,\alpha} \rightarrow (1-\xi_{S}) \psi_{ab} \rho_{a,\alpha}$, 
$\langle {\bf v} \rangle^{\ast,a}_{ab,\,\alpha} \rightarrow \langle {\bf v} \rangle_{a,\alpha}$, 
${\bf S}^{\ast,a}_{ab,\,\alpha}\rightarrow {\bf S}_{a,\alpha}$, ${\bf u}^{\ast,a}_{ab,\,\alpha} \rightarrow (1+\xi_{m})\,(\langle {\bf v} \rangle_{a,\alpha} - {\bf v}_{F,ab})$. 
Here the terms $\xi_{m}$ and $\xi_{S}$ are the upwind numerical dissipation terms; their placement is motivated by anisotropic SPH solvers well-studied in the elastodynamics literature (which use a similar reconstruction, hence our reference; see \citealt{deng:giant.impact.sim.mfm}), and some tedious algebra (for our basis functions below) gives $\xi_{S} = c_{e}/(S_{M}+c_{e})$, $\xi_{m} = c_{e}/v_{A}$, with $S_{M} \equiv v_{A} + c_{e}^{2}/(v_{A} + c_{e})$, $v_{A} \equiv |(\langle {\bf v} \rangle_{a,\alpha} - {\bf v}_{F,ab}) \cdot \hat{\bf A}_{ab}|$, and $c_{e}^{2} \equiv \hat{\bf A}_{ab}^{\rm T} \cdot {\bf S}_{a,\alpha} \cdot \hat{\bf A}_{ab}$. 
If we insert this in Eq.~\ref{eqn:flux.direct} with $\xi_{m}=\xi_{S}=0$, it is both low-order and (still) unstable. If we use this with the above $\xi$ values, it is stable, but extremely diffusive (as expected).

\subsubsection{One-Sided Collisionless Riemann Solution (Default)}
\label{sec:flux.onesided}

A better solution is to consider an actual upwind Riemann problem solution, and there is a simple generalization of known methods. For collisionless or weakly-collisional systems (where the collisional deflection length or mean free path is large compared to the effective thickness of the face/resolution), then on an upwind face, the matched-basis state on the opposite side carries no incoming component along $\hat{\bf A}_{ab}$ that can ``feed back.'' So the symmetric central-difference dissipation built into an HLL or GLF solver is spurious. The natural conclusion is to treat the right state in each per-pair Riemann problem as a ${\bf v}$-space vacuum from the perspective of the outgoing basis on the upwind side. 
Essentially, the upwind solution in the weakly-collisional regime is the linear sum of the upwinded vacuum solutions ``from each side.'' This makes computing an \textit{exact} upwind Riemann problem solution trivial -- one simply computes the exact integral flux ``out of'' each basis $\alpha$ ``into'' each basis $\beta$ (e.g.\ computing the flux integrals from the reconstructed primitives over the velocities projected relative to their position and direction versus the moving face which will flow ``out''). 
This is therefore our default flux formulation adopted. But to compute this rigorously, we \textit{do} require specifying the functional form of the basis functions $\tilde{g}_{\alpha}({\bf v})$.

\subsubsection{Choice of Basis Function \&\ Exact Gaussian Riemann Solution}
\label{sec:flux.gaussian}

We desire a second-order function with defined moments. To be fully closed, the functions must be even in $({\bf v}-\langle {\bf v}_{\alpha} \rangle)$. One obvious function which can be constructed that satisfies these criteria is the three-dimensional Gaussian function:
\begin{align}
\tilde{g}_{\alpha}({\bf v}) &= \left[ {{\rm Det}(2\pi\,{\bf S}_{\alpha})} \right]^{-1/2}\,\exp^{-\mathcal{F}_{\alpha}({\bf v})}  \ , \\ 
\nonumber \mathcal{F}_{\alpha} &\equiv \frac{1}{2}\,\left({\bf v}-\langle {\bf v}_{\alpha} \rangle \right)^{\rm T}\,{\bf S}_{\alpha}^{-1}\,\left({\bf v}-\langle {\bf v}_{\alpha} \rangle \right) \ .
\end{align}
This is a reasonable basis, in principle, and it also has the advantage that it reduces \textit{exactly} in the collisional limit to hydrodynamics or MHD as desired, because it becomes a Maxwellian. 
It is certainly possible that using different basis functions might be more efficient (i.e.\ require fewer evolved variables to represent the same information) or more accurate in some circumstances (if e.g.\ there is a large skew in the real velocity DF), but we defer an exploration of this to future work.

With this, it is possible to evaluate an exact one-sided kinetic flux into the velocity-space vacuum, since for Gaussian basis functions the half-space moment integrals close in terms of standard error functions. 
Let $\hat{\bf x} \equiv \hat{\bf A}$ be the face-normal, and define the upwind normal-flow velocity $u_{x} \equiv (\langle {\bf v} \rangle_{\rm L} - {\bf v}_{f}) \cdot \hat{\bf A}$, as well as ${\bf S}_{x} \equiv {\bf S} \cdot \hat{\bf A}$, and  
\begin{align}
\sigma_{x} &\equiv (\hat{\bf A} \cdot {\bf S} \cdot \hat{\bf A})^{1/2}, \quad 
\tilde{u} \equiv {u_{x} \over \sigma_{x}}, \quad
\tilde{w} \equiv u_{x} \,\zeta + \sigma_{x}\, \omega  , \\
\nonumber \zeta &\equiv {1\over 2}\,{\rm erfc}\!\left(-{\tilde{u}\over\sqrt{2}}\right), \quad
\omega \equiv {1\over\sqrt{2\pi}}\,e^{-\tilde{u}^{2}/2}, 
\end{align}
so that $\zeta$ and $\omega$ are the standard normal CDF and PDF, respectively. Then the exact one-sided fluxes in the standard face-normal component form are given by
\begin{align}
\label{eqn:gaussian.flux}
\mathcal{F}_{\rm G} &= \rho\,
\begin{pmatrix}
\tilde{w} \\
\langle {\bf v} \rangle \,\tilde{w} + {\bf S}_{x}\,\zeta  \\
\langle {\bf v}\otimes{\bf v} \rangle \,\tilde{w} + \left( \langle {\bf v} \rangle \otimes {\bf S}_{x} + {\bf S}_{x}\otimes\langle {\bf v}\rangle \right) \zeta + \dfrac{{\bf S}_{x}\otimes {\bf S}_{x}}{\sigma_{x}}\,\omega
\end{pmatrix}.
\end{align}
In the cold-outflow/converging limit $\tilde{u} \to +\infty$: $\zeta \to 1$, $\omega \to 0$, $\tilde{w} \to u_{x}$, and $\mathcal{F}_{\rm G} \to \rho u_{x}$. 
In the cold-receding limit $\tilde{u} \to -\infty$, $\mathcal{F}_{\rm G} \to {\bf 0}$ (exponentially suppressed Gaussian tail). 

Note that the terms break into three groups with simple physical interpretations. Those in $\tilde{w}$ represent advection (the dependence of $\tilde{w}$ on $\sigma_{x}$ just reflects the finite-velocity width of the basis). The ${\bf S}_{x} \zeta$ terms in momentum and energy are the ``pressure'' and ``PdV'' or ``adiabatic'' (work/stress-tensor energy transfer) terms -- even if the fluid is collisionless, it still has a stress tensor providing support in e.g.\ a Jeans sense. The last term in ${\bf S}_{x} \otimes {\bf S}_{x} \omega$ is akin to an effective conductivity.

We can immediately distinguish the numerical upwinding terms from comparison with Eq.~\ref{eqn:flux.direct}. This is especially simple if we assume a stationary (non-moving) equilibrium: then from ``each side'' $\mathcal{F}_{G} \to \rho\,(\sigma_{x}\omega_{0},\,{\bf S}_{x}/2,\,{\bf S}\sigma_{x}\omega_{0} + \sigma_{x}^{-1} {\bf S}_{x}{\bf S}_{x} \omega_{2})$ (note here we have $\zeta\to1/2$, true for any symmetric basis in this limit, and distinguish the prefactors $\omega_{0}$ and $\omega_{2}$ as these can depend \textit{weakly} on the shape of the basis function, with values $=(0.4,0.4)$ here or $=(0.43,\,0.22)$ for the top-hat basis below), while the \textit{single-valued} $\mathcal{F}^{\ast}_{\rm direct} \to \rho\,(0,\,{\bf S}_{x},\,\mathbf{0})$. These might appear quite different, but note that if we have a sufficiently smooth DF, such that the spatial reconstructions of $f_{ab}$ \textit{agree} from the left and right sides of the face, then the signed terms are such that the mass flux and conduction terms (those in $\omega_{0,2}$) calculated from the matched bases on each side exactly cancel, while the signed ${\bf S}_{x}/2$ term adds from each side, giving $\mathcal{F}_{G,\,L}+\mathcal{F}_{G,\,R} \to \mathcal{F}^{\ast}_{\rm direct}$ exactly as intended. So these terms only act (like any other well-posed numerical dissipation terms) when the reconstruction produces a discontinuity at the face. For smooth flows, since we implement a piecewise-linear reconstruction, this means the dissipation/upwinding enters at second-order, e.g.\ for conduction $\mathcal{O}(\Delta x^{2} \partial^{2}[ \rho\sigma_{x}^{3}]/\partial r^{2})$. 

The exact Gaussian basis has, formally, particles at infinite velocities (since the Gaussian tail extends to arbitrarily large $|{\bf v}|$), but one can generalize this with appropriate basis function choice and construction to the relativistic Riemann problem. For the non-relativistic case here, the contribution of the tail to the fluxes (the only place it appears causally) is bounded by the largest eigenvalue of the Jacobian, which is well-understood for a single-moment Gaussian (Maxwellian) 10-moment method, giving the signal speed $|u_{x}| + \chi\,\sigma_{x}$ with $\chi = \sqrt{3}$ for advective and ``pressure'' terms, \textit{if} we adopted a non-upwinded HLL or GLF-type flux. However for the heat-flux term \textit{with the higher-order reconstruction and less-diffusive (physically minimal) upwinding}, this is not sufficient to always capture the signal speed of communication via the heat flux ``tail'' communicating (especially if there are ``near-empty'' basis functions into which this heat flux channels energy). In \textit{practice}, we find that because the flux diminishes rapidly (by definition) for high-$\sigma$ fluctuations, adopting a ``safety factor'' increasing $\chi$ to $\sim 4-5$ is sufficient to stabilize the method in the CFL criterion below, when also coupled to our conservative realizability repair methods. But this not only requires shorter timesteps, it does pose a risk of unphysical heat fluxes arising: consider a simple 1D Gaussian where only the ``tail'' at some large $v>0$ is lost to heat flux. After some time, the physical basis would be a skew-Gaussian without said tail to $v>0$, and no heat flux. But because we re-project the updated moments onto a new Gaussian, a new tail is ``generated'' by the re-projection, producing an additional (unphysical) heat flux.

\subsubsection{Alternative Approximate Solvers}
\label{sec:flux.approx}

A variety of exact and approximate alternatives can be derived from different assumptions for the basis functions, which we illustrate here. 
As noted above, a disadvantage of Gaussian bases is that they lack compact support. So instead, consider a ``top hat'' distribution in velocity space (essentially a constant-density in $d^{3}{\bf v}$ parallelogram) defined (after rotating into the diagonal coordinate frame) over $-\sqrt{3 S_{ii}} < v_{i} - \langle {\bf v} \rangle_{i} < \sqrt{3 S_{ii}}$. The expressions in arbitrary dimensions for any anisotropy and ``slice'' (direction of $\hat{\bf A}$ and offset ${\bf v}_{F}$) do not have simple representations, but it is worth noting that for second-order-closed bases and a one-sided collisionless Riemann problem, the dimensional form of Eq.~\ref{eqn:gaussian.flux} is complete. The only difference between different assumptions is the form of the ``weight functions'' $\tilde{w}$, $\zeta$, $\omega$ associated to the advective, pressure, and conductive fluxes. So, akin to approximate solvers like HLLC, we can approximate these in higher-dimensions with expressions exact for the 1D Riemann problem. 

Defining $c_{x} \equiv \sqrt{3} \sigma_{x}$ this gives: 
\begin{align}
\label{eqn:hllc.generic}
\mathcal{F}_{\rm TH} &\equiv
\begin{cases}
\mathcal{F}_{\rm TH,0}   \hfill & \ ( c_{x} \le u_{x} ) \\
\mathcal{F}_{\rm TH,1}  \hfill & \ ( -c_{x} < u_{x} < c_{x} ) \\
{\bf 0}   \hfill & \ ( u_{x} \le -c_{x} ) 
\end{cases} .
\end{align}
The piecewise definition of $\tilde{g}$ divides the solution into a fan with: 
\begin{align}
\label{eqn:hllc.F0}
\mathcal{F}_{\rm TH,0} &= \rho\,
\begin{pmatrix}
u_{x} \\
\langle {\bf v} \rangle \,u_{x} + {\bf S}_{x}  \\
\langle {\bf v}\otimes{\bf v} \rangle \,u_{x} + \langle {\bf v} \rangle \otimes {\bf S}_{x} + {\bf S}_{x}\otimes\langle {\bf v}\rangle
\end{pmatrix}, \\
\nonumber 
\mathcal{F}_{\rm TH,1}  &= \rho\, 
\begin{pmatrix}
\tilde{w}^{\prime} \\
\langle {\bf v} \rangle \,\tilde{w}^{\prime} + {\bf S}_{x}\,\zeta^{\prime}  \\
\langle {\bf v}\otimes{\bf v} \rangle \,\tilde{w}^{\prime} + \left( \langle {\bf v} \rangle \otimes {\bf S}_{x} + {\bf S}_{x}\otimes\langle {\bf v}\rangle \right) \zeta^{\prime} + \dfrac{{\bf S}_{x}\otimes {\bf S}_{x}}{\sigma_{x}}\,\omega^{\prime}
\end{pmatrix}, \\ 
\nonumber \tilde{w}^{\prime} &\equiv  \frac{c_{x}}{4}\left(1+u^{\prime} \right)^{2} , \quad u^{\prime} \equiv \frac{u_{x}}{c_{x}} , \\ 
\zeta^{\prime} &\equiv \frac{(2-u^{\prime})}{4} \left(1+u^{\prime} \right)^{2} , \quad \omega^{\prime} \equiv \frac{\sqrt{3}}{8}\left(1- u^{\prime 2} \right)^{2} . 
\end{align}
In other words, $\mathcal{F}_{\rm TH,0} = \mathcal{F}_{\rm G}\left(\tilde{w},\zeta,\omega \to u_{x},1,0 \right)$ and 
$\mathcal{F}_{\rm TH,1} = \mathcal{F}_{\rm G}\left(\tilde{w},\zeta,\omega \to \tilde{w}^{\prime},\zeta^{\prime},\omega^{\prime} \right)$. 

This is useful as a contrast to Eq.~\ref{eqn:gaussian.flux}, given that this is close to the most-compact centrally-peaked distribution with the same moments $\rho$, $\langle {\bf v} \rangle$, ${\bf S}$.  Direct comparison of the fluxes or ``effective'' $\tilde{w},\,\zeta,\,\omega$ (advection/pressure/conduction) terms shows that the advection \&\ pressure terms scale nearly identically up to $\mathcal{O}(\%)$ level in Eqs.~\ref{eqn:gaussian.flux} \&\ \ref{eqn:hllc.F0}, with the major difference being compact support (going to zero outside of domain), which leads to the well-defined maximum ($|u_{x}| + c_{x}$) and minimum signal speeds. However, the ``conduction'' term is quantitatively quite different (albeit qualitatively similar): it is a factor of $\sim 2$ lower at its peak ($u_{x} = 0$), and tapers exactly to nil at $|u_{x}| \ge c_{x}$, while $\omega$ decays as a Gaussian in $u_{x}/\sigma_{x}$. More importantly, the form means that the ``re-projection problem'' of Gaussian bases in \S~\ref{sec:flux.gaussian} is eliminated: once the slice of the parallelogram producing a physical heat flux has moved out of a cell, the updated parallelogram correctly produces no overlap, and this term vanishes. 

Because it retains compact support -- giving a well-defined finite maximum signal speed $|u_{x}| + c_{x}$ (with $c_{x} = \sqrt{3}\,\sigma_{x}$) and truncating the conductive flux exactly for $|u_{x}| \ge c_{x}$ -- while reproducing the advective and pressure fluxes of the exact Gaussian to $\mathcal{O}(\%)$, we adopt this top-hat one-sided solver as our default in the tests and applications below, retaining the exact Gaussian solver of \S\ref{sec:flux.gaussian} as a reference cross-check (the two are compared directly in \S\ref{sec:test.free.slot.ab}). The compactly-truncated conductive flux is important in practice: the non-compact Gaussian conduction term ($\propto {\bf S}_{x}\otimes{\bf S}_{x}\,\omega/\sigma_{x}$, which diverges as $\sigma_{x}\rightarrow0$ for cold streams) drives a stronger core contraction in the 3D equilibrium tests of \S\ref{sec:test.hernquist} that the top-hat truncation removes.

\subsection{Face Velocity ${\bf v}_{F}$ for the Chosen Partition}
\label{sec:vF.policy}

Per \S~\ref{sec:volume.partition.mfm:motion}, we now have what is needed to compute the face velocity ${\bf v}_{F,ab}$. This is trivial for some partition choices, but for our MFM-like volume partition it is defined as the ${\bf v}_{F,\,ab}$ of the mean Lagrangian frame through which net mass flux vanishes (hence requiring a flux definition): 
\begin{align}
\label{eq:mdot.face}
\frac{d m_{a,\,b}}{dt } &\equiv \left[ \sum_{\alpha} {\bf F}_{a\rightarrow b,\alpha}^{\ast,\,(m)} +  \sum_{\beta} {\bf F}_{b\rightarrow a,\beta}^{\ast,\,(m)}  \right] \cdot {\bf A}_{ab} = 0\ .
\end{align}
At second order, the transverse component of ${\bf v}_{F,\,ab}$ never enters the fluxes or evolution, so we only need to calculate the normal component $v_{F,\,n} \equiv {\bf v}_{F,\,ab} \cdot \hat{\bf A}_{ab}$. So we just need to solve Eq.~\ref{eq:mdot.face} given our flux definitions at each face for the scalar $v_{F,\,n}$.

For the exact Gaussian fluxes $\mathcal{F}_{\rm G}$, this is $\sum_{\alpha,\,\beta} \rho\, \tilde{w} = 0$, or
\begin{align}
\label{eqn:vF.analytic}
v_{F,\,n} &= \frac{\sum_{\alpha,\beta}  \rho\,( \sigma_{x} \,\omega + \zeta \langle {\bf v}\rangle_{x})}{\sum_{\alpha,\beta}  \rho\, \zeta}
\end{align}
-- i.e.\ a simple weighted-average of the left and right state velocities (over bases), plus a term in the non-cold limit in $\sigma_{x}$ which represents the upwind ``contact wave-like'' motion. 
The only non-trivial aspect of this is that the face velocity appears implicitly in $u_{x}$, and therefore the weights ($\zeta$ and $\omega$), in a nonlinear fashion. We therefore determine $v_{F,\,n}$ at each face by directly solving $\Sigma F_{m}(v_{F,\,n}) = 0$ via a one-dimensional bracketed Brent's-method root-find \citep{brent:1973.algorithms.minimization.without.derivatives}, which can be applied to any flux definition.

\subsection{Pairing of Bases Across Faces}
\label{sec:pairing}

\begin{figure}
\plotone{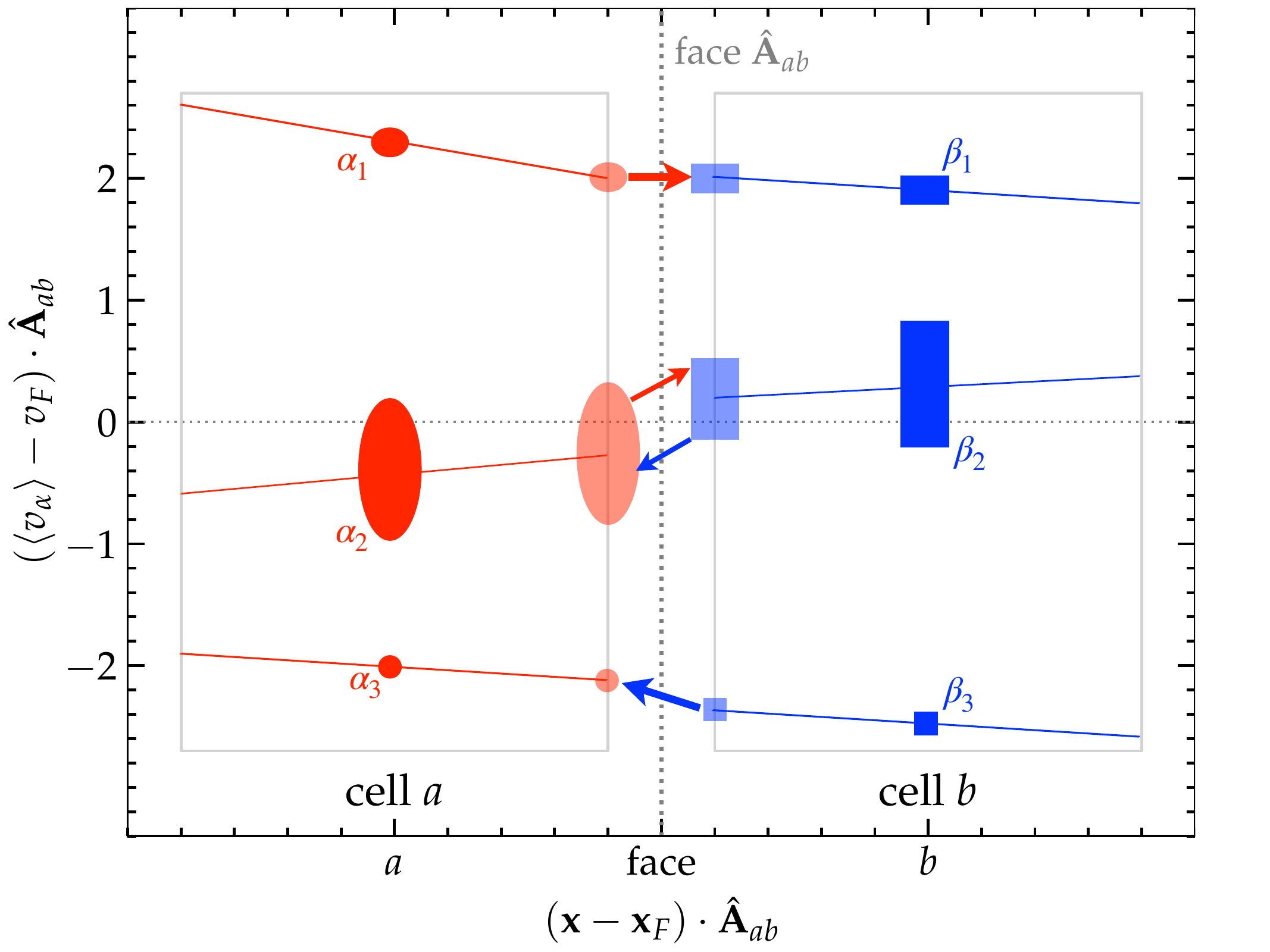}{1.0}
\vspace{-0.5cm}
\caption{Cartoon of the basis-pairing, reconstruction, and fluxes at a face $\hat{\bf A}_{ab}$ between adjacent cells $a$ \&\ $b$ (\S~\ref{sec:pairing}), in the face frame ($\hat{\bf A}_{ab}$ projection).
Axes show the position and velocity along this axis, relative to the face (at ${\bf x}_{F}$, ${\bf v}_{F}$). 
Each cell carries bases: red circles ($\alpha$) in $a$, blue squares ($\beta$) in $b$, with the mean $\langle v \rangle$ shown and vertical spread proportional to the dispersion/width in $v$. 
Lines show the piecewise-linear reconstruction adopted here: gradients are constructed between matched bases in neighbors to extrapolate the DF to the face from both sides ($L$ and $R$ states). 
The pairing is done on the extrapolated states based on continuity in phase-space of the bases. 
Arrows show the sense of the fluxes: when $v > v_{F}$, mass flows $L\to R$ (and vice-versa). For the middle basis, the spread means the basis represents some material with $v > v_{F}$ and some with $v < v_{F}$, so there is flux $L\to R$ and $R\to L$. The fluxes follow the one-sided vacuum Riemann solutions in \S~\ref{sec:flux.onesided}, computed in Godunov fashion over conserved quantities (mass, momentum, stress) ensuring manifest conservation. 
\label{fig:cartoon.pairing}}
\end{figure}

A subtle but important issue, which has no exact analog in standard finite-volume hydrodynamics, arises in the bookkeeping of where the flux at a face goes (Fig.~\ref{fig:cartoon.pairing}). The conservative update transfers mass, momentum, and stress out of some basis component $\alpha$ on side $a$ and into some matched basis component $\beta = \beta(\alpha)$ on side $b$. Like most Godunov methods, conservation is enforced by requiring the $-{\bf F}$ subtracted from $\alpha$ on $a$ (``flux out'') and the $+{\bf F}$ added to $\beta$ on $b$ (``flux in'') to be identical. In an infinite (complete) basis, this is unambiguous: the ``correct'' matched component on $b$ is the one with $\langle {\bf v}_{\beta} \rangle = \langle {\bf v}_{\alpha} \rangle$ and ${\bf S}_{\beta}={\bf S}_{\alpha}$, which simply represents the same parcel of phase space being advected into the neighboring cell, and in this limit the scheme would be exactly dissipationless in velocity space. With a finite basis, however, an exact match is generally not available, and some choice has to be made about which $\beta$ to associate with each $\alpha$.

\subsubsection{Cost Functions on the Basis Pair Space}
\label{sec:pairing.cost}

We therefore introduce a {cost}, $C_{\alpha\beta}$, for pairing $\alpha$ on $a$ with $\beta$ on $b$, designed to be small when $\alpha$ and $\beta$ are smoothly-connected in phase-space, and large otherwise. We define $C_{\alpha\beta}$ on the moments $Q$ of $\alpha$ and $\beta$, so it can compare any two bases: importantly in what follows, it can be applied to the \textit{reconstructed} bases/moments at the inter-cell face $Q^{\ast,a}_{ab,\alpha}$, $Q^{\ast,b}_{ab,\beta}$, not just the cell-centered values. 

Naively, one might consider (and we have) many choices, including: (a) the mean-velocity (squared) distance $C_{\alpha\beta}^{\rm v_{0}} \equiv |\langle {\bf v}_{\alpha} \rangle - \langle {\bf v}_{\beta} \rangle|^{2}$; (b) the simplified-Wasserstein and $L^{2}$ Bures-Wasserstein distances \citep[e.g.][and references therein]{marchisio.fox:2013.qmom.book.multiphase.moments} $C_{\alpha\beta}^{\rm tr} \equiv |\langle {\bf v}_{\alpha} \rangle - \langle {\bf v}_{\beta} \rangle|^{2} + \left( \sqrt{{\rm Tr}\,{\bf S}_{\alpha}} - \sqrt{{\rm Tr}\,{\bf S}_{\beta}} \right)^{2}$ and $C_{\alpha\beta}^{\rm W^{2}} \equiv |\langle {\bf v}_{\alpha} \rangle - \langle {\bf v}_{\beta} \rangle|^{2} + {\rm Tr}\left[ {\bf S}_{\alpha} + {\bf S}_{\beta} - 2\,({\bf S}_{\alpha}^{1/2}\,{\bf S}_{\beta}\,{\bf S}_{\alpha}^{1/2})^{1/2} \right]$, (c) the negative log Bhattacharyya overlap $C_{\alpha\beta}^{\rm BC} \equiv -\log\,{\rm BC}(\tilde{g}_{\alpha},\,\tilde{g}_{\beta})$ \citep[an overlap measure between two probability densities;][]{bhattacharyya:1946.measure.divergence.multinomial}. 
But these share three problems. 
(1) They have units of velocity-squared, so (1a) penalize a fractionally-small mismatch between two bases with large absolute $|{\bf v}|$ much more than an $\mathcal{O}(1)$ mismatch between bases with smaller absolute $|{\bf v}|$, and (1b) make empty-basis thresholds (below) give cell-dependent results. 
(2) ${\bf S}$ does not appear sensibly: the difference between ${\bf S}$ or $\sigma \equiv \sqrt{{\rm Tr}[{\bf S}]}$ is given equal penalty to ${\bf v}$ distances. So a cold basis flowing into $b$, given a choice between a basis (2a) with identical $\langle {\bf v} \rangle$ and large $|{\bf S}|$ (completely overlapping it) or (2b) with cold $|{\bf S}|\to 0$ but significantly different $\langle {\bf v}\rangle$ (zero overlap) could actually choose the latter. 
(3) Most importantly, only velocity information appears, so these cannot say if the DF $f({\bf v},\,...)$ (whose similarity we wish to evaluate) is continuous.
These betray the origins of (abc) as $\chi^{2}$-like measures for unit-normalized probability distribution functions, \textit{not} what we require. As a result, each of these cost functions fails badly on at least one test problem in \S~\ref{sec:tests}, producing obviously incorrect ``matches'' with consequences ranging from severe numerical dissipation to outright instability. This is especially acute in problems where mean velocities are similar and small, like the outer edges/turnaround points of the harmonic breathing test (\S~\ref{sec:test.harmonic}). 

We address these by defining a combined dimensionless cost: 
\begin{align}
\label{eqn:cost.default} C_{\alpha\beta} &\equiv C^{\rm v}_{\alpha\beta} + \lambda_{\rho}\,C^{\rho}_{\alpha\beta}\,, \\ 
\label{eqn:cost.cont} C^{\rm v}_{\alpha\beta} &\equiv \frac{|{\bf u}_{\alpha} - {\bf u}_{\beta}|^{2}}{|{\bf u}_{\alpha}|^{2} + |{\bf u}_{\beta}|^{2} + 2\,|{\bf u}_{\alpha}\cdot{\bf u}_{\beta}| + {\rm Tr}[{\bf S}^{\ast}_{\alpha}] + {\rm Tr}[{\bf S}^{\ast}_{\beta}]}\,, \\
\label{eqn:cost.rho} C^{\rho}_{\alpha\beta} &\equiv \frac{|\rho^{\ast}_{a,\,\alpha} - \rho^{\ast}_{b,\,\beta}|}{\rho^{\ast}_{a,\,\alpha} + \rho^{\ast}_{b,\,\beta} + 0.001\rho^{\ast}_{a}} \, .
\end{align}
Here ${\bf u}_{\alpha} \equiv \langle {\bf v}^{\ast}_{\alpha} \rangle - {\bf v}_{F}$, $\rho^{\ast}$, and ${\bf S}^{\ast}$ refer to the \textit{reconstructed} variables at the face (\S~\ref{sec:reconstruction}),\footnote{The gradient stencil is itself built on matching (\S~\ref{sec:reconstruction.matched}), so we use the previous-step gradients to estimate the reconstruction in this step.} in the frame of the face velocity ${\bf v}_{F}$ (preserving Galilean-invariance), and $\rho_{a}^{\ast} \equiv \sum_{\alpha} \rho_{a,\alpha}$.
This combines a velocity $C^{\rm v}_{\alpha\beta}$ and density $C^{\rho}_{\alpha\beta}$ separation, akin to a leading-order $f(...)$ separation, at the face. Using the reconstructed values means that in smooth flows the distance vanishes perfectly (otherwise we deal with larger resolution-dependent errors). 
Both costs are normalized by their characteristic values (and bounded in $[0,\,1]$) so small \textit{fractional} differences are favored (units-independent); the ``floor'' $0.001\rho^{\ast}_{a}$ is a safety factor to prevent numerical divergence for near-vacuum bases. 
And in $C^{\rm v}$ the {combined} dispersion appears in the denominator, favoring bases whose velocity-centers differ by less than either basis dispersion. 

The relative-weight factor $\lambda_{\rho}$ is an adjustable parameter. Too-small $\lambda_{\rho} \to 0$ gives the problematic velocity-only weights. But too-large $\lambda_{\rho}$ begins to penalize legitimate same-stream matches which happen to straddle steep density gradients. In situations with a real velocity discontinuity (e.g.\ a cold stream jumping cells), it is better to think of $C^{\rho}_{\alpha\beta}$ as a ``tiebreaker,'' for which one desires $\lambda_{\rho} < 1$. In practice, we find the behavior is insensitive in our tests in the range $\lambda_{\rho} \sim 0.05-0.5$, so adopt a value near the middle, $\lambda_{\rho}=0.25$ as a default. But we stress that there are many conceivable improved cost functions worth exploring.

\subsubsection{Empty-Basis or ``free-slot'' Weighting}
\label{sec:pairing.freeslot}

The more interesting question arises when no basis on side $b$ is a close velocity-space match for $\alpha$ -- for example, when a new stream is injected into a cell whose neighbors do not yet carry a matching basis component. In this case the relevant question is not ``which $\beta$ has nearest velocity to $\alpha$?'' but rather ``which $\beta$ can be overwritten by mass and moments from $\alpha$ with the smallest information loss?'' For our purposes the answer is the {emptiest} basis: numerical mixing entropy is maximized when the two paired bases have comparable mass (the merge mixes ``average'' properties of both, losing the identity of each), and is minimized when one of the pair has negligibly small mass (the merge inherits essentially all properties of the heavier basis, while the lighter one is overwritten without loss).

Naively one might therefore change the weights Eq.~\ref{eqn:cost.default} to favor low-mass bases, e.g.\ multiplying by $\rho_{b,\,\beta}/(\rho_{a,\,\alpha}+\rho_{b,\,\beta})$. But that can easily lead to \textit{always} favoring the lowest-mass basis, even when there is a good (just not perfect) fit, and forcing multiple bases to always all write to the same basis, producing extreme numerical dissipation. What we really want is for this to be a ``fallback'' when there is no good \textit{velocity} fit (we also do not want to penalize a good velocity match at a spatial density discontinuity, which is a new basis moving into a cell and the code behaving correctly). There are many re-weightings one might imagine to accomplish this, but we have primarily experimented with the following:
\begin{align}
\label{eqn:cost.free.slot} C^{\rm eb}_{\alpha\beta} &\equiv C_{\alpha\beta} \left( 1 - \frac{\tilde{\Theta}\left[ \frac{C_{\alpha}^{\rm fit} - \epsilon_{\tau}/2}{\epsilon_{\tau} } \right] }{1+\rho_{b,\beta}/\rho_{a,\alpha}} \right) \, , \\ 
C^{\rm fit}_{\alpha} &\equiv \min_{\beta}\left[ \frac{| {\bf u}_{\alpha}-{\bf u}_{\beta}|^{2}}{| {\bf u}_{\alpha}-{\bf u}_{\beta}|^{2} + {\rm Tr}[{\bf S}^{\ast}_{\alpha}] + \epsilon_{v}^{\prime\,2}} +  \frac{\epsilon_{\tau}}{1+\rho_{b,\beta}/\epsilon_{\rm eb}\rho_{b}} \right] \ . 
\end{align}
In $C^{\rm fit}_{\alpha}$, the first term is akin to the usual velocity-cost ($C^{\rm v}$ in Eq.~\ref{eqn:cost.cont}; with $\epsilon^{\prime}_{v}$ (set to $0.01$ times the denominator in $C^{v}_{\alpha\beta}$) here just preventing divide-by-zero errors for good matches of very cold bases); one could simply use $C^{\rm v}$ here as well. The second term in $\epsilon_{\tau}\ll1$ (here $\sim 0.04$ after some trial-and-error) is a function designed to further low-cost-weight the fits in ``well-resolved'' bases with $\rho_{b,\beta}$ larger than some tiny fraction of $\epsilon_{\rm eb}\rho_{b}$ (here $\epsilon_{\rm eb}= 0.01$ chosen somewhat arbitrarily). The function $\tilde{\Theta}$ is some sigmoid function,\footnote{Here (arbitrarily) defined by the spline $\tilde{\Theta}(x<0) \equiv 0$, $\tilde{\Theta}(x>1) \equiv 1$, $\tilde{\Theta}(0\le x\le 1) \equiv x^{2}(3-2x)$.} which addresses whether there is a good fit ($\to 0$) or not ($\to 1$). The denominator $\rho_{b,\beta}/\rho_{a,\alpha}$ then upweights the lowest-mass basis when there is no good fit. If a good fit exists, then $C^{\rm fit}_{\alpha}$ is small, $\tilde{\Theta} \to 0$, and $C^{\rm eb}_{\alpha\beta} \to C_{\alpha\beta}$. If no good fit exists, then $C^{\rm fit}_{\alpha} \gg \epsilon_{\tau}$, $\tilde{\Theta} \to 1$, and $C^{\rm eb}_{\alpha\beta} \to C_{\alpha\beta} [ \rho_{b,\beta} / ( \rho_{b,\beta} + \rho_{a,\alpha})]$, biasing pairing to the lowest-mass available basis. We show below this plays a critical role in dealing with very different bases interacting, as can arise in complex problems with finite basis number.

\subsubsection{One-Sided (Asymmetric) Pairing}
\label{sec:pairing.onesided}

The one-sided Riemann solution of \S~\ref{sec:flux.onesided} pairs naturally with a {one-sided} assignment in which each side $a$, $b$ independently selects its own outgoing pairings $\beta(\alpha)$ and $\alpha(\beta)$. This means $\beta(\alpha)$ as defined by $a$'s outgoing flux need not equal the inverse of $\alpha(\beta)$ as defined by $b$'s outgoing flux. The method remains exactly conservative pair-by-pair, because each outgoing flux is subtracted from its source and added to the other cell. Our default is to simply independently assign each basis the cost-minimizing $\alpha(\beta)$ and $\beta(\alpha)$. We note the two directions are built from two separately-evaluated cost matrices rather than from one matrix and its transpose.

\subsubsection{Pairwise-Symmetric Assignment Algorithms}
\label{sec:pairing.assignment}

Alternatively, one could enforce pairwise-symmetric pairings (e.g.\ ``if $\alpha$ donates to $\beta$, then $\beta$ donates to $\alpha$''). Given a cost matrix there are several well-known algorithms for this ``assignment problem'' \citep[see e.g.][]{kuhn:1955.hungarian.method.assignment.problem,burkard:2009.assignment.problems.book}: the simplest is the {greedy} (or ``one-sided nearest'')\footnote{Note that the greedy rule in this sense is \textit{not} the same as our default: the greedy rule lets one side's cost minimum dictate the pairing for both, whereas in \S~\ref{sec:pairing.onesided} each side minimizes its own cost matrix and the two need not agree.} rule, in which each basis $\alpha$ is paired with its individual cost-minimum $\beta$ on the other side (so setting $\beta$'s choice without evaluating its cost matrix); the {mutual-nearest} rule, in which $\alpha$ and $\beta$ are paired iff each is the cost-minimum of the other; and the optimal-bijection or {Hungarian} rule \citep{kuhn:1955.hungarian.method.assignment.problem}, in which $\beta(\alpha)$ is a permutation of $\{1,\,2,\,\dots,\,N_{\rm basis}\}$ chosen to minimize $\sum_{\alpha} C_{\alpha\,\beta(\alpha)}$. We have also experimented with weighted pairing, where the flux from $\alpha$ can be spread to multiple bases $\beta$.
As shown below, regardless of which rule we adopt, these all introduce significant dissipation compared to our default method, by roughly two orders of magnitude in kinetic energy on the test of \S~\ref{sec:test.pairing}. It is easy to see why -- if $a$ has two cold bases $\alpha_{1,2}$ which differ slightly, and $b$ has one basis $\beta_{1}$ identical to $\alpha_{1}$ but another $\beta_{2}$ very different, our default allows $\alpha_{1}$ and $\alpha_{2}$ to both pair with $\beta_{1}$, minimizing mixing entropy and numerical dissipation.

\subsection{Time-Integration \&\ Averaging Steps}

\subsubsection{Flux Assignment and Updates}
\label{sec:update}

Given the per-pair flux ${\bf F}_{ab,\,\alpha}^{\ast,a}$ from \S~\ref{sec:flux} and the basis pairing $\beta(\alpha)$ from \S~\ref{sec:pairing}, the conservative update follows from Eq.~\ref{eqn:conserved.update.total}: 
\begin{align}
\frac{d ({\bf U} V)_{a,\alpha}}{dt} &= -\sum_{b} \left[ F_{a\rightarrow b,\alpha}^{\ast} + \sum_{\beta: \alpha=\alpha(\beta)} {\bf F}_{b\rightarrow a,\beta}^{\ast} \right] \cdot {\bf A}_{ab}
\end{align}
which can then be advanced in time according to the usual schemes, e.g.: 
\begin{align}
({\bf U} V)_{a,\alpha} \rightarrow ({\bf U} V)_{a,\alpha} + \frac{d ({\bf U} V)_{a,\alpha}}{dt} \Delta t\ . 
\end{align}
Many schemes for this exist and are well-studied, for example higher-order Runge-Kutta-type integrators are common, and for certain types of problems different time-integrators satisfying symplectic or other properties are important. 
Here we adopt a simple second-order leapfrog (kick-drift-kick + Strang-split external force update) scheme matching exactly the MFM hydrodynamics and gravity implementation in \GIZMO, with adaptive hierarchical timesteps  (see \citealt{hopkins:gizmo} for details). But we have tested a simple explicit-Euler update and do not measure any difference on any of the test problems we studied here. 

After the per-step, per-basis conservative update, the averaging step is trivial. The cell-centered specific/primitive variables are recomputed \textit{from the conserved variables} as: 
\begin{align}
{\bf U}_{a,\alpha} &= ({\bf U} V)_{a,\alpha} / V_{a} \\ 
\nonumber \rho_{a,\alpha} &= m_{a,\alpha}/V_{a}, \\ 
\nonumber \langle {\bf v} \rangle_{a,\alpha} &= {\bf p}_{a,\alpha} / m_{a,\alpha} , \\
\nonumber  {\bf S}_{a,\alpha} &= {\bf T}_{a,\,\alpha}/m_{a,\,\alpha} - \langle {\bf v} \rangle_{a,\alpha} \otimes \langle {\bf v}  \rangle_{a,\alpha} \ . 
\end{align}
Any time the conservative or primitive variables are updated (drift or kick), this is followed by the realizability-repair step below.

Because the net flux into/out of any face adds to zero (Eq.~\ref{eq:mdot.face}) for the particular MFM partition adopted here, it is necessarily true that the total cell mass change $d m_{a}/dt = \sum_{\alpha} d m_{a,\alpha}/dt = 0$ to machine precision, and because all bases within a cell $a$ are updated on the same timestep $\Delta t = \Delta t_{a}$, this means $m_{a}$ is manifestly conserved independent of the details of the time integration scheme. But it also means, like with MFM hydrodynamics and standard MC/PIC/NB integration for short and long-range forces, the accuracy of total momentum or energy conservation depends on the flux-accumulation and timestep scheme, and can be reduced to integration accuracy when individual particle-differing adaptive timesteps $\Delta t_{a} \ne \Delta t_{b}$ are adopted.

\subsubsection{Timestep Constraints}
\label{sec:timesteps}

In time integration, all of the usual ``external'' timestep criteria apply: e.g.\ acceleration-based criteria commonly adopted for gravity methods \citep[$\Delta t < C_{a}\,(h_{a}/|{\bf a}_{a}|)^{1/2}$ with $C_{a}\approx 0.4$;][]{power:2003.nfw.models.convergence}. Since we solve a series of hyperbolic equations, we also require the usual Courant condition: 
\begin{align}
\Delta t_{a} < C_{\rm CFL} \frac{\Delta x_{a}}{v_{\rm sig,\,max}^{a}}
\end{align}
where we define the signal velocity $v_{\rm sig,\,max}^{a}$ as the maximum over all $|u_{x}| + \chi\sigma_{x}$ of all interacting neighbor faces and bases with bases in $a$ (as defined in \S~\ref{sec:flux}). We find this (with $C_{\rm CFL}\le 0.5$) is sufficient for stability. 

Many other timestep criteria are possible in addition (taking the minimum over all). One that we do implement, which can occasionally matter, is a per-basis {mass-flux} criterion that bounds the single-basis fractional mass change in a step:
\begin{align}
\label{eqn:dt.massflux} \Delta t_{a} &< C_{m}\, {\rm MIN}_{\alpha}\left[ \frac{m_{a,\,\alpha}^{\rm eff}}{|dm_{a,\alpha}/dt|} \right]\,,
\end{align}
with $C_{m}=0.1$, $m_{a,\,\alpha}^{\rm eff} \equiv {\rm MAX}\left( m_{a,\,\alpha},\ f_{\rm floor}\,m_{a} \right)$, $f_{\rm floor}=0.01$. The floor in $m_{a,\,\alpha}^{\rm eff}$ prevents $\Delta t\to 0$ in negligible-mass/empty bases. In our tests below the CFL condition almost always dominates, but this enforces an effectively stricter CFL when a simulation contains a high-mass basis rapidly flowing ``out'' into a new lower-mass basis (e.g.\ a cold clump moving between cells). This is not strictly necessary, but it reduces the number of realizability repairs required. 

Another common criterion for moving-mesh methods like MFM is that the mesh not deform too heavily in a single timestep, often approximated via a divergence-like criterion: $\Delta t_{a} < C_{\rm mesh} |\langle \tilde{\nabla}\cdot {\bf v}\rangle_{a} |^{-1}$, where $\langle \tilde{\nabla}\cdot {\bf v}\rangle_{a}$ refers to the numerical definition of the partition relative to the ``velocities'' of the mesh-generating points, and is defined for MFM in \citet{hopkins:fire2.methods}. We include this in our implementation as a safety guard but find in our test problems that the CFL condition is basically always stricter.

\subsubsection{Realizability Repair}
\label{sec:repair}

After the conservative update, each basis must remain a valid representation of an underlying phase-space DF: each $m_{a,\,\alpha}$ must be non-negative and ${\bf S}_{a,\alpha}$ must be SPD, and no two basis components should become so degenerate (in mean velocity and dispersion) that they together carry less information than a single component. Numerical roundoff and the upwind diffusion of the Riemann step can occasionally produce mildly non-physical states -- a slightly-negative eigenvalue of ${\bf S}_{a,\alpha}$, or a near-zero $m_{a,\,\alpha}$ which (left alone) would amplify rapidly. We therefore apply a local realizability-preserving correction $D$ after updates. This problem has a long history in the moment-method literature for kinetic theory and multiphase flows \citep[e.g.][]{levermore:1981.fld,fox:2008.eqmom.gaussian.quadrature.moment.method,marchisio.fox:2013.qmom.book.multiphase.moments}, from which we adapt the relevant pieces.

The operations $D$ performs are, in order: 

(i) Any basis whose mass has dropped below a small {relative} floor $m_{a,\,\alpha} < \epsilon_{m}\,m_{a}/N_{\rm basis}$ (we take $\epsilon_{m} = 10^{-12}$) is reset to an inert realizable state at the cell-bulk velocity $\langle{\bf v}_{a}\rangle$ -- $(m_{a,\,\alpha},\,{\bf p}_{a,\,\alpha},\,{\bf T}_{a,\,\alpha}) \rightarrow (\epsilon_{m,{\rm eff}},\,\epsilon_{m,{\rm eff}}\,\langle{\bf v}_{a}\rangle,\,\epsilon_{m,{\rm eff}}\,(\langle{\bf v}_{a}\rangle\otimes\langle{\bf v}_{a}\rangle + S_{\rm floor}\,{\bf I}))$ with $\epsilon_{m,{\rm eff}} = \max(\epsilon_{m,{\rm abs}},\,\epsilon_{m}\,m_{a}/N_{\rm basis})$, and the residual mass, momentum, and stress (the difference between the pre-reset and post-reset basis state) are redistributed to a real-mass basis within the same cell. We have experimented with both (a) simply giving ``all'' of the correction to the most-massive basis, or (b) using a CQMOM/EQMOM-style multi-donor redistribution \citep{fox:2008.eqmom.gaussian.quadrature.moment.method} where we redistribute a fraction $=m_{a,\alpha}/\sum_{\alpha^{\prime}} m_{a,\alpha^{\prime}}$ to each surviving basis $a,\alpha$. We see no significant difference between (a) and (b). The essential property of this {conservative-shift} repair is that it preserves the cell-summed $(m_{a},\,{\bf p}_{a},\,{\bf T}_{a})$ exactly to machine precision while eliminating the floating-point-noise dominated primitive variables associated with the bad basis. 

(ii) Each ${\bf S}_{a,\alpha}$ is symmetrized (${\bf S}_{a,\alpha} \rightarrow ({\bf S}_{a,\alpha} + {\bf S}_{a,\alpha}^{T})/2$) and then projected to the positive-definite cone via a standard symmetric eigendecomposition, with any negative eigenvalues replaced by $\epsilon_{S}\,{\rm Tr}({\bf S}_{a,\alpha})/D$, with $\epsilon_{S} = 10^{-12}$. If the cell has ${\rm Tr}({\bf S}_{a,\alpha}) < 0$ we repair by replacing this with a tiny positive $10^{-30}|\langle {\bf v} \rangle_{a,\alpha}|^{2}$, but we find this is never invoked. Again any change to conserved quantities is distributed to the other bases.

(iii) Any pair of components which have become locally near-degenerate (in mean velocity {and} dispersion) is merged to a single component, preserving $m$, ${\bf p}$, and ${\bf T}$ for the cell exactly, while any single component whose dispersion has grown anomalously large is split into two, again preserving the cell-summed moments. The conservative merge/split algebra follows the EQMOM method in \citet{fox:2008.eqmom.gaussian.quadrature.moment.method,marchisio.fox:2013.qmom.book.multiphase.moments}.

Note $D$ is purely intra-cell and involves no communication. The total drift in $({\bf p},\,{\bf T})$ induced by $D$, summed over all cells and time steps, gives a useful measure of how often and how aggressively the repair operator has been called. In practice, on the test problems we consider here, this cumulative drift is many orders of magnitude smaller than the equivalent drift from the Riemann diffusion itself, so the specific form of $D$ is not a significant source of error.

\subsection{External \&\ Long-Range Forces}
\label{sec:gravity}

\subsubsection{Evaluation and Time-Integration}

External and/or long-range forces can be operator-split from the Riemann problem (or just like with hydrodynamics, they can be included in principle, but this is rarely done as it is generally much less accurate). The evaluation of some external force ${\bf F}_{\rm ext}$ at the position or integrated over the domain $\Omega_{a}$ follows the usual methods. For example, for self-gravity, one can use the usual tree or fast multipole or particle-mesh type methods to obtain the potential $\Phi$ or acceleration $-\nabla \Phi$. 
This can then be applied to update the basis moments according to whatever time integration scheme is used.

For forces which are velocity-independent, this is especially trivial because it is a uniform translation in velocity space ($d{\bf v}/dt = {\bf a}_{\rm ext}$); $m_{a,\alpha}$ and ${\bf S}_{a,\alpha}$ are invariant while ${\bf p}$ and therefore ${\bf T}$ change as: 
\begin{align}
{\bf p}_{a,\,\alpha} &\rightarrow {\bf p}_{a,\,\alpha} + m_{a,\,\alpha}\,{\bf a}_{\rm ext}^{a}\,\Delta t\ ,\\ 
\label{eqn:T.kick.update}
{\bf T}_{a,\,\alpha} &\rightarrow m_{a,\,\alpha}\,{\bf S}_{\alpha} + \frac{{\bf p}_{a,\,\alpha} \otimes {\bf p}_{a,\,\alpha}}{m_{a,\,\alpha}}\ .
\end{align}
In our specific implementation we follow \S~\ref{sec:update} and do this following a leapfrog time-stepper \citep{springel:gadget}. 

For velocity-dependent forces this step is slightly more complicated but still straightforward. For example, in the limit of large (resolved) gyro-radii, a Boris integrator can be used, and one simply needs to calculate the map from the moments at time $t$ to some $t+\Delta t$ (e.g.\ $\langle {\bf v}^{n} \rangle \to \langle {\bf v}^{n+1} \rangle$). Again for a Lorentz force this is particularly simple, as the electric field term is velocity independent and the magnetic force is pure rotation about $\hat{\bf B}$ (with ${\bf v}$-independent gyro-frequency). But even for complicated cases this is a strictly local operation on each basis within each cell, so can be integrated to any arbitrary desired accuracy numerically, or subcycled, in a straightforward manner. 

Note that \textit{short}-range self-forces fall into the ``collision operator'' category, which we discuss in \S~\ref{sec:collisions}.

\subsubsection{Force Softening}

Typically, long-range (e.g.\ gravitational or Coulomb or light-mediator-driven dark/fifth-force) forces often have a divergence at small $r$ that must be softened when a single numerical particle or cell represents many micro-physical ``true'' particles. But in MC/PIC/NB methods, such softening is fundamentally ad hoc and ill-defined, because the discretization of the DF $f({\bf x},\,{\bf v},\,t,\,...)$ for purposes of other dynamics is fundamentally a $\delta$-function (there is no self-consistent volume partition under motion/advection of the particles; see \citealt{hopkins:tidal.softening} and references in \S~\ref{sec:intro}). But in our proposed method, there is a well-defined volume partition and reconstruction. That means that there is a unique, self-consistent, well-defined, and \textit{correct} force softening: the force induced by the collective spatially-distributed DF. For e.g.\ gravitational force softening, the continuous reconstruction of $\rho({\bf x}) = \int f d^{3}{\bf v}$ inserted into the Poisson equation corresponds to a smooth, divergence-free $\Phi$ and ${\bf a}_{\rm grav} = -\nabla \Phi$.

For our adopted MFM partition, this corresponds identically to the adaptive gravitational force softening defined for MFM fluid dynamics \citep[see][]{hopkins:gizmo,hopkins:fire2.methods,hopkins:tidal.softening}. Derived properly \citep[see e.g.][]{price:2007.lagrangian.adaptive.softening} this ensures manifest conservation of momentum and energy under gravitational forces, as well as consistency with the Vlasov equation dynamics, and numerically is identical to the ``$n_{\rm NGB}$'' implementation in \S~3.3 of \citet{hopkins:tidal.softening}.\footnote{We stress that this is qualitatively different from adaptive force softening in standard MC/PIC/NB methods for collisionless fluids, where (as stressed in \citealt{hopkins:tidal.softening}) the choice of softening is an ad hoc numerical parameter. In those applications, the softening is chosen to scale with some other property of the system like a kernel density estimator applied to the point-mass-like distribution of MC/PIC/NB particles. But this leads to contradictions under the actual particle evolution (especially in stream intersection), because there are no mass fluxes or volumes associated dynamically with the particles.}

\subsection{Collision Operators \&\ Local Mixing}
\label{sec:collisions}

As noted above, the collision operator $(\partial f/\partial t)_{\rm collisions}$ can be operator-split from the previous operations. The precise form depends on the physics being modelled -- self-interacting dark matter, annihilation, neutrino-baryon coupling, plasma Coulomb collisions -- but a feature shared by all of these is that the relevant collision operator is {local}: it depends only on $f({\bf x},\,{\bf v})$ within the cell $a$, and not on the neighbors. Because the full phase-space DF $f_{a}({\bf v})$ is carried in each cell, the operator can therefore be applied entirely within a single cell -- no inter-cell communication is required, and (more importantly) the operator can be physically correct rather than artificially coarse-grained over super-particles, as it must be in MC/PIC/NB methods.

\subsubsection{Example: Hard-Sphere Elastic Scattering}
\label{sec:collisions.hardsphere}

To illustrate, we work out explicitly the case of isotropic, elastic, two-body scattering with a constant (hard-sphere like) cross-section $\sigma$ between identical particles of physical mass $\mu_{p}$ (so that an SIDM-like model with cross-section per unit mass $\sigma/\mu_{p}$ specified is recovered). The Boltzmann collision operator in this case is
\begin{align}
\label{eqn:coll.boltzmann}
\nonumber \frac{\partial f({\bf v}_{1})}{\partial t} \Big |_{\rm coll} &= \int\!d^{3}{\bf v}_{2}\,\frac{\sigma/\mu_{p}}{4\pi}\!\int\!d\Omega\,|{\bf v}_{1}-{\bf v}_{2}|\,
\mathcal{H}_{12} \\ 
\mathcal{H}_{12} &\equiv  f({\bf v}_{1}^{\prime})\,f({\bf v}_{2}^{\prime})- f({\bf v}_{1})\,f({\bf v}_{2})\,,
\end{align}
where $({\bf v}_{1}^{\prime},\,{\bf v}_{2}^{\prime})$ are the post-collision velocities of a pair drawn from the isotropic-in-centre-of-mass distribution and $({\bf v}_{1},\,{\bf v}_{2})$ are the corresponding pre-collision velocities. The mass, momentum, and trace of the second-moment tensor (i.e.\ the kinetic energy) of the integrand vanish identically on integration over ${\bf v}$, so the cell-integrated $(m_{a},\,{\bf p}_{a},\,{\rm Tr}({\bf T}_{a}))$ are exactly conserved by Eq.~\ref{eqn:coll.boltzmann}. 
The {anisotropic} part of ${\bf T}_{a}$ is not conserved: isotropic elastic scattering relaxes the local distribution toward a Maxwell-Boltzmann distribution, so the off-diagonal and trace-free parts of ${\bf T}_{a}$ decay on a collisional timescale.

Especially for our Gaussian-basis representation, the standard simplification is the BGK closure \citep{bgk:1954.collision.model}, which replaces Eq.~\ref{eqn:coll.boltzmann} by a relaxation toward the local Maxwell-Boltzmann fixed point:
\begin{align}
\label{eqn:coll.bgk}
\left(\frac{\partial f}{\partial t}\right)_{\rm coll} &= -\nu_{a}\,\bigl[f_{a}({\bf v}) - f_{a}^{\rm eq}({\bf v})\bigr]\,,
\end{align}
which conserves $(m_{a},\,{\bf p}_{a},\,{\rm Tr}({\bf T}_{a}))$ by construction (the loss and gain terms have the same first three moments) and recovers the full Boltzmann scattering rate and equilibrium. The collision rate is
\begin{align}
\label{eqn:coll.nu}
\nu_{a} &\equiv n_{a}\,\sigma\,\langle |{\bf v}_{\rm rel}|\rangle \,=\, \frac{\rho_{a}\,\sigma}{\mu_{p}}\,\sqrt{\frac{16\,k_{\rm B}\,T_{a,\,{\rm kin}}}{\pi\,\mu_{p}}}\,,
\end{align}
with $\rho_{a} = m_{a}/V_{a}$, $n_{a} = \rho_{a}/\mu_{p}$, and the cell kinetic temperature defined from the trace of the centre-of-mass second-moment tensor,
\begin{align}
\label{eqn:coll.Tkin}
k_{\rm B}\,T_{a,\,{\rm kin}} &\equiv \frac{\mu_{p}}{3}\,{\rm Tr}({\bf S}_{a}^{\rm com})\,, \quad {\bf S}_{a}^{\rm com} \equiv \frac{{\bf T}_{a}}{m_{a}} - \langle{\bf v}_{a}\rangle\otimes\langle{\bf v}_{a}\rangle\,,
\end{align}
and $\langle {\bf v}_{a}\rangle \equiv {\bf p}_{a}/m_{a}$ is the bulk velocity of the cell. The Maxwell-Boltzmann fixed point is the single isotropic Gaussian
\begin{align}
\label{eqn:coll.feq}
f_{a}^{\rm eq}({\bf v}) &= \rho_{a}\,(2\pi\,\sigma_{\rm eq}^{2})^{-3/2}\,\exp\!\left[-\frac{|{\bf v}-\langle{\bf v}_{a}\rangle|^{2}}{2\,\sigma_{\rm eq}^{2}}\right]\,, \quad \sigma_{\rm eq}^{2} \equiv \frac{1}{3}\,{\rm Tr}({\bf S}_{a}^{\rm com})\,.
\end{align}

Eq.~\ref{eqn:coll.bgk} integrates over a step $\Delta t$ in closed form, $f_{a}^{n+1} = e^{-\nu_{a}\Delta t}\,f_{a}^{n} + (1 - e^{-\nu_{a}\Delta t})\,f_{a}^{\rm eq}$. The basis representation $f_{a}^{n} = \rho_{a}\sum_{\alpha}\varpi_{\alpha}^{n}\,\tilde{g}_{\alpha}^{n}({\bf v})$ is then re-projected onto the same number of components by a per-basis convex update which holds the basis masses fixed and pulls each $(\langle{\bf v}_{\alpha}\rangle,\,{\bf S}_{\alpha})$ toward the equilibrium values, with relaxation fraction
\begin{align}
\label{eqn:coll.eta}
\eta_{a} &\equiv 1 - e^{-\nu_{a}\,\Delta t}\,.
\end{align}
The resulting per-basis update is
\begin{align}
\label{eqn:coll.update.v}
\langle{\bf v}_{\alpha}\rangle^{n+1} &= (1-\eta_{a})\,\langle{\bf v}_{\alpha}\rangle^{n} + \eta_{a}\,\langle{\bf v}_{a}\rangle\,, \\
\label{eqn:coll.update.S}
{\bf S}_{\alpha}^{n+1} &= (1-\eta_{a})\,{\bf S}_{\alpha}^{n} + \eta_{a}\,\sigma_{\rm eq}^{2}\,{\bf I} + \eta_{a}(1-\eta_{a})\,\Delta{\bf v}_{\alpha}\otimes\Delta{\bf v}_{\alpha}\,, \\
\label{eqn:coll.update.m}
m_{a,\,\alpha}^{n+1} &= m_{a,\,\alpha}^{n}\,,
\end{align}
where $\Delta{\bf v}_{\alpha} \equiv \langle{\bf v}_{\alpha}\rangle^{n} - \langle{\bf v}_{a}\rangle$. The third (variance-of-the-mean) term in Eq.~\ref{eqn:coll.update.S} is necessary: $f^{n+1}$ is a {convex} mixture of the per-basis $\tilde{g}_{\alpha}^{n}$ and the equilibrium $f_{a}^{\rm eq}$, and the variance of a convex mixture includes the squared spread of the component means about their joint mean, which here is exactly $\eta_{a}(1-\eta_{a})\Delta{\bf v}_{\alpha}\otimes\Delta{\bf v}_{\alpha}$. One can verify by direct substitution that the cell-summed $(m_{a},\,{\bf p}_{a},\,{\bf T}_{a}) = \sum_{\alpha}(m_{a,\,\alpha},\,m_{a,\,\alpha}\langle{\bf v}_{\alpha}\rangle,\,m_{a,\,\alpha}[{\bf S}_{\alpha} + \langle{\bf v}_{\alpha}\rangle\otimes\langle{\bf v}_{\alpha}\rangle])$ are individually exactly preserved by Eqs.~\ref{eqn:coll.update.v}-\ref{eqn:coll.update.m}. In the collisionless limit $\nu_{a}\,\Delta t\rightarrow 0$, $\eta_{a}\rightarrow 0$ and the update is the identity; in the strongly-collisional limit $\nu_{a}\,\Delta t \gg 1$, $\eta_{a}\rightarrow 1$ and every basis collapses onto the single isotropic Maxwell-Boltzmann component, which the realizability-repair step (\S~\ref{sec:repair}) will then merge into a single Gaussian via the EQMOM-style algebra, correctly recovering the hydrodynamic limit.

For non-Maxwellian targets -- velocity-dependent $\sigma({\bf v}_{\rm rel})$, anisotropic differential cross-sections, annihilation/inelastic operators, or non-spherically-symmetric scattering kernels -- Eq.~\ref{eqn:coll.boltzmann} generalizes in the obvious way, and the post-step distribution $f_{a}^{n+1}$ can be projected onto the same $N_{\rm basis}$ Gaussian mixture by minimizing the $L^{2}$-Wasserstein cost (\S~\ref{sec:pairing.cost}) between $f_{a}^{n+1}$ and the basis, using the same machinery already employed for inter-cell pairing (\S~\ref{sec:pairing}). The key conceptual point is that all of this is purely intra-cell -- there is no projection back onto neighboring cells, and no operator-coarsening of the kind that {must} be applied in MC/PIC/NB schemes, where collisions can only be represented as collisions between entire super-particles each of which represents some incredibly large number of individual physical particles (e.g.\ $\sim 10^{60{-}90}$ for common dark matter applications).

\subsubsection{Other Collision Models}
\label{sec:collisions.other}

The hard-sphere BGK example above is intentionally a simple case, but the same intra-cell structure accommodates a wide range of microphysical operators with only minor modification. Dissipative or ``hit-and-stick'' interactions (relevant e.g.\ for coagulating dust) differ from elastic scattering only in that kinetic energy is not conserved by the collision, so the trace of ${\bf T}_{a}$ is reduced (rather than merely isotropized) by the operator; internally-excited or radiatively-decaying states that return their energy to bulk motion appear as a heat-injection term, applied by convolving each basis with a broader Gaussian of the appropriate energy each step. Velocity-dependent cross-sections $\sigma({\bf v}_{\rm rel})$ and anisotropic differential kernels are handled by the projection of \S\ref{sec:collisions.hardsphere}; and ``warm'' populations require no special treatment at all, since a finite initial dispersion is simply an initialization of ${\bf S}_{\alpha}$.

The same locality makes the method a natural fit for weakly-collisional plasmas, where Coulomb collisions are velocity-dependent but local, and quasi-neutrality holds on the scales of interest. The additional ingredient there is the electromagnetic force, which enters as the (generally velocity-dependent) external-force operator of \S\ref{sec:gravity} rather than as a collision term; given the local per-cell moments one has the information needed to solve for the fields (e.g.\ the electron pressure and current), so that the magnetized case can be built on top of the same per-basis representation, a direction we return to in \S\ref{sec:application}.

\subsubsection{Difference from Monte Carlo/PIC/$N$-body Methods}
\label{sec:collisions.vs.nbody}

It is worth emphasizing how different this is from the treatment of collisions in MC/PIC/NB schemes (e.g.\ standard SIDM or particulate-collision implementations). There, because the DF is sampled by $\delta$-function super-particles with no attached continuum density, the local scattering rate must be estimated from a neighbor-based ``overlap'' integral or kernel-density estimate -- an inherently noisy, resolution-dependent construction with no unique definition -- and each scattering event pairs entire super-particles, each of which represents some enormous number of micro-particles. Recovering a smooth, converged collision term then requires drawing many Monte-Carlo scattering realizations per particle per step: the sampling error falls only as the square root of the number of realizations, so beating the underlying MC/PIC/NB noise demands very large particle counts and is correspondingly expensive. At low particle number the result is order-unity wrong in the local velocity DF (the scattering samples the wrong distribution), while at high particle number the pairwise-exchange bookkeeping tends to violate energy conservation. In the method here, by contrast, the collision operator acts on the \textit{actual} local DF carried in each cell, at high order, with no sampling noise and exact conservation of the collisional invariants by construction (\S\ref{sec:collisions.hardsphere}).

\subsubsection{How Collisional Can the Method Go?}
\label{sec:collisions.howcollisional}

The operator-split collision operator above recovers the strongly-collisional (hydrodynamic) limit correctly, as we demonstrate below (\S\ref{sec:test.collisions}), because in that limit the intra-cell BGK relaxation drives every cell to a single local Maxwellian each step. There is, however, one regime where the naive operator-split becomes inaccurate: when the collisional mean free path becomes very small compared to the inter-element spacing -- equivalently, when the collision frequency greatly exceeds the inverse stream-crossing time set by the CFL condition -- the inter-cell Riemann problem, which is solved as if the fluid were collisionless over the face, no longer captures the (now fluid-like) collisions at the face within a single step. The scheme remains stable, but the operator-split can introduce large spurious oscillations. Estimating the collision rate $R \simeq \rho\,(\sigma/\mu_{p})\,\delta v$ against the CFL step $\Delta t = C_{\rm CFL}\,\Delta x/v_{\rm sig}$ with $\Delta x \sim (\Delta m/\rho)^{1/3}$, this becomes problematic only for extremely large cross-sections,
\begin{align}
\frac{\sigma}{\mu_{p}} &\gg 2\times 10^{4}\,{\rm cm^{2}\,g^{-1}}\,\left(\frac{\Delta m}{10^{3}\,{\rm M_{\odot}}}\right)^{-1/3}\,\left(\frac{\rho}{10^{8}\,{\rm M_{\odot}\,kpc^{-3}}}\right)^{-2/3}\,,
\end{align}
far outside the range of interest for e.g.\ self-interacting dark matter. 

Even in that regime, however, the fix is straightforward and requires nothing intrinsic to the method: one simply replaces the collisionless face solution with a Riemann solver that carries the partially-collisional (fluid-like) exchange directly. For a given volume partition and flux definition, this is simply the corresponding ``usual'' fluid dynamic, single-basis (e.g.\ hydrodynamic or MHD) flux. For the MFM partition we have adopted as our default, \citet{hopkins:gizmo} show this has an especially simple form in the HLLC approximation: 
\begin{align}
\label{eqn:hllc.fluid}
\mathcal{G}^{c}_{\rm HLLC} &\equiv
\begin{cases}
\mathcal{F}^{c}_{1}  \hfill & \ ( v_{R}-v_{L} \le c^{\ast} ) \\
{\bf 0}   \hfill & \ ( v_{R}-v_{L} > c^{\ast} )
\end{cases}
\end{align}
where $c^{\ast} = \max(c_{L},\,c_{R})$; $v_{K} = {\bf v}_{K} \cdot \hat{\bf A}$. The non-vacuum solution is: 
\begin{align}
\label{eqn:hllc.fluid.F1}
\mathcal{F}^{c}_{1} &= 
\begin{pmatrix}
0 \\
P^{\ast} \mathbb{I} \\ 
(2\,P^{\ast} S^{\ast}/D) \mathbb{I} \hat{\bf A}
\end{pmatrix}, \\
\nonumber S^{\ast} &= \frac{(P_{R} - P_{L}) + (\eta_{L} v_{L} - \eta_{R} v_{R} )}{\eta_{L} - \eta_{R}} \ , \\
\nonumber P^{\ast} &= \frac{ P_{L} \eta_{R} - P_{R} \eta_{L} + \eta_{L} \eta_{R} (v_{R} - v_{L} ) }{\eta_{R} - \eta_{L} } \ , \\
\nonumber \eta_{K} &= \rho_{K} (S_{K} - v_{K}) \ , \\
\nonumber S_{L} &\approx \min(v_{L},\,v_{R}) - c^{\ast} \, , \, S_{R} \approx \max(v_{L},\,v_{R}) + c^{\ast} \ .
\end{align}
Here $\mathcal{G}^{c}_{\rm HLLC}$ is the standard fluid HLLC solution for the cell-summed (isotropic-pressure) state, with $P_{K} \equiv \rho_{K}\,{\rm Tr}({\bf S}_{K}^{\rm com})/D$ the effective pressure, computed on the \textit{sum} over bases (i.e.\ from the total cell state rather than per-basis). The $S_{L,R}$ are approximate/guesses, others can be used.
For intermediate-collisionality cases, one can construct more exact solutions; however since most of the work is being done (correctly) by the actual collision operator on the fluxes, one can (at the desired level of accuracy) simply interpolate between the limiting fluxes based on the local collisionality. In other words, define the per-basis effective flux as
\begin{align}
\label{eqn:coll.blend}
\mathcal{G}^{\rm eff}_{\alpha} &\approx w_{c} \, \mathcal{G}^{\rm collisionless}_{\alpha} + \left(1 - w_{c}\right) \left(\frac{\rho_{K,\,\alpha}}{\rho_{K}}\right) \mathcal{G}_{\rm HLLC}^{c} \ , \\
\nonumber w_{c} &\equiv \exp{\left\{  - \frac{\Delta x_{ab}}{ \lambda_{\rm MFP} } \right\} } = \exp{\left\{  - \Delta x_{ab}\,\rho \, \sigma / \mu_{p} \right\} }\, ,
\end{align}
so that $w_{c}\rightarrow1$ recovers the collisionless per-basis solver of \S\ref{sec:flux} and $w_{c}\rightarrow0$ the fluid solution. The $\rho_{K,\,\alpha}/\rho_{K}$ weight simply distributes the (mass-weighted) collisional flux to each basis in proportion to its fraction of the cell mass; since this weight sums to unity and $\mathcal{G}_{\rm HLLC}^{c}$ is common to all bases, the blend is manifestly conservative.

\begin{table}
\caption{Trade-off between closure choices for the local velocity DF (\S~\ref{sec:moment.order}).\vspace{-0.5cm}
}
\label{tbl:moment.choice}
\begin{center}
\begin{tabular}{p{0.30\columnwidth} p{0.62\columnwidth}}
\hline
Representation & Properties \\
\hline
$10$-moment, many bases (\textit{default}) & Robust for arbitrary non-thermal DFs; converges to the true DF (and its aggregate heat flux) by adding bases; cost is more slots, pairing, and empty-basis bookkeeping. \\
$13$-moment, per-basis closure & Efficient for \textit{bounded} deviations from a Maxwellian (near-thermal plasma); requires a valid skew/kurtosis closure; limited realizability/skewness range. \\
\hline
\end{tabular}
\end{center}
\end{table}

\subsection{Moment Order and the 10- vs.\ 13-Moment Choice}
\label{sec:moment.order}

The default method as tested in this paper evolves, per basis, the full set of second moments (a $10$-moment closure in $D=3$: mass, three momenta, and the six independent components of the symmetric ${\bf T}_{\alpha}$). As discussed in \S~\ref{sec:Q.vs.U.choice}, a natural question of considerable interest for plasma applications in particular, is whether one should instead (or in addition) evolve a per-basis \textit{third} moment -- a heat-flux or skewness vector ${\bf q}_{\alpha}$ -- giving a $13$-moment closure. We have implemented and experimented with such an extension, and sketch its construction below; here we summarize why the multi-basis second-moment method is our default here, and the regimes in which the $13$-moment extension may be useful.

Per \S~\ref{sec:Q.vs.U.choice}, a key point is that \textit{the multi-basis second-moment method already carries an aggregate heat flux}. The cell's total third moment $\int ({\bf v}-\langle{\bf v}_{a}\rangle)^{\otimes 3}\,f_{a}\,d^{D}{\bf v}$ is generically nonzero even when every individual basis is symmetric, because it receives a contribution from the \textit{distribution of basis mean velocities} $\langle{\bf v}_{\alpha}\rangle$ about the bulk $\langle{\bf v}_{a}\rangle$; and because those means are Lagrangian-evolved, this aggregate skewness evolves dynamically. A per-basis ${\bf q}_{\alpha}$ adds a \textit{sub-grid} skew \textit{within} each basis, which is a different (and complementary) piece of information. This has a direct consequence for any practical implementation with a finite ``budget'' of evolved variables. 
Refining the representation of the DF by adding \textit{bases} is asymptotically convergent; a sum of many Gaussian or top-hat bases can reproduce any $f({\bf v})$ (hence any true skew/kurtosis) to arbitrary accuracy in principle (\S\ref{sec:test.convergence.Nbasis}). Refining instead by imposing a per-basis heat-flux term --- especially the reduced third moment evolved in 13-moment methods (${\bf q}_{i}$ which has dimension $D$, not the full third-moment $D\times D \times D$ tensor ${\bf G}_{ijk}$) --- is asymptotically biased: it requires imposing a specific prior or functional form for the shape of the DF (imposing the closure for the reduced third moment, for the un-evolved fourth moment[s], etc.). Even adding arbitrary additional moments (e.g.\ going to a 26-moment closure) can only converge if the shape of the \textit{true} DF is one which can be represented by the particular functional form assumed (e.g.\ if the closure assumption is fundamentally correct). 

The multiple-cold stream problems we consider provide simple illustrations of this. Trying to represent these with fewer 13 or higher-moment skew-kurtosis-Gaussian-like bases will intrinsically fail, because the DF is multi-peaked. A more subtle failure appears in the Hernquist sphere test: the shape of the \textit{equilibrium} DF at $r\to 0$ fundamentally lies outside of the envelope representable by the usual 13-moment (Grad--Hermite or Pearson-IV) closures, and so naively applying those closures to our method drives the DF towards un-representable states, de-stabilizing the method and forcing it to constantly trigger the ``repair'' operation. The same failure mode occurs in the harmonic-oscillator test if we try a simple linear or quadratic DF modification to construct a 13-moment version of the top-hat basis. 

On the other hand, if the DF \textit{is} close to Maxwellian and \textit{does} obey the closure assumptions in the 13-moment model, then it is much more efficient to represent e.g.\ a mildly-skew Gaussian DF with a single skew-Gaussian (the 13-moment approach) than with a series of 10-moment bases. And in many plasma physics applications, with strong ``effective'' collisionality or interactions mediated by magnetic fields on the kinetic scale, multi-peaked DFs are strongly unstable, and the physics imposes constraints on the range of stable DF shapes (which the multi-moment method does not automatically respect, unless one adds these as additional constraints). 

The upshot is a clean dichotomy in when each choice is preferable, which we summarize in Table~\ref{tbl:moment.choice}. For \textit{arbitrary}, far-from-thermal, genuinely collisionless distributions -- cold dark matter, collisionless or weakly-collisional dust and neutral populations, multi-stream flows -- the robust route to the ``correct'' heat flux is to \textit{add second-moment bases}; this is the method we validate and adopt here. For systems held \textit{near} a Maxwellian by frequent collisions or by self-regulating kinetic instabilities (much of collisional and weakly-collisional plasma physics) the deviation from the shape prior is bounded, a per-basis Grad--Hermite or Pearson-IV $13$-moment closure is valid and efficient, and the $13$-moment extension is the natural tool.

\subsection{Implementation \&\ Public Code}
\label{sec:public.code}

Rather than provide a limited ``pseudo-code'' example, we make public a full implementation of the scheme above, with documentation and test problems, in the public version of the code \GIZMO. \GIZMO\ is modular, so the method for integrating e.g.\ dark matter or stellar or plasma or particulate dynamics can be enabled (replacing the usual MC/PIC/NB solvers) independently of the other physics in the code. The implementation provides four options at each dimensionality ($D=1,2,3$) for the basis-moment closure order (e.g.\ $N_{\rm moments} \in \{5,\,10,\,13\}$ in 3D), corresponding to evolving $(m,\,{\bf p},\,{\rm Tr}({\bf T}))$,  $(m,\,{\bf p},\,{\bf T})$, $(m,\,{\bf p},\,{\bf T},\,{\bf q})$ -- i.e.\ the streaming, isotropic-dispersion, and full second-moment limits respectively as well as the 13-moment reduced heat-flux -- and a user-selectable $N_{\rm basis}$ (we have tested $N_{\rm basis}$ from $1$ to $128$). The per-element computational cost is dominated by the basis-pairing cost matrix and the per-pair Riemann solve and face-velocity root-find, all of which scale as $O(N_{\rm basis}^{2})$ for the assignment rules above, so the per-element overhead relative to standard MC/PIC/NB is fixed and small for the modest $N_{\rm basis}$ of interest. The gradient loop reuses the same kernel-moment matrix ${\bf E}_{a}^{-1}$ as the face-area construction (Eqs.~\ref{eqn:mfm.face.area.def.sub2}, \ref{eqn:lsq}), so adds no additional cost beyond what is already paid by the underlying MFM machinery. We refer interested readers to the public release of \GIZMO\ for further implementation details.

\section{Differences \&\ Similarities with Other Methods}

Here we summarize similarities \&\ differences with other methods for treating the Vlasov equation in the weakly-collisional limit. There are many super-families of approaches, so it is impossible to be complete, but these methods are all complementary in different regimes and some discussion is useful. 

\subsection{``Standard'' Monte-Carlo, PIC, or N-body Methods}
\label{sec:compare:nbody}

The method here is fundamentally distinct from ``standard'' MC/PIC/NB methods, in that we evolve the continuous phase-space DF, rather than sampling it discretely. This has a number of advantages. 
(1) It has easily defined convergence properties, with a clear convergence order (because the continuous equations are being solved), and can be combined with higher-order spatial or velocity or time reconstruction/evolution methods. 
(2) The method carries an actual local phase-space DF with each element, giving an enormous increase in information on the phase-space distribution. In MC/PIC/NB methods this must be computed in ensembles, requiring averaging over many thousands of particles to match (at $\sim10\%$ error-tolerance) the information content carried by a single resolution element in our new method.
(3) With a local, well-defined continuous phase-space DF, one can easily and rigorously apply local collision operators smoothly and at high order.  This is especially useful for models of complicated particle interactions or scattering, or highly-anisotropic or strongly velocity-dependent interactions which depend on the tails of the velocity DF. 
(4) The Riemann problem being solved between elements allows for collisional or other fluid-like effects, as well. Crucially, this means the method can interpolate \textit{between} the weakly and strongly collisional limits smoothly -- it reduces to fluid dynamics in the strongly collisional limit, without requiring completely different numerical limits. 
(5) Mass injection or loss, and particle destruction (e.g.\ annihilation) can be defined in a rigorous manner without a discontinuous element-creation or splitting on the grid.
(6) Force softening becomes a well-defined operation with a specific physical meaning (there is a ``correct'' formulation), because the mesh-generating points define an actual volume partition with a continuous density reconstruction well-defined over the domain.
(7) Some spurious discreteness, collisional and relaxation effects are removed \citep[see e.g.][]{melott:1997.discreteness.error.nbody,binney:2004.discreteness.in.nbody,chavanis:2006.violent.relaxation.dynamics,campa:2009.stamech.of.finite.nbody,dehnen:2011.nbody.review}. For example, spurious short-range MC/PIC/NB heating is not only suppressed but the sign of the error actually reverses: at too-low resolution, the DF is ``too smooth,'' i.e.\ real heating will be suppressed.

The tradeoff for these advantages is primarily twofold. 
(1) The method entails a significant increase in computational complexity and cost, since the per-element work is multiplicatively larger -- each face requires a per-pair Riemann solve and the construction of an $N_{\rm basis} \times N_{\rm basis}$ pairing cost matrix (plus memory cost of these variables). In the test problems we run, the larger cost is the timestep condition: a CFL-like condition must be enforced for stability because our method includes intercell fluxes, while a \textit{strictly} collisionless MC/PIC/NB method does not; but this cost difference is much smaller in situations with even weak collisionality. 
(2) The method introduces numerical diffusivity/dissipation. While not a problem for some situations, and controllable (this decreases with resolution and order of the method in a well-defined manner), this is a challenge for some problems evolved to very long timescales.

\subsection{``Eulerian'' \&\ ``Semi-Lagrangian'' Direct Vlasov Methods}
\label{sec:compare.eulerian}

Among methods which directly numerically evolve the DF in phase-space, by far the most popular methods follow the ``splitting algorithm'' of \citet{cheng:1976.vlasov.split} and discretize $f$ on a 6+1 dimensional grid. Our algorithm is fundamentally a sub-class of these but with (as implemented) a Lagrangian mesh-free discretization in both configuration and velocity space. Compared to ``Eulerian'' or ``Semi-Lagrangian'' or ``Integer Lattice'' methods \citep[see e.g.,][for a few examples]{shoucri:1978.vlasov.splitting.schemes,syer:1995.lattice.dynamics,sonnedrucker:1999.semi.lagrangian.vlasov,filbet:2001.conservative.vlasov,umeda:2008.vlasov.non.oscillatory.scheme,crouseilles:2010.semi.lagrangian.vlasov,yoshikawa:2013.6d.vlasov.direct,mocz:2017.integer.lattice.vlasov}, in which there is an explicit, regular mesh in $\Delta x$, $\Delta v_{x}$, etc., the method here has several advantages. (1) Most important, use of a Lagrangian, basis-function representation for the velocity distribution which is independently evolved for each mesh element allows us to achieve good accuracy without an enormous fixed velocity ``grid.'' This means it is computationally feasible to run large, high-resolution simulations -- the method is many orders-of-magnitude less expensive than a ``uniform grid'' in 6D phase-space as adopted in most Eulerian direct-Vlasov or lattice-Boltzmann methods \citep[compare e.g.][]{tanaka:2017.6d.phase.space.solvers.finite.volume}. (2) Because, as implemented, our grid is Lagrangian in \textit{both} position and velocity space (the mesh-generating points and velocity basis functions move with the flow in position and velocity space, respectively), we automatically achieve higher resolution in regions of interest (where structure forms and phase-space densities are high). (3) Because we maintain fixed-mass elements moving with the center-of-mass, the scheme couples naturally and accurately to standard (well-studied and efficient) MC/PIC/NB long-range force (e.g.\ gravity, Coulomb, Lorentz) solvers (maintaining conservative, symplectic and Hermitian properties of those solvers), regardless of their nature or algorithmic structure. (4) The method here ensures machine-accurate mass conservation even under variable/adaptive timestepping. 

Our method does share common ground with the RHD approach in \citet{jiang:2014.rhd.solver.local}, itself a generalization of ``short characteristics'' schemes \citep{olson:1987.short.char.rhd}. There, transport is solved on a set of rays of fixed directions on a fixed grid, for a fixed frequency, via fluxes between finite-volume elements. Our method generalizes this to an arbitrary moving mesh in space, with the ``ray'' directions ($\hat{\bf v}$) and ``frequency'' ($|{\bf v}|$ or momentum) allowed to vary in a locally-adaptive manner and treated as basis functions of arbitrary order, with additional moments to represent the shape of the DF per-basis (as compared to a $\delta$ function in $d^{3}{\bf p}$). 

Since one can, in principle, make the velocity ``grid'' as dense as desired in our scheme, there is no inherent tradeoff here, except for a modest increase in complexity of the scheme compared to the most naive Eulerian 6+1D grid schemes. However many practical implementations (e.g.\ the tests above) use a relatively limited velocity grid, so the ``velocity space'' resolution \textit{at a given spatial position} may be more limited, although across the entire simulation it may be much better.

\subsection{``Exact'' Phase-Space (Tessellation \&\ ``Waterbag'') Methods}
\label{sec:compare.waterbag}

Elegant methods exist which evolve the boundaries of phase-space distributions exactly; these have the advantage over our method here of exactly conserving phase-space density (e.g.\ they can be exactly dissipationless). However, thus far these have proven either too expensive for applications like cosmological simulations, or too restrictive for the general cases that our method is designed to simulate. For example, the ``waterbag'' method \citep{depackh:1962.waterbag.method,janin:1971.waterbag} can treat arbitrary phase-space density distributions by evolving their boundaries geometrically -- however, it is super-exponentially more expensive than our method here (and the CPU and memory costs scale super-exponentially with the number of ``phase wrappings'' and number of phase-space dimensions). This has prohibited its application in all but highly-idealized, one-dimensional problems \citep[see][]{alard:2005.cloudy.vlasov,besse:2008.wavelet.vlasov,colombi:2014.waterbags}. \citet{abel:phase.space.sheets.for.dm,shandarin:2012.tesselation.cold.flows,hahn:2013.dm.phase.space.sheet.sims} have developed a specialized application of this method -- valid for very specific initial conditions and assumptions about the collisionless fluid -- into a much more efficient tessellation method \citep[extended in][]{hahn:2016.tesselation.cdm.scheme,sousbie:2016.vlasov.poisson.tesselation,kaehler:2017.dm.tesselation.sims} for adapting traditional N-body methods to evolve phase space ``sheets'' directly (also adapted as the ``simplex-in-cell'' or SIC technique in PIC simulations; \citealt{kates.harbeck:2016.SIC.plasma.sims}). However, as these authors caution, their method requires initial conditions which form an infinitesimally thin sheet in phase space (a highly restrictive case) and does not allow general collision/mixing operators.\footnote{A collision operator in the tessellation method of \citet{hahn:2013.dm.phase.space.sheet.sims} would distort the ``sheets'' in phase space where they intercept other sheets, and could scatter matter out-of-plane of the sheet, violating the fundamental assumptions of the reconstruction.} In addition, this method requires expensive reconstruction of all intersecting sheets to evaluate phase-space distributions, and again becomes super-exponentially more expensive when multiple phase-wrappings occur in gravitational collapse (prohibiting many applications; adaptive refinement of the phase-space elements, as in \citealt{hahn:2016.tesselation.cdm.scheme,stucker:2020.hybrid.dm.phase.sheet.where.simple.switches.to.nbody.in.dense.haloes.where.sheet.complexity.explodes}, can follow strong mixing, but the number of elements and hence the cost still grows rapidly with the number of wraps).

\begin{figure*}
\plottwo{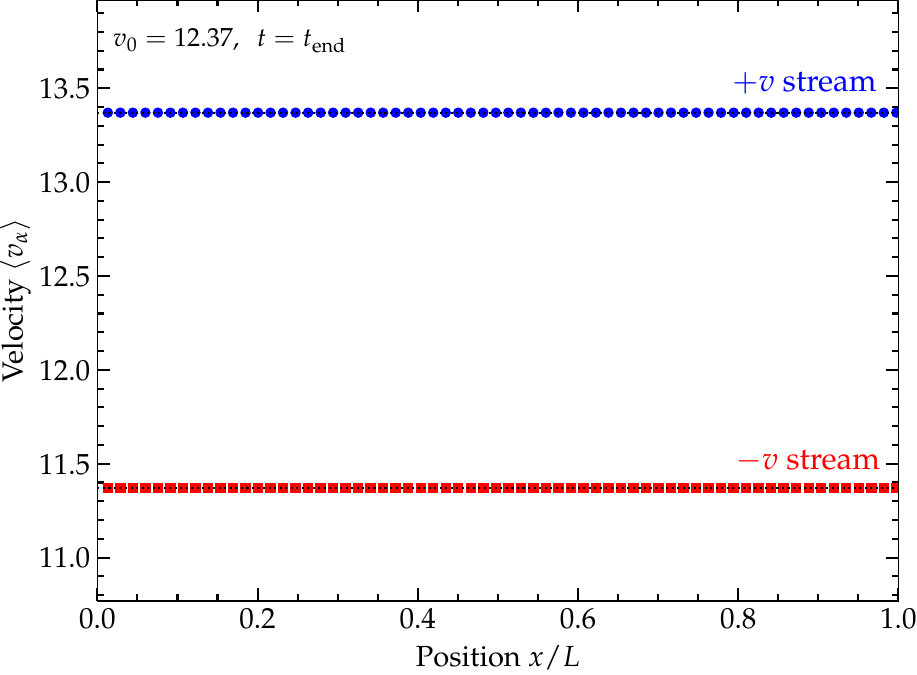}{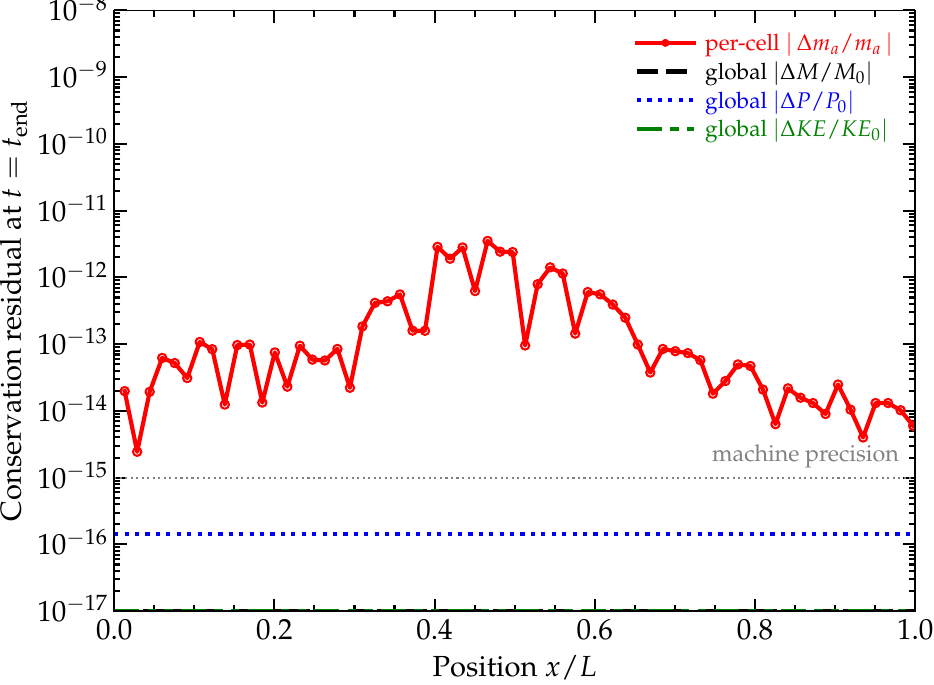}\vspace{-0.5cm}
\caption{Two-stream test of \S~\ref{sec:test.2stream}, after many crossing times in 1D with $N=64$ mesh-generating points under a Galilean boost $v_{0} = 12.37$.
\textit{Left:} final phase-space distribution $f(x,\,v_{x},\,t_{\rm end})$. The exact solution is two horizontal lines at $v = v_{0} \pm 1$ (blue: $+v$, red: $-v$); the two streams are preserved to integration accuracy with no visible diffusion in velocity space.
\textit{Right:} cumulative conservation residuals at the end of the run. The per-cell mass drift $|\Delta m_{a}/m_{a}|$ (red, sampled at every mesh cell) is everywhere below $10^{-11}$. The globally-conserved scalars all sit at or below the double-precision floor (machine-precision reference indicated): $|\Delta M / M_{0}|$ and $|\Delta KE / KE_{0}|$ are exactly zero in the floating-point arithmetic, while the momentum drift $|\Delta P / P_{0}|$ is $\sim 10^{-16}$. The same test repeated in the rest frame is identical at every cell to machine precision after subtracting the boost shift (Galilean invariance; see also Fig.~\ref{fig:test.galilean}).
\label{fig:test.2stream}}\vspace{-0.3cm}
\end{figure*}

\subsection{``Moments'' \&\ Wavefunction Methods}
\label{sec:compare:moments}

Obviously there is a significant methods overlap with 10 and 13-moments methods for evolving the DF in plasma physics. We extend and generalize these methods (for the first time, to our knowledge) in several ways including the application to moving meshes, mass-conserving MFM-like grids, new DF shapes (top-hat vs.\ Gaussian-like), HLLC-like and exact Riemann problem solutions coupled to higher-order reconstruction, etc. But the most important qualitative extension is the key conceptual difference: instead of a \textit{single} set of moments describing the entire DF (the standard ``moments'' method approach), we generalize to a set of basis functions. This removes the arbitrary ``closure'' choice from the moments methods: provided sufficient basis number, \textit{any} DF can be represented in our method. In contrast, traditional 13-moments methods, by assumption, can only represent a single Gaussian DF (with a single skewness parameter for all directions). This means simple collisionless phenomena modeled here, like cold-stream-crossing, cannot be correctly captured (though to be fair, for a charged, magnetized plasma with ``effective'' collisionality owing to kinetic effects, those types of DFs are often unstable).

It is worth being more specific about this lineage, since the plasma and multiphase-flow moment-closure literature directly informs our work here. The single-Gaussian-basis closures at the 10/13-moment level are traditionally designed for close-to-thermal-equilibrium DFs, as they descend from Grad's Hermite expansion \citep{grad:1949.kinetic.theory.rarefied.gases.hermite.13.moment}. Historically this has the pathology that thermal-equilibrium does not actually lie in the hyperbolic region \citep[see][]{cai.fan.li:2014.globally.hyperbolic.regularization.grad.moment.system}, which has motivated various hyperbolicity-restoring closures, e.g.: Pearson-IV and related interpolations that make the fourth moment[s] a function of the reduced-third moment \citep{torrilhon:2010.hyperbolic.moment.equations.pearson.iv.kinetic.gas,mcdonald.torrilhon:2013.affordable.robust.maximum.entropy.moment.closures}; the 13-moment finite-volume/discontinuous-Galerkin plasma solvers \citep{miller.shumlak:2016.multispecies.13.moment.moderately.collisional.plasma}; and globally-hyperbolic quadrature closures such as HyQMOM \citep{fox.laurent.vie:2018.conditional.hyperbolic.quadrature.method.moments}. 
Our multi-basis representation sidesteps the closure-hyperbolicity problem entirely by construction: each basis is individually a realizable DF, so the mixture is realizable, and hyperbolicity is inherited from the single-basis one-sided kinetic fluxes (\S\ref{sec:flux}) rather than imposed through a moment-inversion closure. 
The closest relatives to the multi-basis idea are the extended and multi-Gaussian quadrature methods of moments \citep[EQMOM/MG-EQMOM;][]{fox:2008.eqmom.gaussian.quadrature.moment.method,marchisio.fox:2013.qmom.book.multiphase.moments} developed for turbulent disperse multiphase flows, in which the DF is represented as a weighted sum of kernel densities (Gaussians, in the multivariate case) reconstructed from a finite set of moments \citep{chalons.fox.laurent.massot.vie:2017.multivariate.gaussian.eqmom.multiphase}. These are closely related quadrature-moment approaches with different assumptions and applications: they are typically applied on fixed Eulerian grids without spatial reconstruction, at fixed low node count, with isotropic or restricted-anisotropy kernels, and are aimed at the moderate-node-number closure-inversion problem rather than at collisionless multi-streaming. The method here differs in adopting a Lagrangian (moving, mass-conserving MFM) mesh with second-order spatial reconstruction, full arbitrary anisotropy of each ${\bf S}_{\alpha}$ in $D=3$, an arbitrary user-set $N_{\rm basis}$, a distinct one-sided kinetic Riemann flux and face-velocity construction, and the basis-pairing and empty-basis machinery of \S\ref{sec:pairing} that has no analog in the fixed-node quadrature methods. One can therefore regard it as a member of the same broad quadrature-moment family, but generalized to truly collisionless or weakly-collisional, multi-streaming, anisotropic complex flows. 

Other ``moments'' methods in the literature tend to be even simpler single-basis examples. For example, in radiation-hydrodynamics (RHD), a single basis (per frequency, so the DF in $|{\bf p}|$ is a $\delta$-function) is followed with only 4 moments (energy density and flux) in e.g.\ M1 (and 1 moment in flux-limited diffusion), so the closure is even more restrictive of the shape of the DF and its anisotropy \citep{levermore:1981.fld}. This type of simple closure has also been applied in ``Boltzmann moment methods'' for stellar and dark matter systems \citep{mitchell:2013.boltzmann.moments.methods.for.stars.dm}. But it is well-known that these cannot describe even simple two-stream or anisotropic-DF situations, let alone opposing streams or complicated phase-wrapping.

A number of authors \citep{widrow:1993.schrodinger.for.collisionless.dm,schaller:2014.schrodinger.for.dm.structure,schive:2014.interference.scalar.field.dm,briscese:2017.schrodinger.for.nbody,kopp:2017.schrodinger.vlasov.equation} have also noted that pressure-free fluids can (\textit{assuming} the local velocity DF is trivial) be represented as complex scalar fields $\Psi$ obeying a Schr{\"o}dinger-Poisson-type set of equations. This makes the density field ($\rho \propto |\Psi|^{2}$) smooth and continuous, but again collapses the velocity DF to be single-valued at each ${\bf x}$ (${\bf v} \propto \nabla({\rm arg}[\Psi])$), akin to the ``moments'' methods above. This is a good approximation for e.g.\ Bose-Einstein condensate dark matter, where the system is a true superfluid (the group velocity \textit{is} single-valued, locally), but for collisionless matter one must extend this with a series of wavefunctions \citep{schaller:2014.schrodinger.for.dm.structure} akin to our basis functions (but each evolved globally, requiring vastly larger expense to achieve comparable velocity-space resolution). Moreover these methods do not allow collisions, nor couple easily to standard $N$-body/Poisson/Lorentz schemes, nor necessarily conserve mass or momentum, and they entail a steep (quadratic) timestep penalty.

\section{Numerical Tests}
\label{sec:tests}

We now consider a suite of numerical tests of our particular implementation in \GIZMO. We stress that different core implementation choices can give significantly different results. Volume partition, spatial reconstruction, slope-limiting, basis-function shape, number-of-moments, number-of-bases, time-integration scheme, pairing cost assignment, etc. -- these are not ``universal'' and there is no single ``correct'' choice. The purpose here is to highlight qualitative features of the scheme, its ability to handle some situations which are highly non-trivial for 6D Vlasov/Boltzmann or ``moments'' methods or MC/PIC/NB methods, and to demonstrate that it exhibits the stability and convergence properties expected. Unless otherwise stated all implementation choices follow \S~\ref{sec:implementation}, and initial conditions (ICs) are constructed with equal-mass cells. Every test in $D=1$ (1D) has also been run in 3D (with a periodic box and uniform ICs along the extra dimensions), and a couple tests in $D=2$, to validate the method works across dimensions.

\subsection{Two-Stream and Galilean Invariance}
\label{sec:test.2stream}

We first consider a simple 2-stream test. In a periodic 1D box of size $L=1$ and total mass $M=1$, we assign $N_{\rm cells}=64$ uniformly-spaced mesh-generating points (so $\Delta x = L/N_{\rm cells}$) the initial distribution $f(x,\,v,\,t=0) = (\delta[v-1] + \delta[v+1])/2$ ($N_{\rm basis}=2$ per cell), i.e.\ uniform density with half the mass moving at $v=+1$ and half at $v=-1$, then boost the entire box by an arbitrary $v_{0} = 12.37$ in the $\hat{x}$-direction and evolve for many crossing times (Fig.~\ref{fig:test.2stream}). The method reproduces the exact solution to integration accuracy: the two streams are preserved with no detectable diffusion in either position or velocity space; total mass, momentum, and kinetic energy are conserved to machine precision; and the element masses $m_{a}$ are constant in time to the same precision. The repair operator (\S~\ref{sec:repair}) is never triggered. The cost per timestep is independent of how far forward in time the integration is run -- in particular it does not grow as phase wrappings accumulate. 

For comparison, in MC/PIC/NB methods this is a trivial test if half the particles are assigned $v = \pm 1$ (although it is impossible to represent both streams at a single spatial point ${\bf x}$, which is the actual physical content of the initial condition; the Monte-Carlo realization spreads the streams across positions). In moments-based or Schr{\"o}dinger methods (\S~\ref{sec:compare:moments}), it is impossible to ever converge to the correct solution for even this simple problem. Tessellation and waterbag methods, in turn, have difficulty evolving this configuration arbitrarily far forward in time as phase-wrappings accumulate in position-space and the sheet topology becomes pathological.

\begin{figure}
\includegraphics[width=0.965\columnwidth]{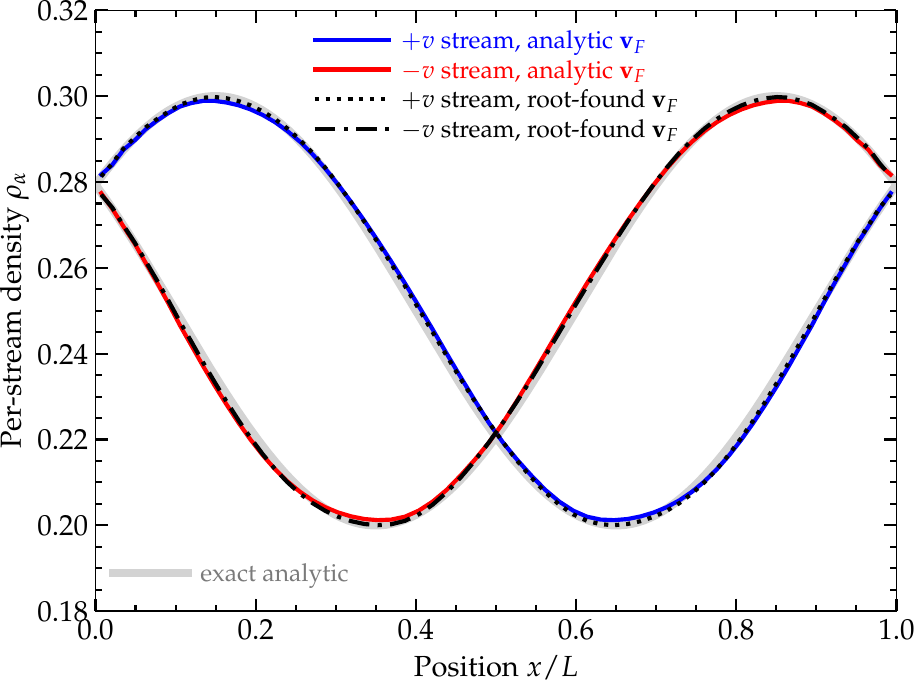}
\includegraphics[width=0.97\columnwidth]{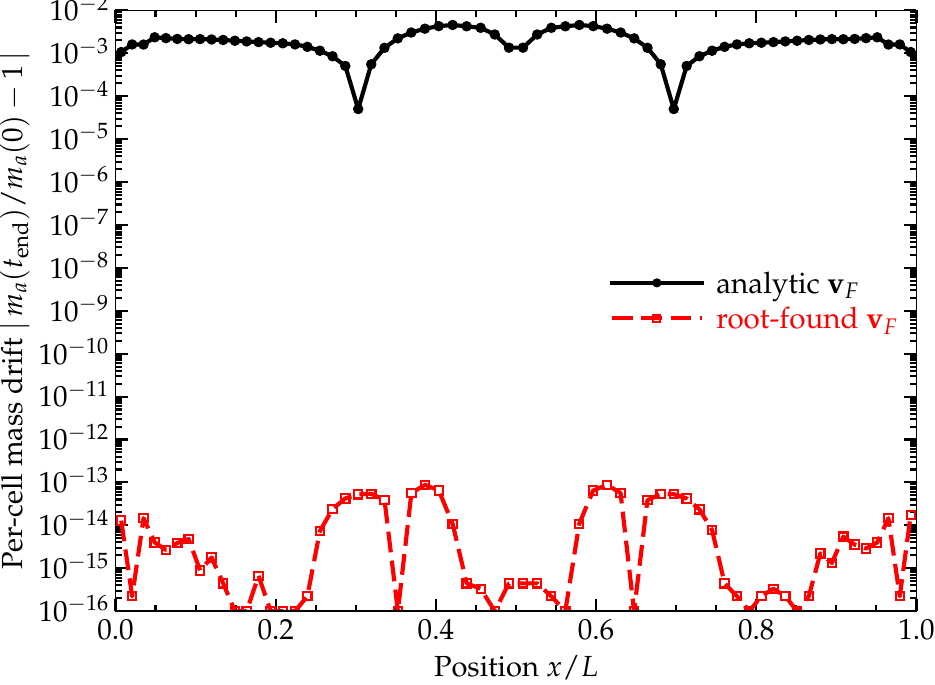}
\vspace{-0.2cm}
\caption{Density-wave test of \S~\ref{sec:test.density.wave.vF}, comparing two choices for the face velocity ${\bf v}_{F}$: the continuum expression (Eq.~\ref{eqn:vF.analytic}) and the per-face root-find on $\Sigma F_{m}(v_{F,n}) = 0$ described in \S~\ref{sec:vF.policy}.
\textit{Top:} per-stream density profiles $\rho_{\alpha}(x,\,t_{\rm end})$ at the end of the run, with the $+v$ (blue) and $-v$ (red) streams from the analytic ${\bf v}_{F}$ shown solid, and the corresponding $+v,\,-v$ streams from the root-find shown as black dotted and dash-dotted lines on top. The two-stream initial condition has the exact analytic solution $\rho_{\pm}(x,\,t) = \tfrac{1}{2}\,[1 + \epsilon\,\cos(2\pi(x\mp t)/L)]$ (each stream rigidly advecting at $v = \pm 1$); this is plotted as the heavy grey curve and lies under both numerical curves to within line thickness.
\textit{Bottom:} the per-cell relative mass drift $|m_{a}(t_{\rm end})/m_{a}(0)-1|$ as a function of position across the mesh, for the two configurations. The analytic ${\bf v}_{F}$ (black) leaves a per-cell mass drift of $\sim 10^{-3}$; the root-find (red) collapses it to machine precision ($\sim 10^{-14}$), $\sim\!11$ orders of magnitude lower, while not visibly changing the integrated per-stream solution.
\label{fig:test.density.wave.vF}}\vspace{-0.cm}
\end{figure}

\subsection{Face Geometry on Irregular Meshes}
\label{sec:test.mesh}

We turn next to the face geometry. The face-vector construction (Eqs.~\ref{eqn:mfm.face.area.def}-\ref{eqn:h.ngb}) follows the standard MFM definition \citep{hopkins:gizmo}, but it enters the CBE machinery in two non-trivial ways: directly through the per-face unit normal $\hat{\bf A}_{ab}$ which appears in the Riemann solution and the basis-pairing participation gate, and indirectly through $V_{a}$ which sets the $Q = U/V$ primitive variables. On a uniform mesh in 1D, the discrete divergence identity $(1/V_{a})\sum_{b} F_{\rm face}({\bf x}_{ab})\,{\bf A}_{ab} = \nabla F({\bf x}_{a})$ holds to machine precision for any linear $F$, and the antisymmetry ${\bf A}_{ab} = -{\bf A}_{ba}$ and row-sum vanishing $\sum_{b}{\bf A}_{ab} = {\bf 0}$ both hold identically. On non-uniform meshes -- here, sampled from $\rho(x) = 1 + \epsilon\cos(2\pi x/L)$ via inverse-CDF at $N_{\rm cells}=64$ (so that the mean cell spacing is $\Delta x \sim L/N_{\rm cells}$), the same density-wave configuration we will use below -- the local divergence-identity error is at the level of the standard MFM linear-reproduction truncation. We stress that {this is purely the local-reconstruction truncation, not a violation of global or per-cell mass conservation}: the row-sum vanishing $\sum_{b}{\bf A}_{ab} = {\bf 0}$ and the pair-symmetric flux assignment ensure that the cell-integrated mass update remains exactly conservative for any face-area definition with these properties. The error here is the same one which appears in the underlying MFM hydrodynamic method and which converges in the usual way with resolution and kernel breadth, and the limit is set by the underlying MFM, not by anything specific to the CBE scheme.

\subsection{Density Wave (\&\ Face Velocity ${\bf v}_{F}$)}
\label{sec:test.density.wave.vF}

A more demanding test of the per-face mass-flux invariance discussed in \S~\ref{sec:vF.policy} is provided by a simple density-wave initial condition. We sample $N_{\rm cells} = 64$ mesh-generating points from $\rho(x) = 1 + \epsilon\cos(2\pi x/L)$ with $\epsilon = 0.2$ via inverse-CDF in 1D, giving each cell $N_{\rm basis} = 2$ equally-weighted basis components at $v = \pm 1$, and evolve to $t = 1$ in units of the box-crossing time at the stream speed. In the continuum limit the solution is exactly the per-stream advection of the two density waves at $v = \pm 1$; in the discretization, however, the spatial variation of $V_{a}$ drives the integration through the $Q = U/V$ frame (\S~\ref{sec:Q.vs.U}), so the test exercises the full reconstruction, Riemann, and pairing machinery non-trivially. We run the test in two configurations identical in every respect except for the face velocity ${\bf v}_{F}$, set either to the analytic continuum expression of Eq.~\ref{eqn:vF.analytic}, or determined per-face by the bracketed Brent root-find on the basis-summed mass flux as described in \S~\ref{sec:vF.policy}, and compare to the exact analytic solution (each stream rigidly advecting at $v = \pm 1$).

The per-stream densities are essentially identical between the two configurations and indistinguishable from the analytic solution (Fig.~\ref{fig:test.density.wave.vF}, top). The per-cell mass drift, however, differs dramatically (Fig.~\ref{fig:test.density.wave.vF}, bottom): with the analytic ${\bf v}_{F}$ the maximum element-mass drift accumulates to a few$\times 10^{-3}$ by the end of the run, while with the root-found ${\bf v}_{F}$ it stays at the double-precision floor ($\sim\!10^{-14}$). The two configurations evolve essentially identical per-stream profiles only because the integration is short enough that the per-cell mass drift has not yet been fed into the gravity or basis-pairing operators in a way that produces visible secondary errors; in a cosmological run with $\sim\!10^{6{-}7}$ steps, the corresponding drift in the analytic configuration would accumulate into a serious source of spurious heating in the gravity solver. We therefore use the root-found ${\bf v}_{F}$ by default in all following tests.

In MC/PIC/NB methods this is mechanically similar to the 2-stream case but much more difficult to represent -- equal-mass particles (as typically adopted) mean that representing a wave with fractional $\epsilon \ll 1$ requires an enormous particle number. Moreover there simply is no unique definition of the $\rho$ reconstruction -- since it can only be defined in post-processing via some kernel-density estimator, if one chooses too small/large a kernel the density fluctuations are over-or-under-estimated, and the subtle velocity gradient is noise-dominated. Even for the ideal kernel, convergence in MC/PIC/NB problems requires a kernel neighbor number $N_{\rm ngb}\to \infty$ while particle number $N_{\rm tot} \to \infty$ slightly faster, giving best-case convergence rates roughly like $N^{-1/4}$, whereas our method scales as $N^{-2}$. 
In moments or Schr{\"o}dinger methods this configuration is simply impossible to represent, and tessellation or waterbag methods break almost immediately (after a couple crossings the cost becomes super-exponential in time; see \citealt{hahn:2016.tesselation.cdm.scheme,stucker:2020.hybrid.dm.phase.sheet.where.simple.switches.to.nbody.in.dense.haloes.where.sheet.complexity.explodes}). 
While conceptually this can be represented just as accurately by ``direct'' or Eulerian/lattice-Boltzmann or Vlasov methods, the motion over the configuration-space grid produces serious damping (this is hydrodynamically equivalent to a soundwave problem with an infinite-Mach-number boost over the grid), which wipes out the wave in a couple of box-crossing times \citep{hopkins:gizmo}; our method produces zero damping.

\begin{figure}
\includegraphics[width=0.98\columnwidth]{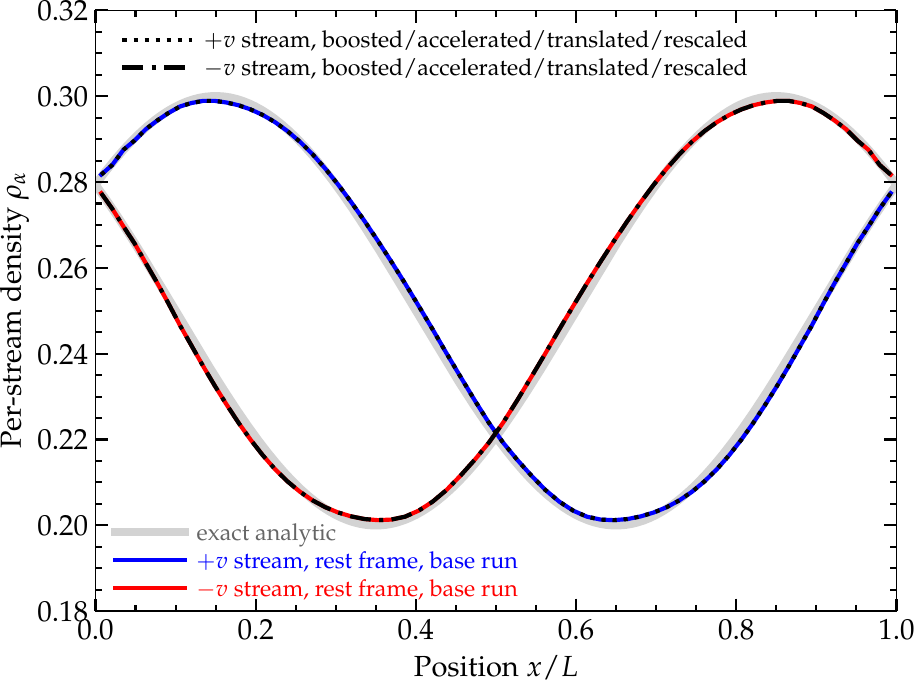}\vspace{-0.2cm}
\caption{Galilean-invariance test of \S~\ref{sec:test.density.wave.galilean}. 
We compare the results of the 2-wave test as Fig.~\ref{fig:test.density.wave.vF}, to a version which is rotated, has units rescaled, translated, boosted, has a uniform acceleration added, has $N_{\rm basis}$ increased, and adds a constant non-moving background density (the latter is translated and boosted back to the same frame for comparison here). 
The two are indistinguishable as expected.
\label{fig:test.galilean}}
\end{figure}

\subsection{Invariance: Translation, Rotation, Units, Galilean, Equivalence, and More}
\label{sec:test.density.wave.galilean}

We validate the invariances our method should exhibit by taking the 2-stream and density wave test ICs, and 
(1) rotating by a random angle; 
(2) multiplying by different unit masses and lengths (rescaling the time unit accordingly); 
(3) adding a large translation/coordinate offset; 
(4) boosting the velocities by a random $\mathcal{O}(1)$ vector; 
(5) adding a uniform different random vector acceleration; 
(6) increasing the basis function number by $12$; 
(7) adding a constant non-moving background density which is much larger than the moving densities. 
After shifting back to a comoving frame, the per-stream relative densities agree to a relative $L_{1}$ error of $\sim 10^{-13}$ between the simulations (Fig.~\ref{fig:test.galilean}). 
This is consistent with the continuum equations, but not automatic in discretization. 

MC/PIC/NB methods can also be invariant to 1-6. However many violate (5) because (lacking a well-defined and Galilean-invariant Courant criterion) they use timestep criteria explicitly based on the absolute acceleration scale \citep[see][]{power:2003.nfw.models.convergence,grudic:2020.tidal.timestep.criterion}. They are explicitly \textit{not} invariant to (7), as adding a background density at fixed particle number directly increases the shot noise in representing a small perturbative velocity flow ``on top'' of this (the error increases with [7]). 
Direct/Eulerian/lattice-Boltzmann Vlasov methods fail badly under 1/3/4/5, as these require both a much larger grid, but also involve much stronger advection at oblique angles across both the real and velocity-space grids, dramatically increasing damping. 
Waterbag/tessellation/moments/wavefunction methods already fail to evolve these problems and that failure becomes more pronounced with these variations.

\subsection{Convergence with $N_{\rm cells}$}
\label{sec:test.convergence}

\begin{figure}
\includegraphics[width=0.98\columnwidth]{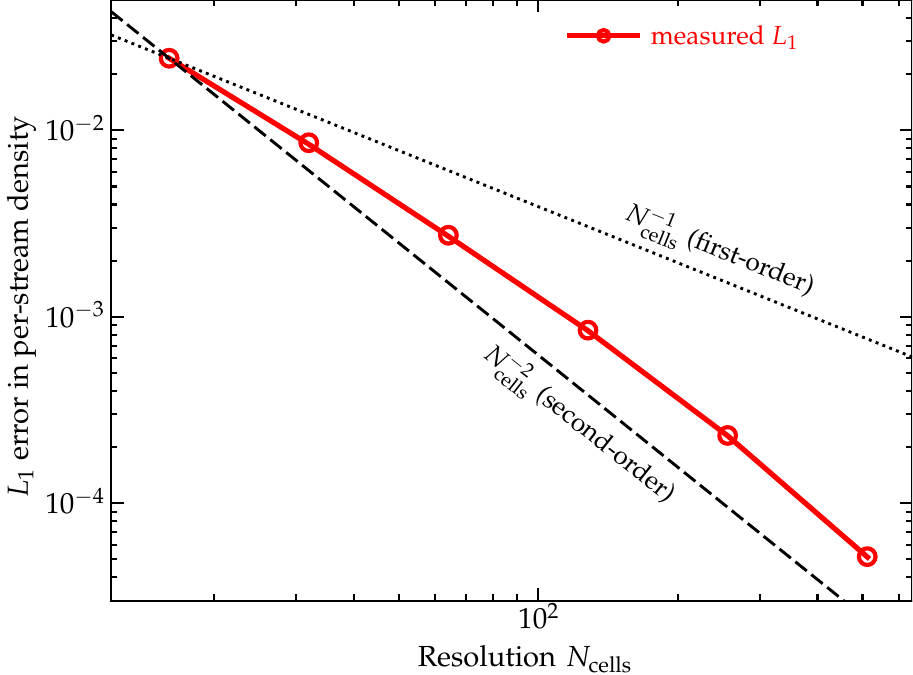}\vspace{-0.2cm}
\caption{$L_{1}$ convergence on the two-stream density-wave test of \S~\ref{sec:test.density.wave.vF}, evaluated at $t = 0.15$ against a same-kernel, same-scheme high-resolution reference run at $N_{\rm cells}^{\rm ref} = 1024$. Red circles: measured $L_{1}$ error in the per-stream density (averaged over the two streams), normalized by the mean density, as a function of $N_{\rm cells}$ at fixed $N_{\rm basis} = 2$ (the minimal-sufficient basis for two physical streams). Dashed and dotted lines mark the formal $N_{\rm cells}^{-2}$ (second-order) and $N_{\rm cells}^{-1}$ (first-order) slopes for reference; the measured slope steepens from $\approx -1.7$ at low $N_{\rm cells}$ to $\approx -2.1$ over the $N_{\rm cells}=256{-}512$ interval, recovering the formal second-order accuracy of the underlying MFM reconstruction as the TVD slope limiter becomes inactive at finer resolutions.
\label{fig:test.convergence}}
\end{figure}

Fig.~\ref{fig:test.convergence} shows a simple convergence study. Fixing $N_{\rm basis} = 2$ on the two-stream density-wave configuration of \S~\ref{sec:test.density.wave.vF} (which carries exactly two physical streams, so $N_{\rm basis} = 2$ is already the minimal sufficient basis), we vary $N_{\rm cells} \in \{16,\,32,\,64,\,128,\,256,\,512\}$ and integrate to a fixed $t = 0.15$ in box-crossing units. We then compute the $L_{1}$ error in the per-stream density profiles against a same-kernel, same-scheme reference run at $N_{\rm cells}^{\rm ref} = 1024$ (rather than against an external semi-Lagrangian reference, whose own kernel and density-reconstruction biases would otherwise dominate the residual at high $N_{\rm cells}$ and artificially flatten the measured slope; though the results are in fact similar if we just compare to the analytic solution), normalized by the mean density. The measured error decreases as $L_{1} \propto N_{\rm cells}^{-p}$ with $p$ approaching $2$ at the highest resolutions tested ($p \approx 2.1$ between $N_{\rm cells} = 256$ and $512$), confirming the formal second-order accuracy of the underlying MFM reconstruction. The slight steepening with $N_{\rm cells}$ is the expected behavior of a TVD slope limiter (\S~\ref{sec:reconstruction}): the BJ limiter activates more rarely as the local density profile becomes better-resolved, and the second-order MFM stencil is recovered everywhere away from the limited region.

\begin{figure}
\includegraphics[width=0.98\columnwidth]{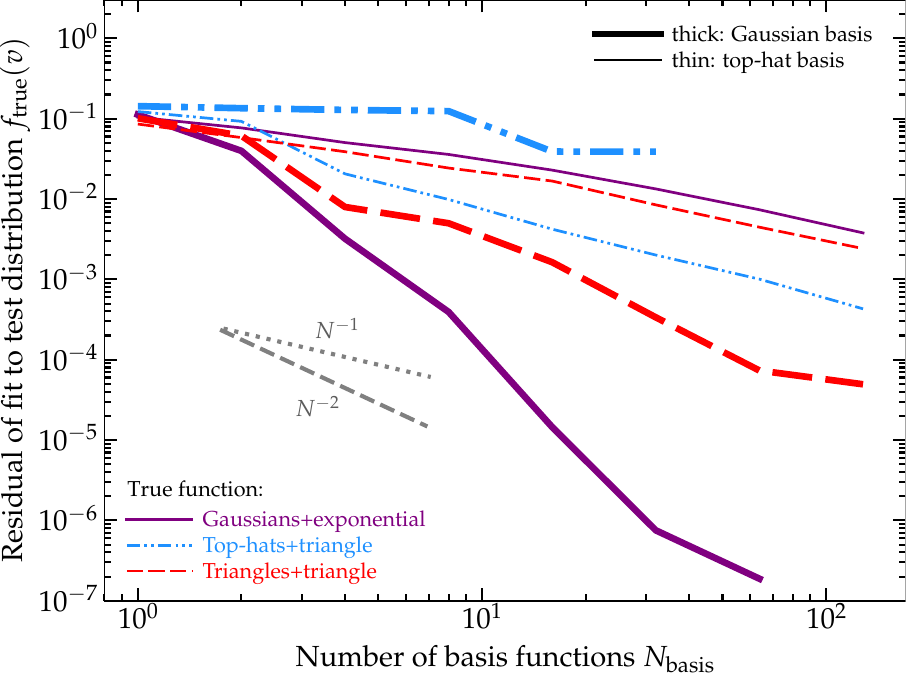}\vspace{-0.2cm}
\caption{Convergence in velocity-space on representation of a deliberately non-trivial target DF $f_{\rm true}(v)$ (Eq.~\ref{eqn:test.f.true}; two stream functions $\mathcal{H}_{s}$ with different DFs themselves plus a broad background component $\mathcal{H}_{b}$), as a function of $N_{\rm basis}$. We plot the rms fit residuals vs.\ $N_{\rm basis}$, assuming either top-hat or Gaussian shaped basis functions in velocity-space, and assuming $f_{\rm true}$ is composed of a mix of either: Gaussians + exponential Laplace functions (non-compact, non-differentiable); top-hats+triangle functions (compact but discontinuous and non-differentiable); or different-sloped-triangular wedges (much broader and non-differentiable). 
Top-hat bases converge as $N_{\rm basis}^{-1}$ as expected. Smooth Gaussian bases converge much faster at low-to-intermediate basis number for smooth $f_{\rm true}$, but slower for truly discontinuous $f_{\rm true}$.\label{fig:test.convergence.Nbasis}}
\end{figure}

\subsection{Convergence with $N_{\rm basis}$: Reconstruction of a Non-Trivial Velocity DF}
\label{sec:test.convergence.Nbasis}

A finite-cardinality Gaussian or top-hat basis is an {exact} representation of any underlying $f({\bf v})$ that is itself a finite Gaussian or top-hat mixture, respectively. 
We consider streaming tests below with cold or matched Gaussian/top-hat streams and $N_{\rm streams} \le N_{\rm basis}$ and verify the reconstructed stream agrees at machine-precision with its analytic counterpart. 
Reconstruction of the DF in \textit{space} is a question of convergence of the spatial reconstruction order, e.g.\ \S~\ref{sec:test.convergence}. 
So the non-trivial question of convergence in velocity-space is the question of the accuracy with which an arbitrary (non-Gaussian/top-hat) target DF can be reconstructed by a sum of $N_{\rm basis}$ finite bases. 

By definition, the total integrals and desired (10/13) moments of the DFs are exactly represented to machine accuracy by our method. 
So a more meaningful error is something like the $L_{n}$ norm of the velocity DF.\footnote{Denote the true DF $f=f_{\rm true}({\bf v})$ and the reconstructed DF $g=\sum_{\alpha} \varpi_{\alpha} \tilde{g}_{\alpha}\left( {\bf v} \right)$. We can define the velocity DF error norm like the usual norm, at the ``quadrature points'' for the reconstruction, i.e.\ mean $\langle {\bf v}_{\alpha} \rangle$ of each basis, as $L_{n} \equiv \left[ N_{\rm basis}^{-1} \sum_{\alpha} \| f( \langle {\bf v}_{\alpha} \rangle) - g( \langle {\bf v}_{\alpha} \rangle)\|^{n} \right]^{1/n}$. But this tends to over-estimate the fit quality for cases like Eq.~\ref{eqn:test.f.true}, where the domain is non-compact, and ignores the non-trivial additional basis function information. We can therefore instead measure an ``rms-like'' $\Delta_{n} \equiv \left[ \Delta v^{-1} \int \|f({\bf v}) - g({\bf v})\|^{n} d^{D}{\bf v} \right]^{1/n}$, but on an unbounded domain this requires some sensible $\Delta v$, which here we have taken to be either the size of the interval over which $99\%$ of the integral/mass in $f$ is contained, or a constant $(-10,\,10)$. Testing all these choices for $n=1,\,2,\,4$, none changes the qualitative scalings in Fig.~\ref{fig:test.convergence.Nbasis}, just the normalization.}
For top-hat basis functions of variable width and independent centers, this is just a straightforward piecewise-constant reconstruction problem and the $L_{n}$ or $\Delta_{n}$ error norms scale $\propto N_{\rm basis}^{-1}$. This is true even for discontinuous functions, though for the maximally non-smooth case (discrete cold streams), we find it useful to instead think of $N_{\rm basis}$ as the number of independent cold streams which can be captured before numerical phase-mixing ensues. 
For a Gaussian basis this is radial basis function interpolation, which is more complicated, but 
if the DF is smooth to degree $k$ and the fit is restricted to a finite domain, then the theoretical convergence limit is $\propto N_{\rm basis}^{-k}$ (exponential as $k\to\infty$) as $N_{\rm basis} \to \infty$. 

Consider an illustrative (challenging) example, with some numbers in Fig.~\ref{fig:test.convergence.Nbasis}. Let the true DF have the form 
\begin{align}
\label{eqn:test.f.true}
f_{\rm true}(v) &= 0.35\,\mathcal{H}_{s}(\mu_{1},\,\sigma_{1}) + 0.35\,\mathcal{H}_{s}(\mu_{2},\,\sigma_{2}) + 0.3\,\mathcal{H}_{b}(\mu_{3},\,\sigma_{3})
\end{align}
where $\mathcal{H}_{s,b}(v;\,\mu,\,\sigma)$ are unit-normalized functions with means $\mu_{1,2,3}=(7/3,0.33,0)$ and dispersions $\sigma_{1,2,3}=(4/3,0.1,1.42)$, which we vary. 
We intentionally consider functions which are non-Gaussian/top-hat, non-differentiable/discontinuous, and non-compact: experimenting with $\mathcal{H}_{s}$ as a Gaussian, or top-hat, or triangular wedge function; and letting $\mathcal{H}_{b}$ similarly be a Gaussian, top-hat, triangle, smooth $\propto \exp{(-|v/v_{0}|^{4})}$ function, or exponential Laplace tail function $\propto e^{-2|v|/3}$. 
In every case, the top-hat convergence approaches $\propto N_{\rm basis}^{-1}$ as expected over $N_{\rm basis} = 1-128$. 
As $N_{\rm basis}$ becomes large, most of the work goes into fitting out the non-compact tails of the distribution. 
For the Gaussian bases, we confirm exponential convergence if we fit only a restricted domain (say $|v|<5$) and pick an infinitely-differentiable $f_{\rm true}$ (e.g.\ Gaussian or $\exp{(-|v|^{4})}$ forms). 
However if we fit the entire domain, then even for these smooth configurations, or for the exponential/Laplace or triangular $\mathcal{N}$ (continuous but non-differentiable) cases, the scaling is often rapid or exponential at low-$N_{\rm basis}$ but becomes $\propto N^{-1}$ at large-$N$ (albeit with a lower normalization than the top-hat case), with the residual dominated in the exponential case by the slowly-falling Laplace tails. 
Not surprisingly (given its simplicity in ${\bf v}$-space), in this particular reconstruction metric (other measures of ``quality'' below differ), the only case where top-hat bases outperform smooth Gaussian bases is if the DF is discontinuous ($k=0$, e.g.\ top-hat $\mathcal{H}_{s,b}$).

\begin{figure}
\hspace{0.07cm}\includegraphics[width=0.97\columnwidth]{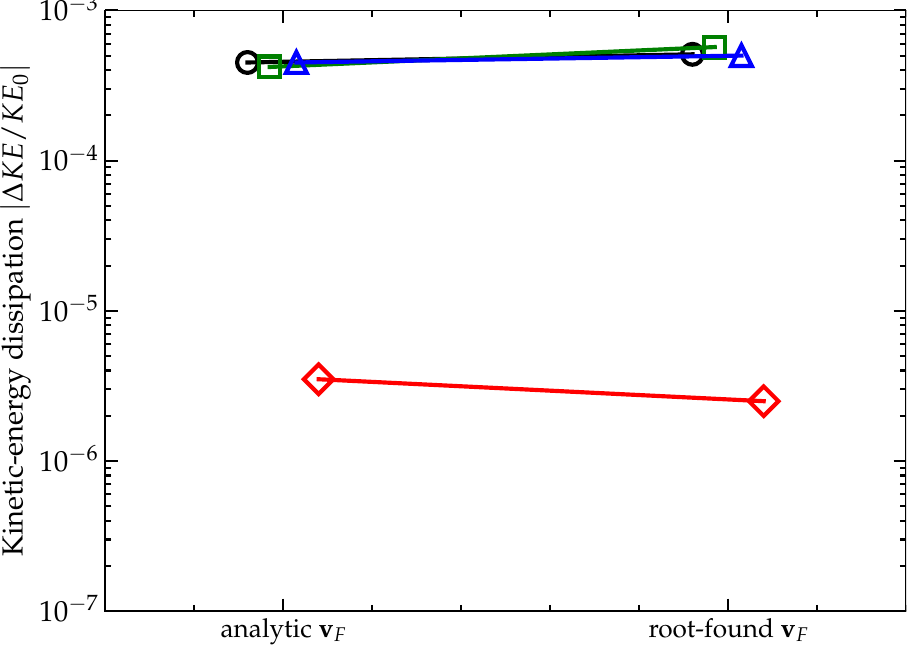}
\includegraphics[width=0.98\columnwidth]{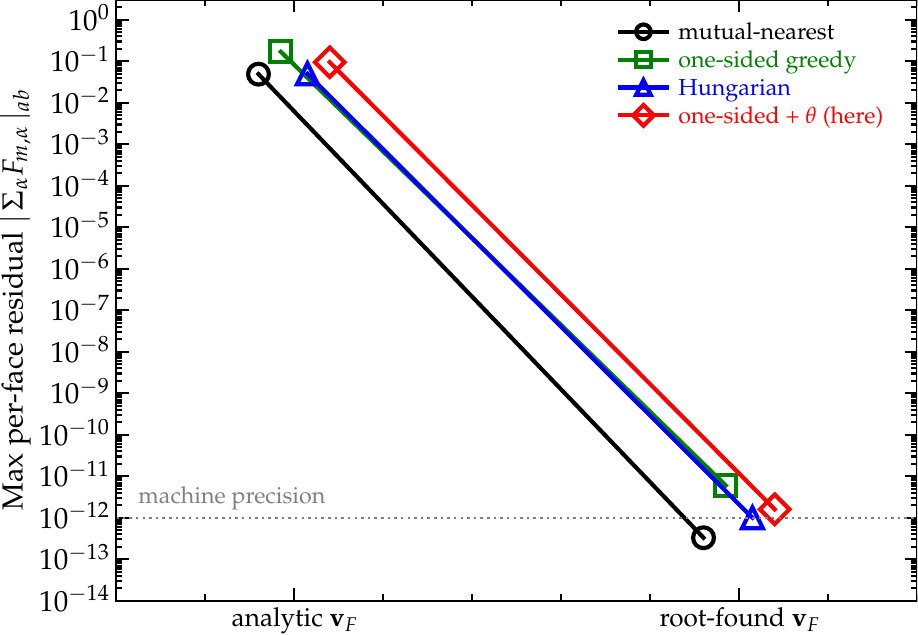}
\caption{Properties of distinct basis-matching/pairing rules (symbols) of \S~\ref{sec:pairing} on the asymmetric-pairing test of \S~\ref{sec:test.pairing}, under either the analytic or the per-face root-found face velocity ${\bf v}_{F}$ (\S~\ref{sec:vF.policy}). The one-sided+$\theta$ (vacuum) rule is the default adopted. 
\textit{Top:} relative kinetic-energy dissipation $|\Delta KE/KE_{0}|$. The one-sided vacuum assignment combined with the corresponding one-sided collisionless Riemann solution (\S~\ref{sec:flux.onesided}) dissipates kinetic energy at roughly $1\%$ of the rate of the symmetric assignments combined with the bilateral solver, independently of the choice of ${\bf v}_{F}$.
\textit{Right:} maximum per-face residual in the basis-summed mass flux $\Sigma_{\alpha} F_{m,\alpha}$. The continuum ${\bf v}_{F}$ gives a residual at the level of the Riemann diffusion ($\sim 10^{-1}$), while the per-face root-find collapses it to machine precision for every pairing rule. Total mass and momentum are conserved to machine precision in every configuration.
\label{fig:test.pairing}}
\end{figure}

\subsection{Asymmetric Pairing}
\label{sec:test.pairing}

We next test the conservation and dissipation properties of the pairing rules described in \S~\ref{sec:pairing}, on an initial condition designed to expose the case where different pairing rules give qualitatively different behavior. We initialize $N_{\rm cells} = 4$ cells (deliberately small, to make the per-face residuals trivially visualizable) each carrying $N_{\rm basis} = 3$ basis components with $v_{\alpha} \in \{+1,\,0,\,-1\}$ and per-cell mass fractions arranged so that multiple components on side $a$ all have the same nearest-velocity component on side $b$ -- the configuration where ``mutual'' pairing would freeze some transfers (since not every $\alpha$ is reciprocally paired), while the one-sided rules of \S~\ref{sec:pairing} make all transfers. We run the test under the four pairing rules of \S~\ref{sec:pairing} (mutual-nearest, one-sided-nearest, Hungarian, and the one-sided vacuum rule), each under both the analytic and root-found ${\bf v}_{F}$ choices of \S~\ref{sec:vF.policy}, and measure the resulting drifts in $M$, ${\bf p}$, $KE$, and the per-face $\Sigma_{\alpha}F_{m,\alpha}$ residual (Fig.~\ref{fig:test.pairing}).

All eight configurations conserve total mass and momentum to machine precision -- a useful confirmation of the formal claim in \S~\ref{sec:pairing} that the antisymmetric $\pm F$ crediting between source and target basis components is exactly conservative regardless of which pairing rule is used, including non-mutual rules. Under the root-found ${\bf v}_{F}$, the per-face residual likewise collapses to machine precision for every pairing rule, confirming that the root-find decouples the per-face mass-flux invariance from the pairing-rule choice. The configurations do differ substantially, however, in kinetic-energy dissipation: the one-sided vacuum pairing combined with the corresponding one-sided collisionless Riemann solution dissipates kinetic energy at roughly $1\%$ of the rate of the symmetric assignments combined with the bilateral solver. The interpretation is geometrical -- the bilateral solver computes a (nonzero) stabilizing dissipation term at every face, including those where one side has no matched outgoing component, while the one-sided form recognizes that in such cases the upwind side carries the entire signal and no symmetric dissipation is required.

\begin{figure}
\centering
\includegraphics[width=0.98\columnwidth]{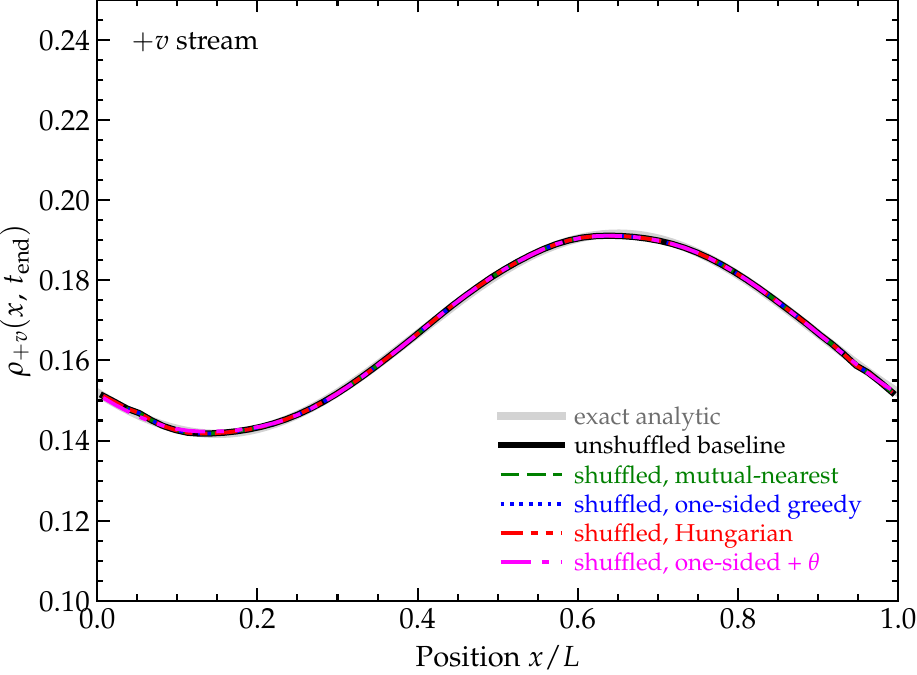}\vspace{-0.2cm}
\caption{Three-stream test with random per-cell basis re-labelling (\S~\ref{sec:test.shuffled}). 
We show the density profile $\rho_{\alpha}(x,\,t_{\rm end})$ of one of the three physical streams (the others are similar). We compare the unshuffled baseline to the four pairing rules of Fig.~\ref{fig:test.pairing} applied to a configuration where the basis indices are independently random-permuted in each cell. 
For every stream and every pairing rule, the shuffled and unshuffled runs agree well with the exact solution and each other. 
A naive ``same-index pairs with itself'' rule fails catastrophically in the shuffled case, not just mixing all bases incorrectly but breaking the slope-limiting and reconstruction so the system becomes numerically unstable.\label{fig:test.shuffled}}
\end{figure}

\subsection{Robustness Against Basis Re-Labelling}
\label{sec:test.shuffled}

A more subtle test of the basis-pairing machinery comes from the fact that the basis index $\alpha$ which identifies a particular component within cell $a$ does not have any physical meaning across cells -- the labelling can be permuted within a given cell without any change in the underlying phase-space DF. In practice, such re-labellings arise naturally in the merge/split step (\S~\ref{sec:repair}), as well as from the order in which streams happen to enter different cells from the initial conditions, so the pairing rule must be insensitive to them. We test this directly by initializing a three-stream configuration ($N_{\rm cells} = 64$, $N_{\rm basis} = 3$) with sinusoidal density modulation, drawing a random permutation of the three basis indices independently in each cell, and comparing the resulting evolution to an otherwise-identical run without the random permutation. Any of the pairing rules of \S~\ref{sec:pairing} (mutual-nearest, one-sided-nearest, Hungarian, or one-sided vacuum) recover the unshuffled result to machine precision, as expected. The naive ``same-$\alpha$ pairs with same-$\alpha$'' rule, in contrast, fails catastrophically -- random basis relabelling drives the equivalent of basis aliasing in the flux equation, and the resulting unphysical fluxes cause kinetic energy to diverge until the integration aborts on a timestep floor within a few tens of steps. This is the failure mode the per-pair cost matrix is designed to prevent, and the test verifies that all of the rules considered in \S~\ref{sec:pairing} successfully do so (Fig.~\ref{fig:test.shuffled}).

\subsection{Cold-Stream Injection: Testing Basis-Matching and Empty-Basis Assignment}
\label{sec:test.free.slot}

\begin{figure*}
\centering
\includegraphics[width=0.33\textwidth]{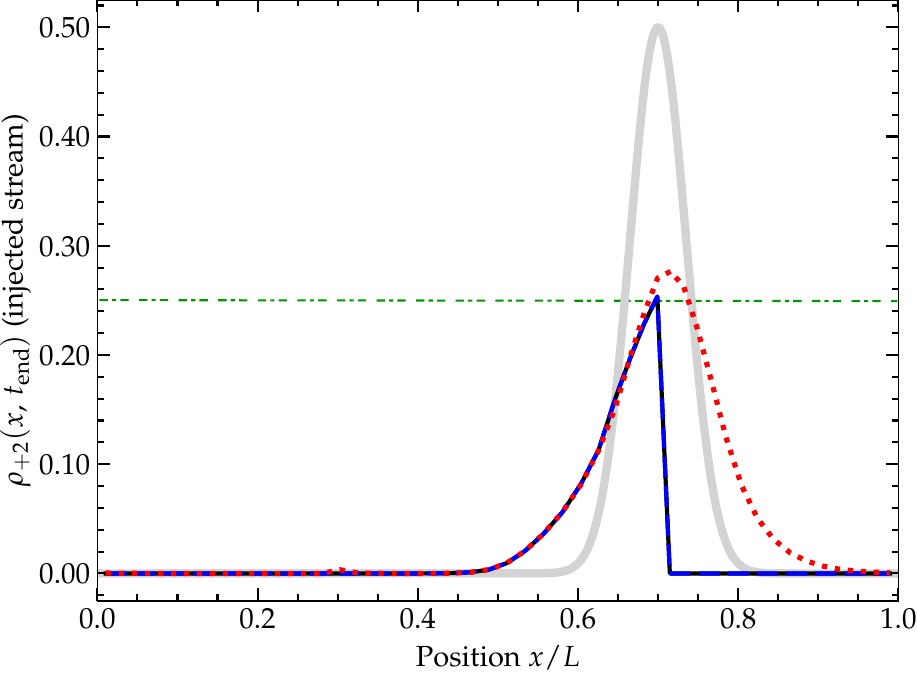}
\includegraphics[width=0.33\textwidth]{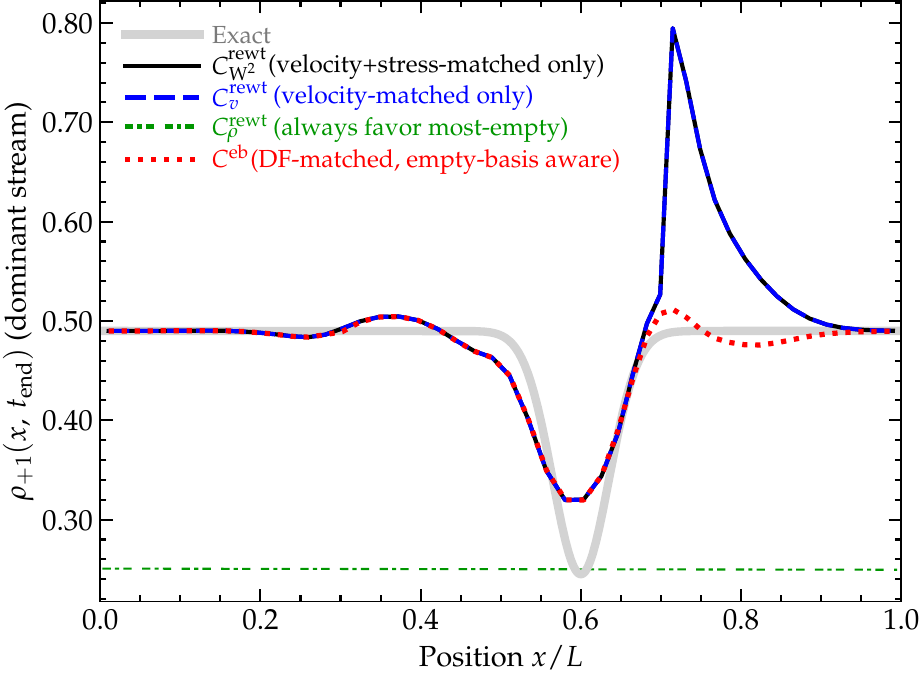}
\includegraphics[width=0.33\textwidth]{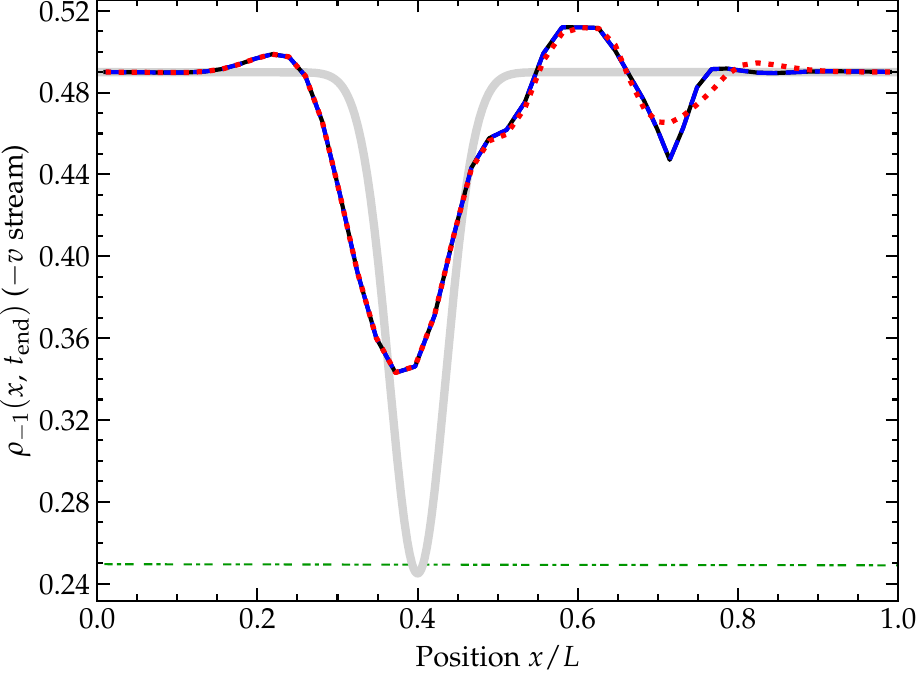}\vspace{-0.5cm}
\caption{The empty-basis test of \S~\ref{sec:test.free.slot}: a non-linear, spatially local $v = +2$ stream is injected at $x=0.5$, $t=0$, and propagates through cells which carry no matching basis component but do carry some near-empty bases. We plot the per-stream density profiles $\rho_{i}(x\,|\,v_{i},\,t=t_{\rm end})$ (integrating $f$ over the cold streams at different velocities, $v_{i}=+2,+1,-1$), compared to the exact analytic solution. 
We contrast four basis-matching cost variants (\S~\ref{sec:test.free.slot.cost}): 
(a) our default $C^{\rm eb}$ that matches based on the full phase-space sheet information and weights to empty bases when no good match is available; or 
(b) reweighting based only on Wasserstein-$2$ cost $C^{\rm rewt}_{\rm W^{2}}$  (matches on distance in $\langle {\bf v}\rangle$ and ${\bf S}$-space), 
(c) the same but using just $\langle {\bf v}\rangle$ ($C^{\rm rewt}_{v}$); 
or 
(d) using our default but always weighting towards the empty bases $C_{\rho}^{\rm rewt}$. 
Without using the full phase-space sheet, (b) and (c) assign mass to the wrong bases: the $+2$ injection routes into the $+1$ basis in the direction of propagation, so it never moves forward and instead simply modifies that basis. 
Always weighting to empty bases (without some threshold for whether there was a good fit or not) leads to the system rapidly distributing the mass from all bases uniformly (so each $N_{\rm basis}$ has equal mass). 
Our default method captures all the qualitative features correctly. The peaks/valleys are smeared out and lower-amplitude, but this is because of coarse resolution ($N=48$ cells) and converges away rapidly. 
\label{fig:test.free.slot}}\vspace{-0.3cm}
\end{figure*}

\begin{figure*}
\centering
\includegraphics[width=0.33\textwidth]{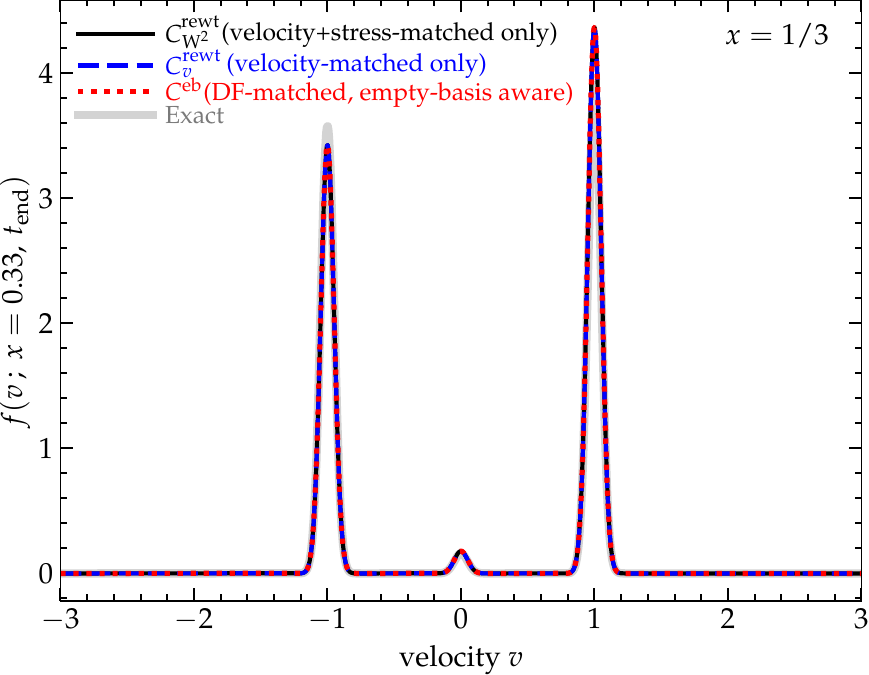}
\includegraphics[width=0.33\textwidth]{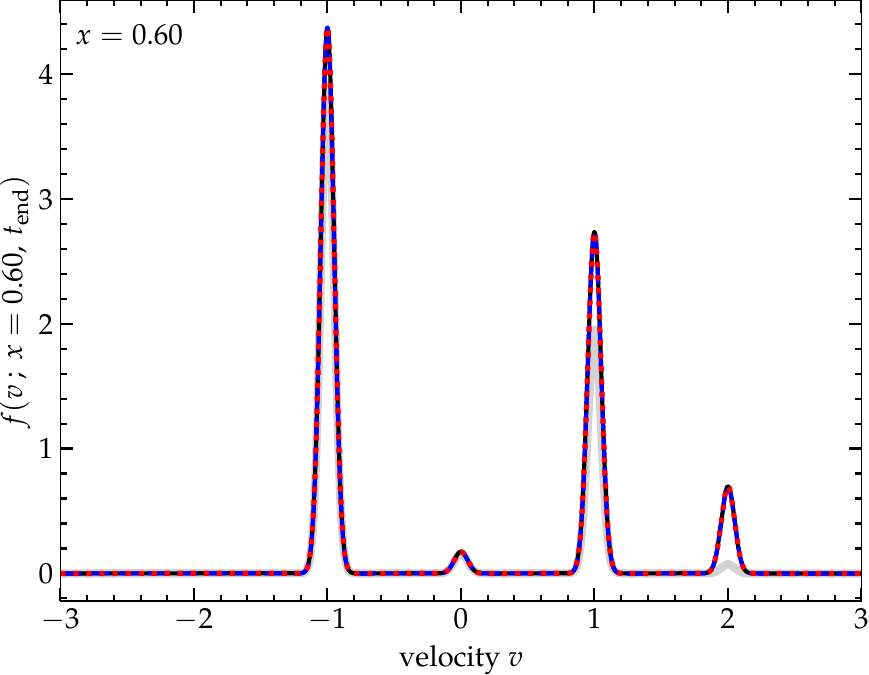}
\includegraphics[width=0.33\textwidth]{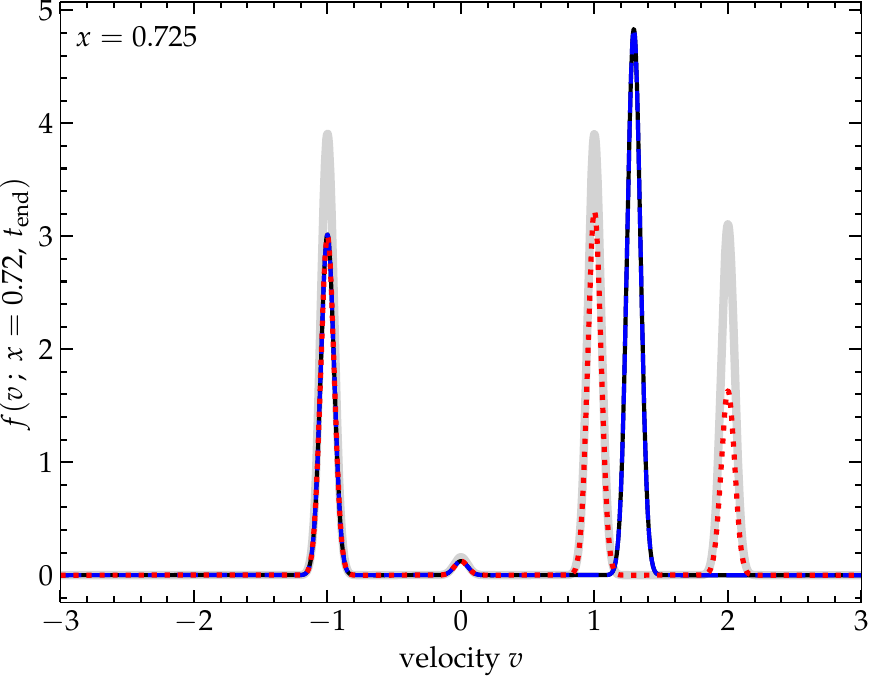}
\vspace{-0.5cm}
\caption{Reconstructed per-cell velocity distribution function $f(v\,;\,x,\,t_{\rm end})$ at three positions -- each panel a \textit{single} cell -- in the empty-basis test from Fig.~\ref{fig:test.free.slot}. 
\textit{Left:} $x=1/3$ (behind injection side), no injection has propagated here yet, so all methods agree with exact. 
\textit{Middle:} $x = 0.6$ (the position of the $v = +1$ depression's analytic centre). The methods agree but all suppress the $v=+1$ dip (over-estimate $f$ here) owing to low resolution: the basis needs to spread into adjacent cells to build a resolved stream.
\textit{Right:} $x=0.725$ (just past the analytic leading edge of the rigid-$+2$-advected perturbation). Our default method preserves a substantial $v = +2$ peak albeit with numerical spreading of the peaks in position-space (lower amplitude here). The velocity-space matching has already merged the $v > 0$ peaks together into a spurious new peak. 
\label{fig:test.free.slot.fv}}\vspace{-0.3cm}
\end{figure*}

We next consider the ``empty-basis test'' -- an extreme, discontinuous and non-linear test of independent cold streams reaching and crossing new regions in the spatial domain. 
In a periodic 1D box (size $L = 1$, mass $M = 1$), we place $N_{\rm cells}$ ($=48$ by default) uniformly-spaced cells, each carrying $N_{\rm basis}$ ($=4$ by default) components at $v_{\alpha} \in \{+1,\,0,\,-1,\,-2\}$. The unperturbed background mass fractions are $\varpi^{\rm bkg}_{\alpha} = (0.49,\,0.02,\,0.49,\,\epsilon_{\rho})$ with $\epsilon_{\rho} = 10^{-8}$ (a numerical floor representing the near-empty fourth slot), so that the bulk fluid is dominated by the two streams at $\pm 1$, plus a tracer stream at $0$ and a near-empty slot at $-2$. All basis components are initialized with the same small isotropic dispersion ${\bf S}_{\alpha} = \sigma_{0}^{2}\,{\bf I}$ with $\sigma_{0}^{2} = 10^{-6}$ (this is the same per-basis dispersion floor used in the realizability repair of \S~\ref{sec:repair}, well below the inter-stream velocity gap, so streams remain distinguishable). We then overlay a localized nonlinear injection of a new stream with mass fraction $\pi(x) = \pi_{\rm max}\,\exp[-((x-x_{0})/\sigma_{\pi})^{2}]$ with $\pi_{\rm max} = 0.5$, $x_{0} = 0.5$, $\sigma_{\pi} = 0.05$, which {both} (i) relabels the $v_{\alpha=4}$ velocity from $-2$ to $+2$ in the cells where $\pi(x) > 10^{-6}$ (i.e.\ $|x-x_{0}| \lesssim 0.18$; in all other cells this basis keeps $v=-2$ and a negligible $10^{-8}$ mass fraction), and (ii) reassigns the per-basis masses in those cells to $\varpi_{4} = \pi(x)$ and $\varpi_{1,2,3} = (1-\pi(x))\,\varpi^{\rm bkg}_{1,2,3}/\sum^{3}\varpi^{\rm bkg}$ so that the total cell mass remains $m_{a} = 1/N_{\rm cells}$ exactly. The total box mass and momentum are therefore unchanged by the injection -- it is an internal redistribution of basis mass within each cell. The $v = +2$ component then propagates across the domain into neighbor cells which have no matching $+2$ basis. 

\begin{figure*}
\centering
\includegraphics[width=0.315\textwidth]{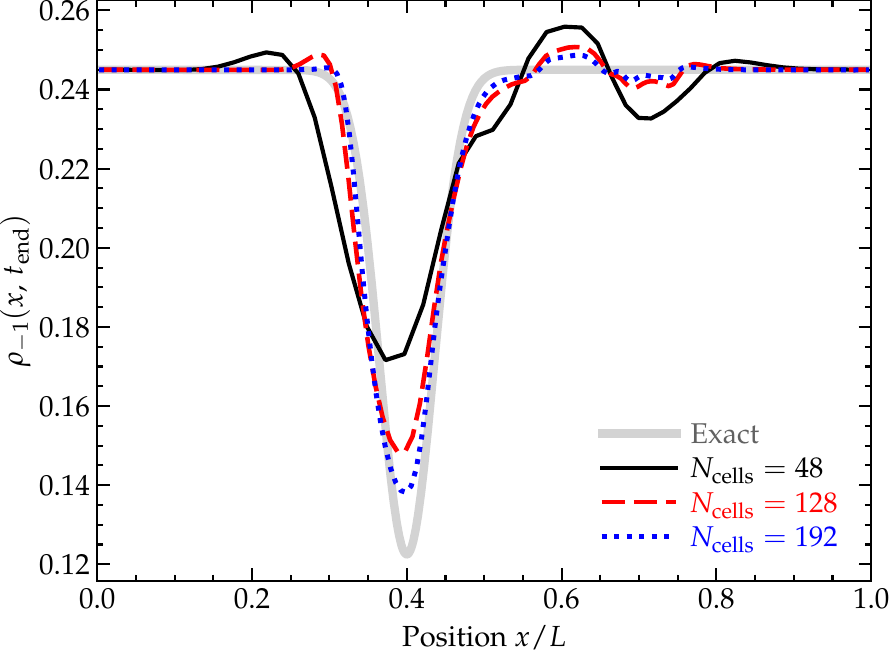}
\includegraphics[width=0.33\textwidth]{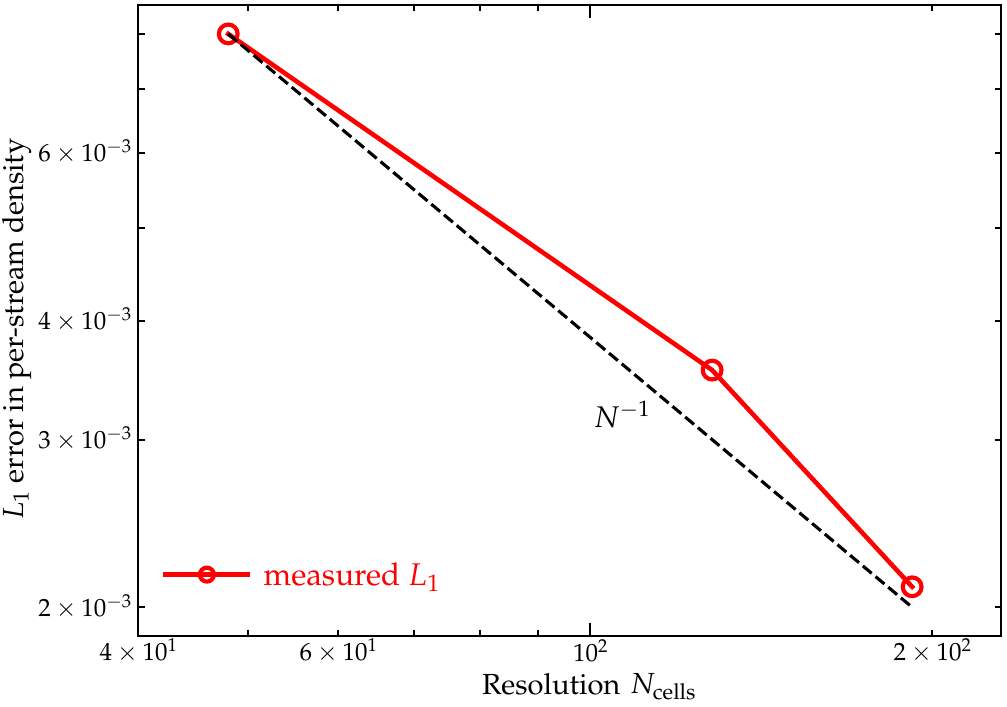}
\includegraphics[width=0.33\textwidth]{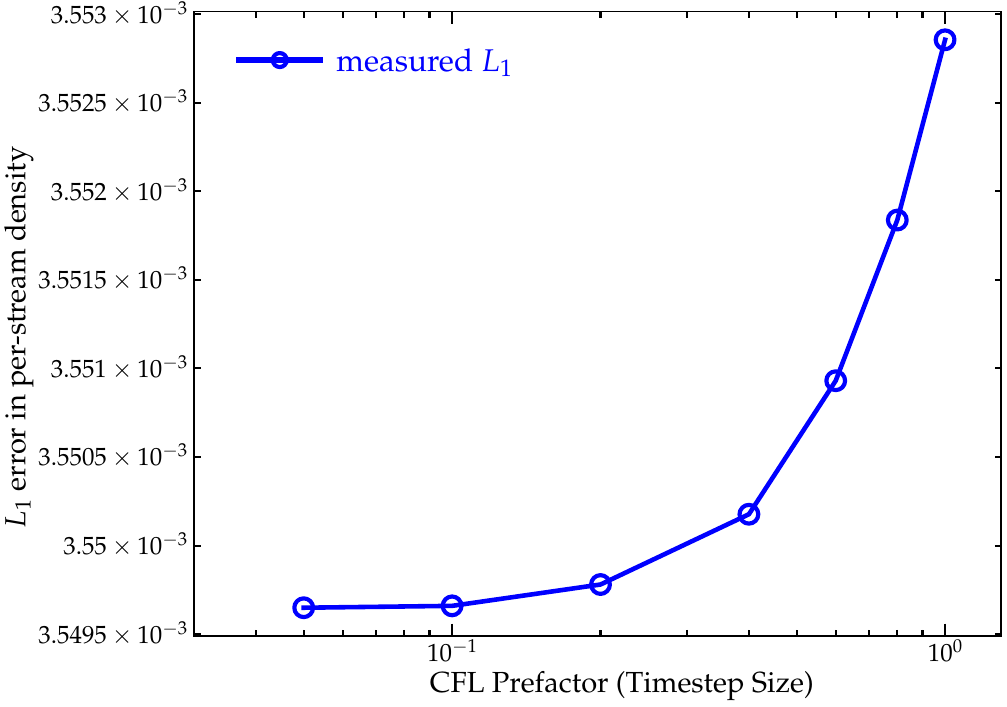}\vspace{-0.3cm}
\caption{Convergence and time-step stability on the empty-basis test of \S~\ref{sec:test.free.slot} (Fig.~\ref{fig:test.free.slot}). 
\textit{Left:} Per-stream profile $\rho_{-1}(x,\,t_{\rm end})$ for one of the streams (the convergence of the others is similar) vs.\ $N_{\rm cells}$. 
\textit{Centre:} The corresponding $L_{1}$ error vs.\ $N_{\rm cells}$, log-log; the convergence is between first and slightly super-first order in $N_{\rm cells}^{-1}$, consistent with a Lagrangian scheme advecting a sharp (discontinuous) feature through discrete basis reassignment at its leading edge.
\textit{Right:} $L_{1}$ error vs.\ $\mathrm{CFL}_{\rm safety}$ at fixed $N_{\rm cells}=128$.
The error is constant to $\lesssim 1\%$ across the full range, i.e.\ the solution is fully temporally converged even at $\mathrm{CFL}_{\rm safety}=1$ and the error is entirely spatial-truncation-limited.
\label{fig:test.resolution}}\vspace{-0.3cm}
\end{figure*}

This is a demanding test: the ICs have $\mathcal{O}(1)$, grid-scale sharp discontinuities in every quantity in the DF, they are highly non-equilibrium/non-time-steady, the new basis must propagate into cells that by construction \textit{do not} have any well-aligned pre-existing basis functions, and the stream must cross into new cells (re-aligning basis functions and ``filling in'' an empty basis from nothing before it can move). And all of this at intentionally coarse spatial and velocity resolution.

\subsubsection{Empty-Basis Cost/Assignment Comparisons}
\label{sec:test.free.slot.cost}

Figs.~\ref{fig:test.free.slot} \&\ \ref{fig:test.free.slot.fv} illustrate the sensitivity of this test to the empty-basis cost/weighting, comparing our adopted basis-assignment machinery from \S~\ref{sec:pairing} including the empty-basis assignment weighting ($C^{\rm eb}$ of Eq.~\ref{eqn:cost.free.slot}), to alternatives: (d) $C^{\rm rewt}_{\rho}$, where we simply multiply the weights by $\rho_{b,\,\beta}/(\rho_{a,\,\alpha}+\rho_{b,\,\beta})$ (as discussed in  \S~\ref{sec:pairing.freeslot}) or (c) $C^{\rm rewt}_{v}$ using \textit{only} velocity-space information to match bases (otherwise keeping our assignment identical) or (b) $C^{\rm rewt}_{W^{2}}$ using the velocity-distance-overlap Wasserstein-2 pairing assignment matrix. 
Here $C^{\rm rewt}_{W^{2}}$ and $C^{\rm rewt}_{v}$ are very similar, because the bases are cold, and both fail in the same way: they identify the dominant $v=+1$ stream as the closest match in velocity space, routing the entire new stream into it, which creates a completely different DF (seen in $\rho(x,t)$ and $f(v\,|\,x,t)$). 
$C^{\rm rewt}_{\rho}$ \textit{also} fails but in a different way: it over-routes everything to the lowest-mass basis creating phase mixing between all streams. And these errors for the $C^{\rm rewt}$ variants do not converge away at higher cell count.

\subsubsection{Resolution and Time-Step Stability}
\label{sec:test.free.slot.resolution}

Fig.~\ref{fig:test.resolution} shows the convergence and time-step robustness of the default scheme, comparing the density profile and $L_{1}$ residual in density vs.\ the exact analytic solution at fixed time $t=t_{\rm end}=0.1$ -- purposely chosen so that the injected stream is still reaching new ``empty'' cells and the system is far from equilibrium -- for $N_{\rm cells} \in \{48,\,128,\,192\}$. We see convergence visually, and measure approximate first-order $L_{1} \propto N^{-1}$ convergence. This is expected: as noted, we intentionally chose a \textit{discontinuous} DF: $f(v = +2,\,r,\,t)$ is spatially-discontinuous at all times $t$ we evolve here, akin to a shock propagating across a grid, so our second-order reconstruction necessarily becomes piecewise-constant owing to the slope-limiters around the discontinuities. 
We also show the $L_{1}$ error at fixed $N_{\rm cells}=128$ as a function of the CFL safety factor (timestep size) across the full range $\mathrm{CFL}_{\rm safety} \in [0.05,\,1.0]$: the variation is $\lesssim 1\%$ relative, indicating the solution is fully temporally converged for any $\mathrm{CFL}_{\rm safety} \lesssim 1$ and that the error is entirely spatial-truncation-limited even at the largest stable timestep. We adopt $\mathrm{CFL}_{\rm safety}=0.2$ as the default for the rest of this paper, as standard in the \GIZMO\ MFM implementation.

\subsubsection{Gaussian vs.\ Top-Hat Basis Functions}
\label{sec:test.free.slot.ab}

\begin{figure}
\centering\includegraphics[width=0.95\columnwidth]{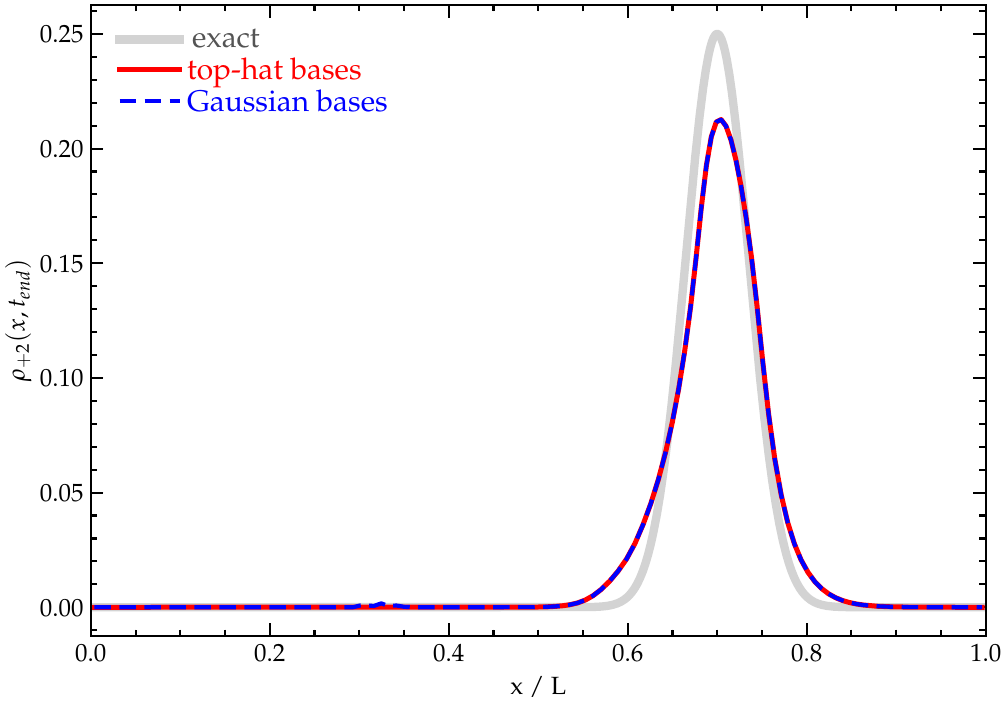}\vspace{-0.3cm}
\caption{Comparison of the top-hat vs.\ Gaussian basis function solutions for the empty-basis test (as Fig.~\ref{fig:test.free.slot}) at $N_{\rm cells}=128$. 
We show the per-stream density profile at $t_{\rm end}=0.1$ for the injected $v=+2$ component (all streams agree similarly between top-hat and Gaussian). 
The stream-integrated densities and mass fractions at all $x,\,t$ agree to $<1\%$.\label{fig:test.ab.hllc.gauss}}
\end{figure}

Fig.~\ref{fig:test.ab.hllc.gauss} compares the results for a top-hat vs.\ Gaussian basis function assumption. 
Because the velocity DF is cold in either case, the two solutions agree to within line thickness in the per-stream density profiles (Fig.~\ref{fig:test.ab.hllc.gauss}), and integrated stream mass-fractions for each stream agree to better than $1\%$.
The maximum mass-weighted dispersion of any basis or any cell is $S_{\rm max}^{\rm real} \sim 2$ times larger with the Gaussian bases (median dispersion is within $\sim 20\%$), for an understandable reason: the Gaussian admits the warm-thermal flux through receding states (in particular $u_{x} \le -c_{x}$, and with a larger conductive weight throughout $-c_{x} < u_{x} < 0$) which the top-hat solver reduces and, for $u_{x} \le -c_{x}$, truncates to zero. But the difference is much smaller in a hot or warm basis as we find below for the harmonic oscillator test -- here it is accentuated by the very cold bases (both of which remain very cold, they are just slightly-differently represented in velocity-space). 

\begin{figure*}
\includegraphics[width=0.98\textwidth]{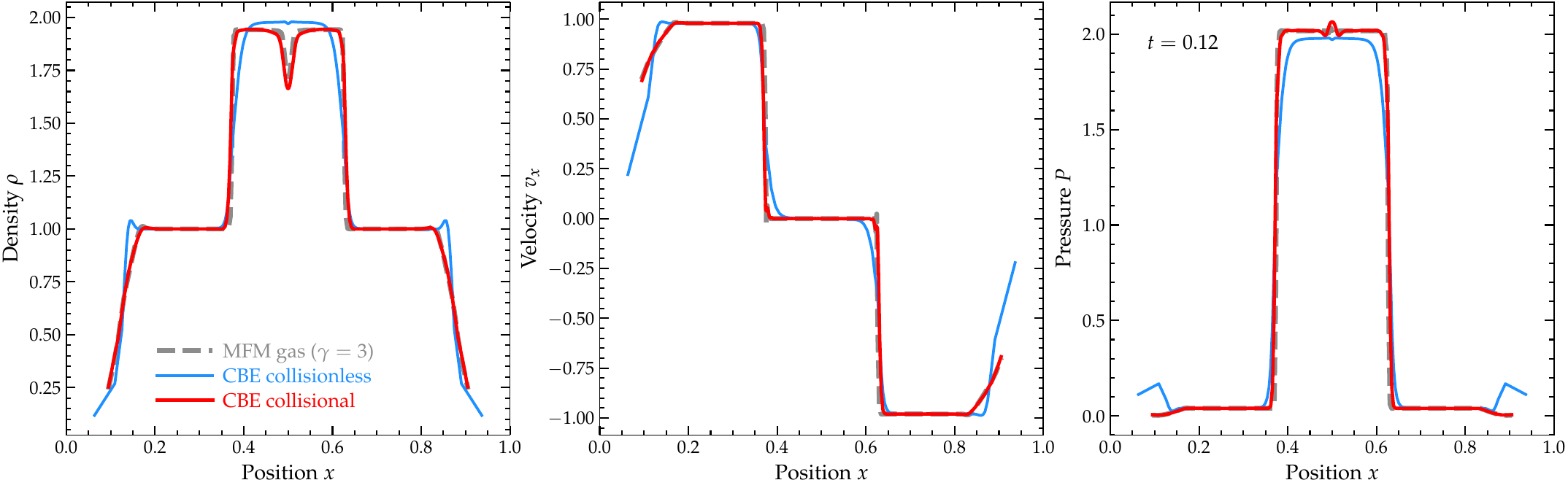}\vspace{-0.2cm}
\caption{Colliding-stream test of the collisional limit (\S\ref{sec:test.collisions}): two cold streams at $v=\pm1$ meet at the box centre. Profiles of density $\rho$, bulk velocity $v$, and pressure $P$ at fixed time. \textit{Collisionless} ($\sigma/\mu_{p}=0$; blue): the streams interpenetrate, giving $\rho=2$ in the overlap with each basis retaining its $\pm1$ velocity (the physically-correct collisionless solution). \textit{Strongly collisional} (large $\sigma/\mu_{p}$; red): the streams thermalize into a symmetric double-shock with a central slab at the $\gamma=3$ strong-shock compression $\rho=(\gamma+1)/(\gamma-1)=2$ and $v\rightarrow0$, matching a standard MFM hydrodynamics run of the moment-matched initial condition (grey dashed) to $\sim1\%$ in $\rho$, $v$, and $P$. The collision operator thus interpolates between free-streaming and fluid dynamics on the \textit{same} problem.\vspace{-0.5cm}
\label{fig:test.collisions}}
\end{figure*}

\subsubsection{Limiter and Predictor Variants}
\label{sec:test.free.slot.limpred}

We further test the robustness of this problem to the slope limiters (\S~\ref{sec:reconstruction}) or the choice to apply or omit \GIZMO's default primitive predictor \citep{hopkins:gizmo}. 
Turning either on/off, at $N_{\rm cells}=128$ with our method otherwise default (full basis-matching) we find that all runs complete to $t_{\rm end}=0.1$, conserve mass and momentum and total energy to machine precision, and give almost identical per-stream profiles and bulk physical solutions. So these are not critical for this problem. However without the limiters we do find that realizability in the reconstructed state is frequently (at least in a couple of cells at every timestep) violated (albeit ``weakly'') and has to be almost constantly repaired by the post-hoc check we apply to the reconstructed states, which, as discussed in \S~\ref{sec:reconstruction}, silently injects $\mathcal{O}(\rho\,S)$ corrections per-event into the second moment. This just ends up being a small impact in the final stress total and solution because $S$ is small, i.e.\ because the streams are cold. With our default slope-limiters, this SPD repair never triggers throughout the run, for any cell.

\subsubsection{In Other Methods}
\label{sec:test.free.slot.altmethods}

MC/PIC/NB methods can certainly handle this test, but they carry all the problems discussed for 2-stream tests, like noise in $\rho(x)$ and $f(v,x,t)$, and the ``new stream injection'' introduces additional ambiguities. In a volume/subregion of the domain where the number of test MC particles with $v=v_{\rm new}=+2$ matched to the new stream value is nil, then $f(v_{\rm new},\,x,\,t)$ only has two values: zero or undefined. But saying this $=0$ for any region without a particle at $v=v_{\rm new}$ means that if the DF were continuous in $v$ but discretely sampled, one would inevitably require that the DF be noisy, jumping between large values and $0$ in any subvolume containing a finite particle number. Smoothing over a large enough spatial kernel to sample $f(x,v,t)$ at all velocities spanned (the ``kernel density estimation in phase space'' solution) forces the neighbor kernel in the part of the volume the new stream has not-yet reached to extend to the volume already-reached, introducing an arbitrarily large, acausal forward-propagating ``tail'' with implied $f(v=v_{\rm new},x,t) > 0$. 
Direct/Eulerian/lattice-Boltzmann Vlasov methods can handle this test but still have the diffusion issues noted above. 
Waterbag/tessellation/moments/wavefunction methods fail even more severely here than in the steady-state multi-stream problem.

\subsection{Colliding Streams: The Collisional Limit}
\label{sec:test.collisions}

The tests so far are purely collisionless. To verify that the collision operator of \S\ref{sec:collisions} correctly interpolates between the collisionless and hydrodynamic limits, we use a 1D colliding-stream problem: in a periodic box of unit length we initialize uniform density $\rho_{0}=1$ with all the mass in the left half moving at $v=+1$ and all the mass in the right half at $v=-1$ (represented by two basis components per cell at distinct $\pm1$ velocities with the appropriate mass split), a negligible thermal floor, and $N_{\rm cells}=256$. We run the same initial condition at cross-sections $\sigma/\mu_{p}$ spanning the collisionless and strongly-collisional regimes, and in the strongly-collisional case compare against a matched MFM hydrodynamics run of the \textit{moment-matched} initial condition (the same density, bulk velocity, and pressure per cell) with adiabatic index $\gamma=3$, the effective index of a one-dimensional collisional gas closing at the second moment.

The two limits are recovered cleanly (Fig.~\ref{fig:test.collisions}). With no collisions ($\sigma/\mu_{p}=0$) the two streams interpenetrate freely: the density in the overlap region rises to exactly $\rho=2$, each basis retains its $\pm1$ velocity flat across the box, and the solution is the expected collisionless two-stream interpenetration. In the strongly-collisional limit ($\sigma/\mu_{p}$ large, so the mean-free-path weight $w_{c}=e^{-\Delta x/\lambda_{\rm MFP}}\rightarrow0$) the same initial condition instead produces a symmetric double-shock: the two streams thermalize at the interface, the basis representation collapses to a single Maxwellian per cell (via the intra-cell relaxation of \S\ref{sec:collisions.hardsphere} and the merge step of \S\ref{sec:repair}), and a central shocked slab forms at the $\gamma=3$ strong-shock compression $(\gamma+1)/(\gamma-1)=2$ with $v\rightarrow0$. This CBE result matches the matched MFM hydrodynamics run to $\sim1\%$ in the $L_{1}$ norm of $\rho$, $v$, and $P$, with co-located shock fronts, confirming that the operator-split scheme reproduces the collisional (fluid) limit of the identical problem it solves collisionlessly.

MC/PIC/NB methods cannot represent this limit. Collisionality and density are ill-defined, but are often represented via MC super-particle scattering with some ``overlap kernel'' (treating each MC particle as a ``cloud'' of micro-particles moving at the same speed and uniform density, with size of order the inter-particle separation; \citealt{rocha:2013.sidm.sims,meskhidze:2022.gizmo.arepo.code.comparison.sidm.sims.very.similar,arido:2024.sidm.particle.methods.anisotropic.scattering}), but it is well known that this becomes demonstrably wrong (it converges to an unphysical solution) in the short mean-free-path limit (see those references). The failure is independent of the implementation: fundamentally, MC/PIC/NB discards the internal degrees of freedom of the DF, so there is no thermal energy (particles can macroscopically scatter, but this means that instead of e.g.\ a hot, shocked, stationary gas, one has at whatever resolution scale a collection of macro-scopic gas masses moving ``left'' or ``right''), and it is impossible for them to simultaneously conserve momentum and energy in this limit as well. 
Direct/Eulerian/lattice-Boltzmann Vlasov methods can in principle handle this but it is very sensitive to the form of the operator in collisions: it is not trivial that a Maxwellian DF can be retained (as compared to advecting away). 
Waterbag/tessellation methods immediately fail (they become ill-posed) in this regime. 
Moments/wavefunction methods can reproduce the infinitely-collisional limit: indeed this is the one regime where they do work well.

\begin{figure}
\hspace{-0.1cm}\includegraphics[width=0.98\columnwidth]{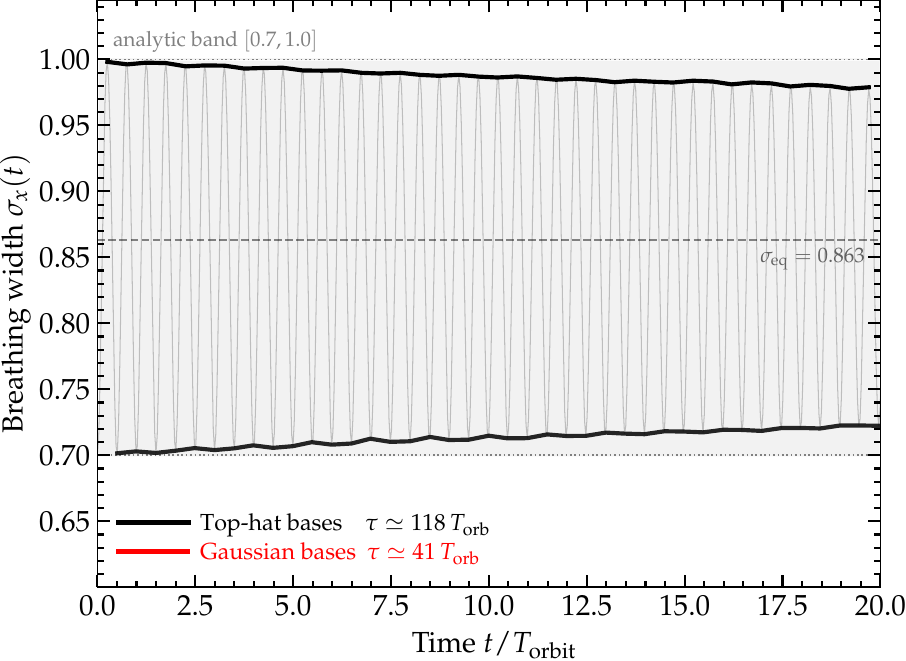}
\includegraphics[width=0.965\columnwidth]{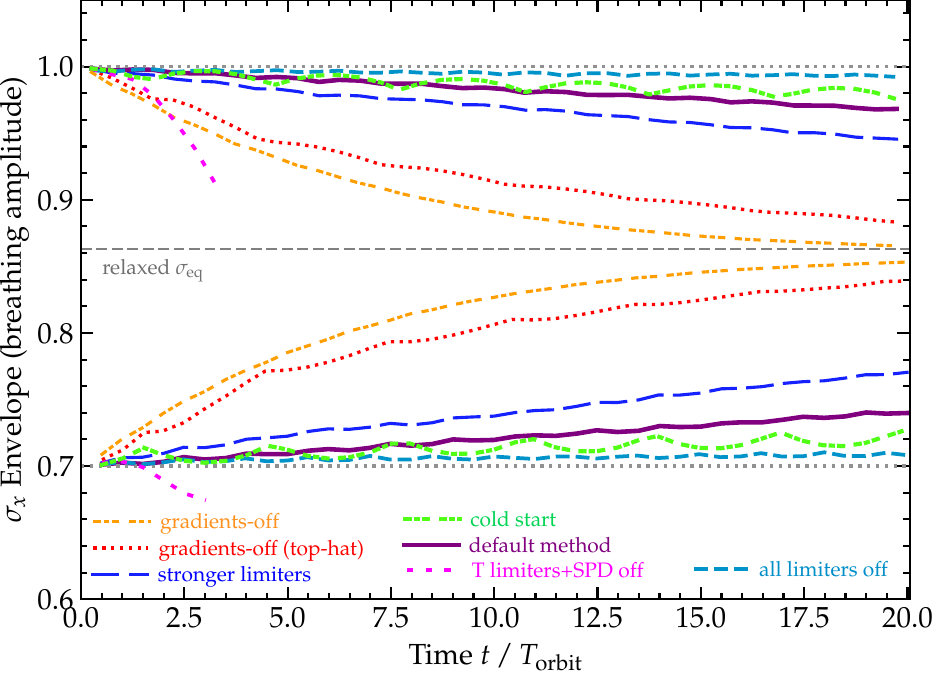}\vspace{-0.2cm}
\caption{Oscillator test (\S\ref{sec:test.harmonic}): a Gaussian phase-space packet with mismatched widths ($\sigma_{x,0}=0.7$, $\sigma_{v,0}/\Omega=1.0$) in a harmonic potential, whose breathing envelope is analytically constant ($[0.7,\,1.0]$; all orbits are isochronous). The mass-weighted spatial width $\sigma_{x}(t)$ is plotted ({\em top; thin}) in orbital periods $t/T_{\rm orbit}$ (each breath is half an orbital period), with its envelope ({\em thick}). 
The correct behavior is recovered but numerical dissipation slowly damps the breathing width (still conserving energy), with damping time $\tau \simeq 115\,T_{\rm orbit}$ for the top-hat bases ({\em top}) at this resolution (the damping decreases rapidly with higher resolution). 
For the Gaussian bases ({\em bottom}) we find the damping is faster ($\tau\simeq41\,T_{\rm orbit}$), owing to the larger stress-dissipation/conductivity term in the Riemann problem. Making slope limiters stronger/weaker increases or decreases the damping (removing them gives almost no damping -- it is numerical dissipation from the imperfect piecewise-linear reconstruction), but removing the stress+SPD checks entirely makes the problem numerically unstable (result diverges and then crashes). A piecewise-constant reconstruction (``gradients off'') damps much more rapidly. The ``cold start'' also damps more slowly.\label{fig:test.harmonic}}
\end{figure}


\subsection{Harmonic Oscillator: Breathing-Mode Damping}
\label{sec:test.harmonic}

The tests above do not directly measure the method's long-term \textit{dynamical} dissipation in the presence of self-consistent forces. For that we use the ``breathing'' mode of a 1D harmonic potential $\Phi(x)=\tfrac{1}{2}\Omega^{2}x^{2}$, in which all orbits share the same period ($T_{\rm orbit}=2\pi/\Omega$, independent of amplitude), so a collisionless phase-space distribution that is initially Gaussian with mismatched position- and velocity-space widths ($\sigma_{x,0}\neq\sigma_{v,0}/\Omega$) oscillates \textit{forever}, with the spatial width breathing between $\sigma_{x,0}$ and $\sigma_{v,0}/\Omega$ at twice the orbital frequency, with no secular change. We take $\sigma_{x,0}=0.7$, $\sigma_{v,0}=1$ (so the envelope oscillates in $[0.7,\,1.0]$, with relaxed width $\sigma_{\rm eq}=[\tfrac{1}{2}(\sigma_{x,0}^{2}+\sigma_{v,0}^{2})]^{1/2}=0.863$ should the mode fully damp) and integrate for $20\,T_{\rm orbit}$ (forty full breaths). Any decay of the breathing envelope is then a direct, physically-normalized measure of numerical dissipation in the method: we fit the envelope to $\propto e^{-t/\tau}$ and quote the damping time $\tau$ in orbital periods, with larger $\tau$ (slower damping) indicating a less dissipative scheme.

The result is shown in Fig.~\ref{fig:test.harmonic}. With the fiducial compact-support top-hat solver ($\chi=\sqrt{3}$), the breathing envelope is essentially undamped over the full integration, decaying from $1.00$ to $\simeq0.98$ over forty breaths (at $N=256$ shown), corresponding to $\tau\simeq115\,T_{\rm orbit} \sim 720\,\Omega^{-1}$; the exact-Gaussian solver damps roughly $2.5{-}3\times$ faster ($\tau\simeq40{-}45\,T_{\rm orbit}$) on the identical initial condition, because its heat-conduction flux term (the ${\bf S}_{x}\otimes{\bf S}_{x}\,\omega/\sigma_{x}$ term in Eq.~\ref{eqn:gaussian.flux}) does not compactly truncate and therefore continues to transfer second-moment energy through weakly-receding faces. That the residual damping is genuinely a property of the inter-cell flux, and not of the leapfrog gravity integration, is already established by the solver comparison itself: the top-hat and Gaussian runs share the identical initial condition and gravity integrator and differ \textit{only} in the face Riemann solution, yet damp at rates differing by a factor of $\sim2.5{-}3$, so the dissipation must originate in the flux and not the force update. Decomposing the dissipation across the reconstruction and limiting choices of \S\ref{sec:reconstruction}, we find the velocity-slope limiter is the dominant contributor to the residual damping -- relaxing it (while retaining the second-moment limiter needed for stability) roughly doubles $\tau$ -- while the density-slope limiter and the realizability-cone tightening are essentially neutral for this diagnostic. Removing slope-limiters entirely makes the dissipation unmeasurable, but destabilizes other problems. Making those limiters much stronger increases the damping. This motivates the small opposite-sign tolerance we adopt in the velocity limiter (\S\ref{sec:reconstruction}), which recovers most of the anti-damping benefit while preserving monotonicity and stability. 
Other choices influence the dissipation as expected. A piecewise-constant reconstruction is much more dissipative, with $\tau\to10\,T_{\rm orbit}$. Replacing our exact Riemann solver with a less-accurate and more-dissipative HLL or global Lax-Friedrichs method similarly damps more strongly. The damping rate decreases ($\tau$ increases) with resolution (roughly $\tau \propto N$ or slightly steeper at higher resolution over $N=128,\,256,\,512$) consistent with a higher-order method, so it is a controllable truncation error, rather than a fixed property of the method.

We have also run several variants of this problem. For example, we can vary $\sigma_{x,0}$ or $\sigma_{v,0}$ but this just rescales the problem units. 
More interesting, we can consider instead a ``cold start'' where at every position we initialize half the mass each with $v=\pm \sigma_{x}\Omega$ (or some multiple thereof), so we have an oscillating version of the two-streams problem. We have also approximated the Gaussian $\sigma_{v,\,0}$ as a sum of $4$ or $8$ independent cold-streams (essentially breaking it up into a simple histogram of cold streams with no dispersion), with equal numbers left-and-right propagating in the initial conditions. All of these oscillate with the same period and produce the same qualitative behaviors. The ``cold stream'' numerical dissipation \textit{first} appears as a blurring of the initially-infinitely-cold basis functions with finite $N_{\rm basis}$, with no envelope damping, then once the bases are non-linearly broadened, the envelope damping appears (so it looks overall like slower damping, but measured just in this fully-non-linear phase, the rates are similar). So the numerical dissipation acts first primarily in velocity-space in our implementation, and only secondarily in configuration space.

This problem is quite sensitive to incorrect basis-pairing (\S~\ref{sec:pairing}), especially for a ``cold start'' IC with bases moving in different directions and very different masses. If we match bases between neighbors purely in velocity-space or with the Wasserstein-2 weights, especially if we ignore the reconstructed velocities extrapolated to the faces (match only on cell-centered quantities) then the problem is that near the turnaround points at the extrema of the breathing mode (where the velocities all approach zero), the second-order errors from linear gradient extrapolation can lead to incorrectly matching bases moving into vs.\ out-of the turnaround. But since these can have very different densities, this leads to either excessive dissipation (with strong slope limiters) or large noise that can even destabilize the problem if no slope-limiters or prior-step gradient information is used (see Fig.~\ref{fig:test.free.slot.fv} and discussion in \S~\ref{sec:pairing}). An accurate and robust second-order reconstruction depends on matching bases in the reconstructed quantities, on the full phase-space sheet. Within the context of our full weights \S~\ref{sec:pairing.freeslot} using the gradients, we see much less sensitivity. 

In this problem, MC/PIC/NB methods give clean evolution of the orbits/oscillations by-particle (provided an accurate second-order or higher force integration scheme), with zero dissipation, but like the 2-stream problem, the noise is quite severe if one attempts to define the DF or $\rho$ anywhere, the density at the \textit{edges} of the envelope is especially ill-defined by definition at any resolution, and convergence in these is very slow. 
Direct/Eulerian/lattice-Boltzmann Vlasov methods can integrate this but the variable acceleration in ${\bf v}$ and ${\bf x}$ produces much stronger damping, even with a high-order scheme.
Waterbag/tessellation and moments/wavefunction methods fail catastrophically everywhere on this problem (it is something of a worst-case scenario).

\subsubsection{Relation to Zeldovich/Plane Wave Collapse}
\label{sec:test.zeldovich}

Briefly, we note that our harmonic-oscillator test is similar to the collisionless \citet{zeldovich:1970.pancakes} pancake test. That follows collapse of a 1D self-gravitating potential $\phi(x)=\bar{\phi}\,\cos{(k x)}$ in a periodic cosmological Einstein-de Sitter box with cold initial conditions, as in e.g.\ \citet{abel:phase.space.sheets.for.dm,hahn:2013.dm.phase.space.sheet.sims} and others. Since we have not implemented cosmological integration in 1D  in \GIZMO\ (only 3D), we cannot directly compare to those authors. But it is worth briefly noting that the pre-shell crossing evolution in our method is trivially identical to standard $N$-body methods, while the post shell-crossing behavior tests the exact same physics and numerical levers as our oscillator problem (especially for a ``cold start''). Indeed in our default case all the cold-start streams intersect at the same time (the system has $\sim \mathcal{O}(N_{\rm cells})$ wraps per period). We can also initialize the oscillator differently to be even more akin to the pancake: if we initialize constant density from $-1<x<1$ and initial velocity $v_{0} = \Omega\,\sin{(m \pi x)}$, then for integer $m>0$ there will be $2m+1$ streams overlapping at the origin every half-period. Obviously if $N_{\rm streams} > N_{\rm basis}$, our method coarse-grains adjacent streams in velocity-space via the basis function pairing; in our default case this is always happening (there is always finite dispersion) by construction. But the total energy and behavior of the envelope is conserved, and of course \textit{any} method without infinite resolution in the Zel'dovich problem must eventually coarse-grain since the intersecting stream number tends to infinity with more phase-wrapping. Our damping rate measures the rate at which \textit{numerical} damping blurs the sharp edges in phase-space (again, a phenomenon generic to basically all methods, but usually just shown qualitatively). 



\begin{figure*}
\includegraphics[width=0.9\textwidth]{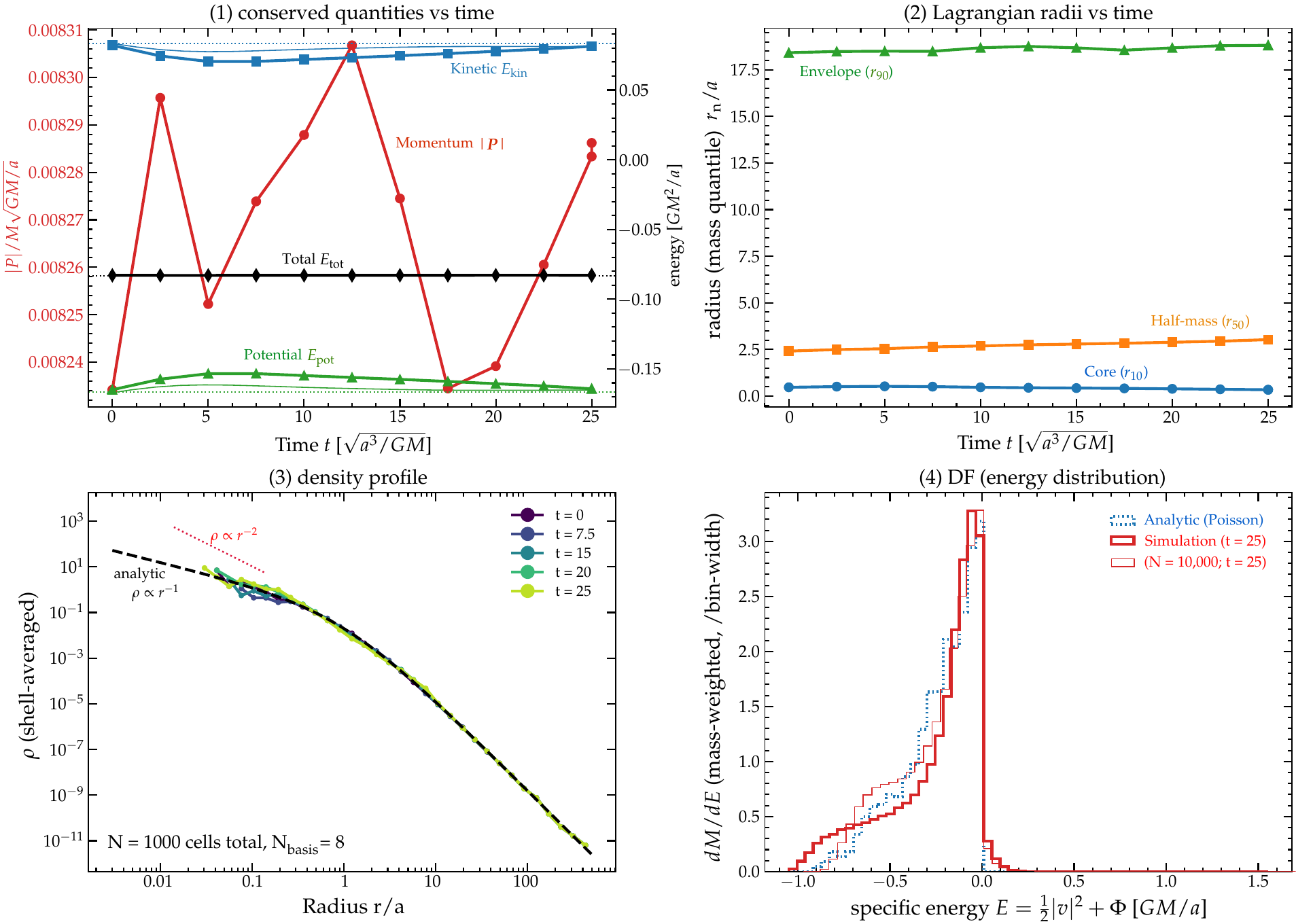}\vspace{-0.2cm}
\caption{Collisionless Hernquist self-gravitating orbit mixing and equilibrium test (\S~\ref{sec:test.hernquist}). 
A set of $1000$ particles is initialized with equal masses and infinitely-cold streams from the Poisson-sampled Hernquist DF (total mass+scale-length+gravitational constant $M=a=G=1$), and allowed to relax for $25$ global dynamical times, with top-hat bases.
We plot evolution of (1) total energies and momenta; (2) radii $r_{n}$ enclosing $n$ percent of the mass; (3) density profile averaged in spherical shells; (4) total (integrated) isotropic DF in terms of the specific energy, $dM/d|\mathcal{E}|$ (thin line shows $N=10^{4}$). 
The system relaxes to the correct equilibrium, is stable, conserves total energy and momentum (despite some initial transients in potential-kinetic energy). At this (low) resolution there is a very slow numerical-dissipation driven relaxation in the core, causing the tail to slightly lower specific energies and some shrinking of $r_{10}$, but its effects on the density profile are minor even at $r \lesssim 0.1\,a$ where the system has evolved $\sim 250$\,dynamical times.\label{fig:test.hernquist.demo}}\vspace{-0.5cm}
\end{figure*}

\begin{figure}
\includegraphics[width=0.98\columnwidth]{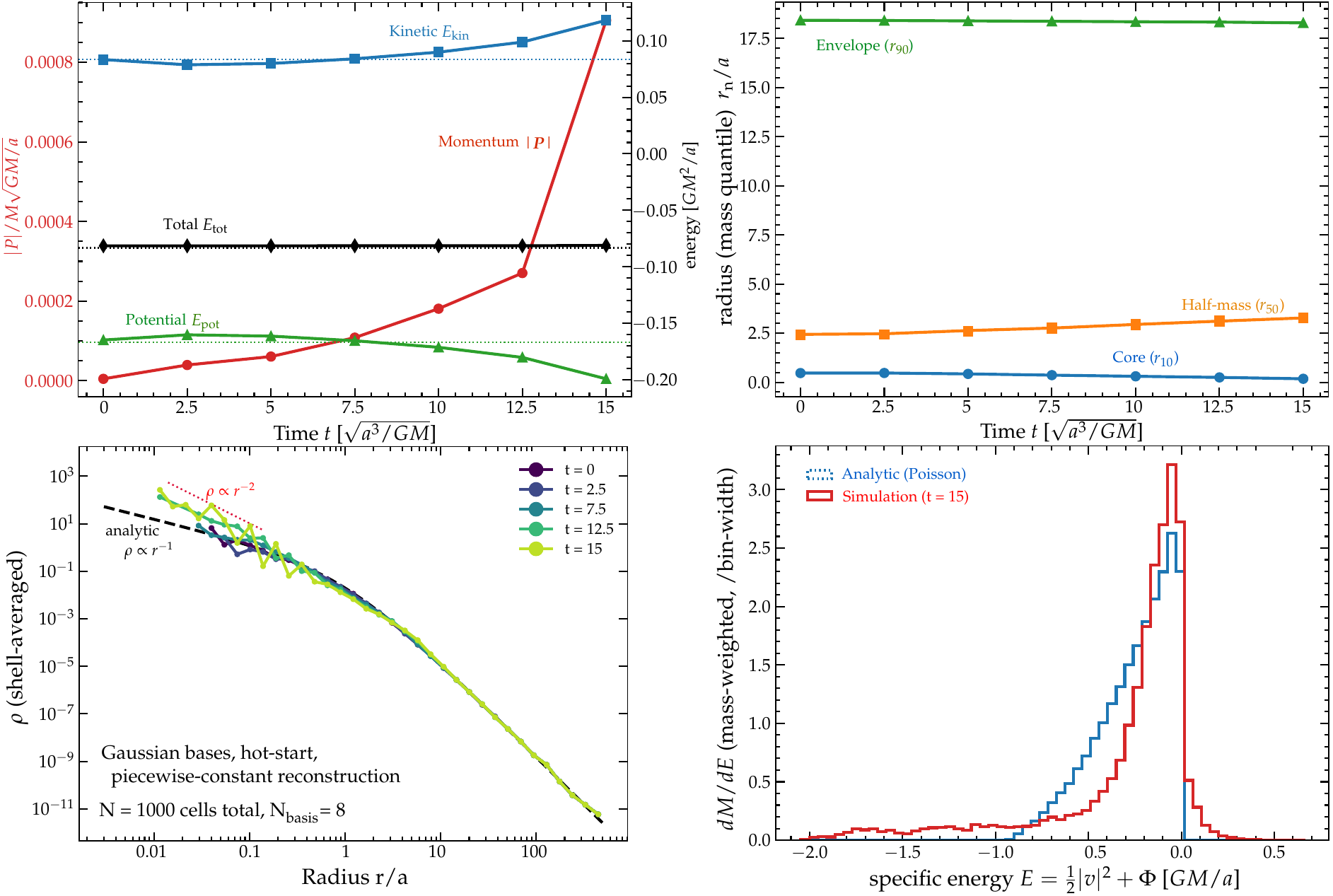}\vspace{-0.2cm}
\caption{As Fig.~\ref{fig:test.hernquist.demo} but a maximally-diffusive ``worst case'' scenario. We initialize with a ``hot start,'' use Gaussian bases, and (most importantly) use a piecewise-constant reconstruction (1st-order instead of our default 2nd-order). All of these increase the numerical dissipation (see Fig.~\ref{fig:test.harmonic}). The gravothermal relaxation is much more pronounced: the core ($r_{10}$) shrinks, $r_{50}$ grows, $|E_{\rm pot}|$ and $|E_{\rm kin}|$ increase together as the tail of the DF extends, and the cusp in $\rho$ steepens to $r^{-3/2}$ after $\sim 100$ local dynamical times.\vspace{-0.2cm}
\label{fig:test.hernquist.diffusive}}
\end{figure}

\begin{figure}
\includegraphics[width=0.98\columnwidth]{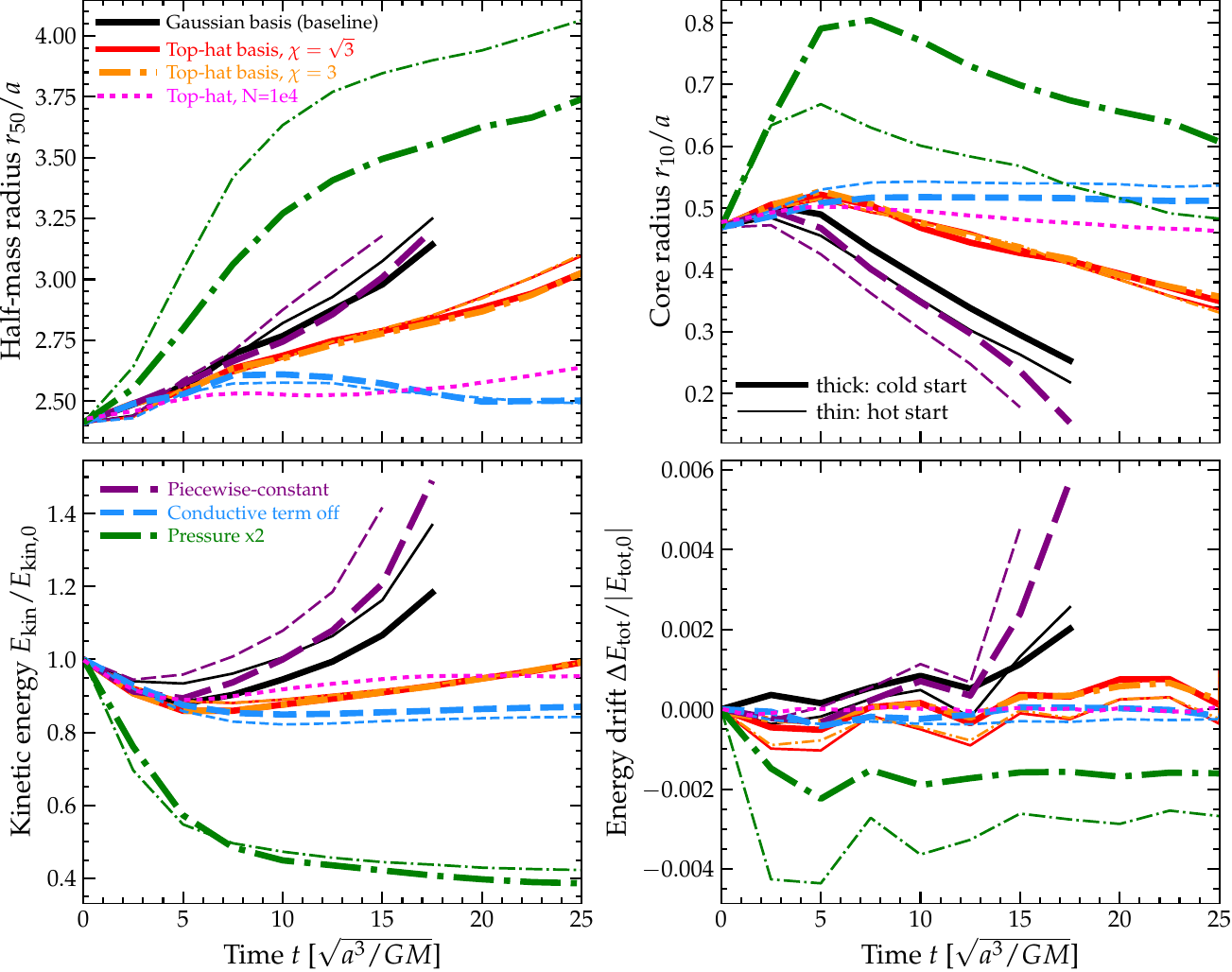}\vspace{-0.2cm}
\caption{Effects on the relaxation of the Hernquist problem (Figs.~\ref{fig:test.hernquist.demo}-\ref{fig:test.hernquist.diffusive}) with different numerical choices. 
We show the half-mass ($r_{50}$) and core ($r_{10}$) radii, kinetic and total energies, as Fig.~\ref{fig:test.hernquist.demo} but zoomed-in so the evolution is more clear. 
All runs here are at $N=1000$ unless otherwise noted, and all $N=1000$ runs include both a ``cold-start'' ({\em thick}) and ``hot-start'' ({\em thin}) variant. 
We compare the default scheme with Gaussian bases (more diffusive); top-hat bases (Fig.~\ref{fig:test.hernquist.demo}) with the default ($\chi=\sqrt{3}$) and larger signal speed ($\chi=3$, so the timesteps are smaller; this has no measurable effect); and Gaussian-basis runs with piecewise-constant (lower-order) reconstruction (Fig.~\ref{fig:test.hernquist.diffusive}, much more diffusive); an artificially modified Riemann pressure (this ``fights'' the core contraction but is simply incorrect and blows up); and turning off the numerical conductivity term arising from upwinding the stress equation in the Riemann problem when the reconstructions do not match. The latter removes the numerical dissipation entirely without destabilizing the problem. The errors also converge away rapidly with resolution.
\label{fig:test.hernquist.variants}}\vspace{-0.3cm}
\end{figure}

\subsection{Non-Linear Virialization: Self-Gravitating Equilibria with Hot and Cold Starts}
\label{sec:test.hernquist}

A more demanding 3D test is the long-time stability of an isotropic, self-gravitating, spherical collisionless equilibrium. 
To realize this we initialize a \citet{hernquist:profile} model with isotropic velocities, using gravitational constant, total mass, and scale-length $G=M=a=1$.
In these units the equilibrium density profile is $\rho=(M/2\pi)\,r^{-1} (r+1)^{-3}$, potential $\phi=-1/(r+1)$ and the DF $f(v\,|\,r)$ as a function of $\mathcal{E} \equiv -E = -\phi - v^{2}/2$ at a given $r$ is entirely analytic. Self-gravity is integrated using the tree method in \GIZMO, operator split as we describe above. We use the standard adaptive individual (unequal) timesteps in \GIZMO, which means momentum and energy conservation drops from machine-accurate to integration-accurate, from both gravity and cell-flux operators. We evolve the system to $t=25\,(GM/a^{3})^{-1/2}$ global dynamical times, which for the inner radii ($r\lesssim 0.1$) on which we focus corresponds to $\gtrsim 250$ crossing/phase-wrapping timescales, and if we assume this represents a typical dark matter halo this corresponds to $\sim 5-10$\,Gyr (i.e.\ roughly a Hubble time). Our fiducial realization uses just $N=1000$ elements and is shown in Fig.~\ref{fig:test.hernquist.demo}, though we have tested up to $N=10^{6}$.

We compare a ``hot'' and ``cold'' start. The ``cold'' start is exactly the IC one would use for traditional MC/PIC/NB methods: we discretize $M$ into $N$ equal masses of $\Delta m = 1/N$, probabilistically assign them radii according to the density profile (drawing a unit random number and equating it to $M(<r)/M = (1+1/r)^{-2}$ to get $r$, then drawing random Euler angles for spherical $\cos{\theta}$, and $\phi$), with discrete, single-valued-per-particle, infinitely cold velocities drawn probabilistically from $f(v\,|\,r)$. 
This is in many ways the ``worst case scenario'' for a method like ours: the cells begin randomly laid out so the mesh is maximally disordered/anisotropic (maximizing flux errors), the cells are all infinitely cold so immediately begin to flow into each other, and at our tested $N=10^{3}-10^{6}$ element resolution this means there are vastly more ``cold streams'' initially than there are basis functions (here $N_{\rm basis} = 4,\,6,\,8,\,10$ were tested). 
The ``hot'' start uses the same (badly-disordered) cell positions from the start, but initializes a much-closer-to-\textit{correct} initial $f(v)$ in each cell based on its position (still intentionally over-simplified: we just use one initial basis function but set it to the correct 3D isotropic velocity dispersion from the Jeans second moment of the equilibrium, with zero mean velocity). 
We have also tried one more case, a ``hot and ordered'' start -- that is much closer to how one would simulate e.g.\ a self-gravitating fluid, with the initial cells laid out in ordered shells rather than randomly \citep[see][]{kegerreis:2019.seagen.stretched.equal.area.spherical.shell.ics.giant.impact.convergence}. We stress that while the cold start is the standard \textit{statistical} representation of the system in MC/PIC/NB simulations, it is in fact very far from equilibrium: at every position, the mass is in large bulk motion, and the DF is infinitely cold. This is basically the opposite of the correct behavior: the mass density should be stationary and the DF dispersion-supported. Essentially, our cold-start case is more akin to trying to resolve the initial collapse and virialization of an object from cold initial conditions. 

We extract the density profile, DF in concentric shells and in total, $\sim 10,\,50,\,90\%$ mass inclusion radii $r_{10,50,90}$, total and kinetic and potential energies, total bulk momentum, and total mass, at various times. This is a demanding test: the mesh is highly disordered, the reconstruction is anisotropic, timesteps are highly unequal and vary by factors up to $\sim 1000$ over the system, the initial DF is far-from-equilibrium and even in equilibrium is strongly spatially-varying, shot noise is large at low resolution, the effective ``pressure'' force is operator-split from gravitational forces, and the system must be evolved for hundreds of shell-crossings and dynamical times per-particle. 

With our default method, the results are encouraging even at very low resolution in Fig.~\ref{fig:test.hernquist.demo}. Despite unequal timesteps and operator-split tree-gravity, momentum and total energy are conserved to expected integration accuracy, $\mathcal{O}(10^{-4}-10^{-3})$ in the units here over $\sim 100$ orbits -- comparable in momentum and only slightly larger in total energy than obtained for pure $N$-body evolution. Mass is conserved to machine precision as expected. The equilibrium density profile, DF (integrated and at different radii), and $r_{10,50,90}$ are stable. And it approaches DF from both hot+cold start. 

\subsubsection{Effects and Convergence in Numerical Dissipation} 

We find this is a sensitive test of any subtle errors or excess dissipation in the method, as shown in Figs.~\ref{fig:test.hernquist.diffusive} \&\ \ref{fig:test.hernquist.variants}. Changing the stress terms in the Riemann problem incorrectly (using e.g.\ a simpler global Lax-Friedrichs-type solution) can over-or-under pressurize the halo, causing it to collapse or blow up. Failure to use a ``wakeup'' scheme with individual timesteps (as in standard hydrodynamics, for any interacting neighbor cells with very different timesteps) leads to large $\mathcal{O}(1)$ energy conservation errors. Models without sensible basis-matching fail catastrophically in the cold-start regime. Ignoring standard stability conditions: e.g.\ removing all slope-limiters or SPD constraints or allowing steps much larger than Courant all lead to catastrophic numerical instability (the total energy changes by orders of magnitude within a percent or so of the total runtime shown here). 

But even if one avoids this, the method is sensitive to the dissipation and the conductive flux term: e.g.\ $\omega\,{\bf S}_{x} \otimes {\bf S}_{x} /\sigma_{x}$ in Eq.~\ref{eqn:gaussian.flux} or \ref{eqn:hllc.F0}. 
Because our methods do involve some dissipation/diffusion, there is a potential heat-transport term. 
Since the system is self-gravitating, it has negative heat capacity, and this produces core contraction, which heats the core further, leading to a potential gravothermal runaway. 
In our implementation, if we use Gaussian bases with the exact-upwind Gaussian fluxes defined in Eq.~\ref{eqn:gaussian.flux} ($\mathcal{F}_{G}$), we see that the core radius $r_{10}$ contracts while half-mass $r_{50}$ expands, and the central density builds up a steeper-than-Hernquist cusp. The evolution is not catastrophic but it is significant at low resolution: at $t=20\,(GM/a^{3})^{-1/2}$ with $N=1000$, $r_{50}$ increases by $\sim 30-40\%$, while $r_{10}$ drops nearly in half (so the density profile steepens from $\rho\propto r^{-1}$ to close to $r^{-1.2}$ at $r\ll 1$). The system still conserves total energy to $\sim 10^{-3}$ accuracy, so this is a heat-transfer effect, not a conservation error. 
The dominant term driving this is the conductive (third-moment) term of the Gaussian flux: disabling just this term eliminates the effect entirely (the core radius stays flat, kinetic+potential energy and profile do not change). We can make the evolution {worse} if we enable a much more dissipative method: using e.g.\ a piecewise-constant method, consistent with what we saw in the harmonic oscillator test, but for our default method it is really dominated by this conductive term. 
This means that the compact support top-hat solver, whose conductive term is $\sim2\times$ smaller at its peak and truncates identically to zero for $|u_{x}|\geq c_{x}$ (\S\ref{sec:flux.approx}; $\omega^{\prime}$ vs.\ $\omega$ in Eq.~\ref{eqn:hllc.F0} for $\mathcal{F}_{\rm TH,1}$), suppresses this quite significantly: the growth in $r_{50}$ or contraction in $r_{10}$ is secular, shows no sign of accelerating, and over the full $t=25\,(GM/a^{3})^{-1/2}$ with $N=1000$ $r_{50}$ only increases by $\sim 10-15\%$ and cusp stays $r^{-1}$ to high precision as shown. Again, removing that conductive term removes even this small residual. 

Note that this arises because of the upwinding of the stress fluxes in the Riemann problem (i.e.\ numerical dissipation) -- it does not appear in the ``direct'' fluxes ${\bf F}$ in Eq.~\ref{eqn:flux.direct}. One way to cure this strictly, at any resolution, is to remove the conductive term. While not strictly self-consistent, we actually find this is not destabilizing the same way that removing the upwind terms (e.g.\ the $\omega \ne 0$ and $\zeta \ne 1/2$ terms) in the mass or momentum fluxes are, so it might be that unconditional stability simply does not require this additional dissipation and it can be removed without harm. But we are cautious since we have no proof of this (only the empirical result that it does not cause any detectable errors in our other test problems). 
One might think that identifying basis functions that better represent the actual DF in more detail would help, but this is a minor effect, since as discussed in \S~\ref{sec:flux.gaussian} the dissipation is only weakly sensitive to the form of the basis (top-hat and Gaussian already nearly bracket the maximal range of monotone bases) but reflects inconsistency in the spatial reconstruction at the order of the method. So a better solution would be a higher-order reconstruction. 

But even at our (second) order, this means the problematic behavior \textit{converges away} at higher resolution. 
In practice, we find a modestly sub-linear decrease in the rate of energy transfer with $N$: consistent with our expectation for a second-order method that the rate scales $\mathcal{O}(\Delta x^{2}) \sim \mathcal{O}(N^{-2/3})$. Even by just $N \sim 10^{4}$, it is very difficult to measure any effect on the timescale for which we simulate.
We stress that this convergence is not in ``particle number at a given radius,'' as one typically requires for $N$-body codes: the errors scale as $\sim \Delta x^{2}$, so increasing resolution helps everywhere. For example, by the common \citealt{power:2003.nfw.models.convergence} criterion $N$-body density profiles are ``converged'' interior to the radii enclosing $\sim 2000$ particles, but our density profiles are similarly converged at $r\sim 0.1\,a$ even at $N=1000$ total, where this radius encloses just $8$ \textit{total} particles.

\begin{figure}
\includegraphics[width=0.98\columnwidth]{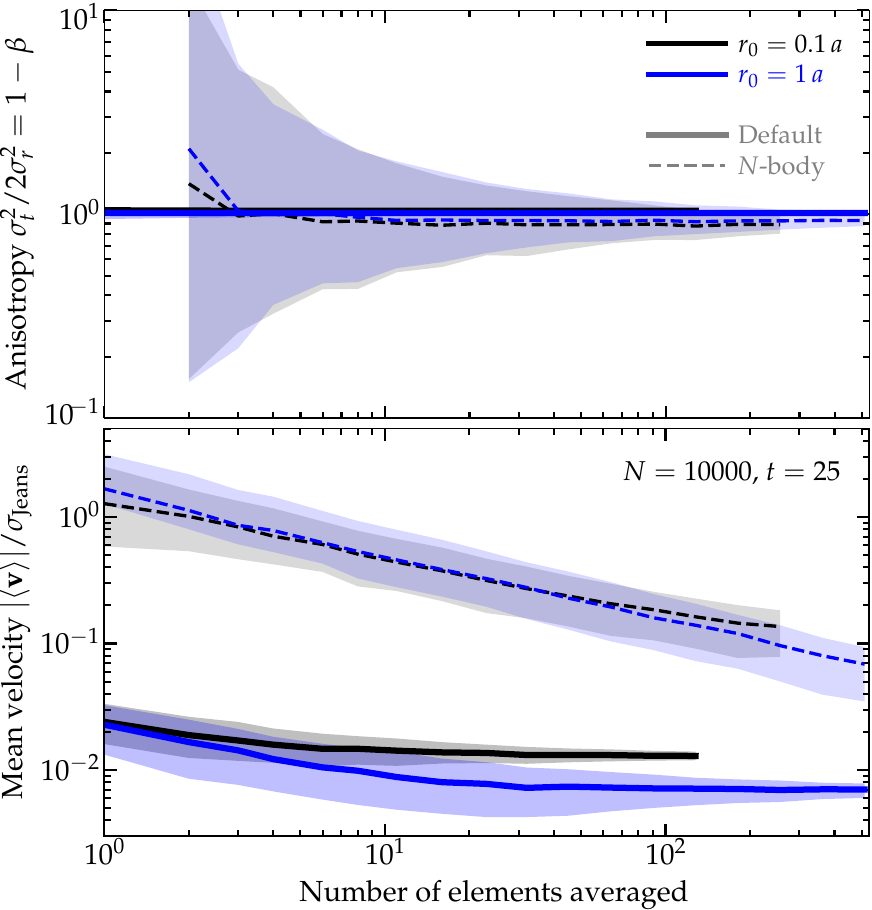}\vspace{-0.2cm}
\caption{Illustration of the convergence in phase-space for the method here (``default'') compared to traditional $N$-body methods on the Hernquist-sphere problem (Fig.~\ref{fig:test.hernquist.demo}). We randomly select a number of resolution elements $N_{e}$ within a shell at some radius $r_{0}=(0.1,\,1)\,a$ at the final $t=25$ time (results here for total $N=10^{4}$, but at a given $N_{e}$ they are similar for any $N$), and build the combined $v_{r,\theta,\phi}$ velocity-space DF from them, measuring two simple moments: the mean velocity $\langle {\bf v} \rangle$ and velocity anisotropy $\beta$ (plotting $1-\beta$ so deviations are more symmetric). We repeat the selection $100$ times and plot the median (lines) and $\pm 1\sigma$ range (shaded). Convergence of other moments (e.g.\ $\sigma_{r,\,\theta,\phi}$ individually, skew, etc.) is similar. The correct equilibrium $|\langle {\bf v}\rangle|=\beta=0$. In traditional $N$-body methods, the envelope of $\beta$ and mean velocities average-down slowly with Poisson statistics, because it is fundamentally a Monte-Carlo method. In our method, a \textit{single} cell measures the values of the DF moments at a given position more robustly and more accurately, compared to an average over hundreds of $N$-body particles in a traditional $N$-body simulation. 
\label{fig:test.hernquist.vsample}}\vspace{-0.2cm}
\end{figure}

\subsubsection{Comparison to Other Methods}

In MC/PIC/NB methods, the numerical dissipation problem above vanishes for truly collisionless systems -- this is a genuine advantage of NB methods. 
But even on a simple problem like this where NB is by far the most popular approach, there are failure modes. 
For example, the force softening $\epsilon_{0}$ must be arbitrarily chosen, and there is no single ``optimal'' $\epsilon_{0}$ (it is usually taken to be the inter-particle spacing, but for equal-mass particles here that varies as $\rho^{-1/3} \propto r^{4/3}$ in the outer regions). Under-softening leads to artificial collisional gravitational heating which causes a gravothermal collapse through an entirely different mechanism, while over-softening flattens the cusp artificially and corrupts solutions out to much larger $r$ while artificially disrupting real bound structures \citep{ludlow:2019.spurious.equipartition.heating.galaxy.sizes,ludlow:2021.disk.heating.by.dm.particles,dehnen:2001.optimal.softening,zhang:2019.optimal.cosmo.sim.softenings,wilkinson:2023.collisional.disk.heating.only.need.1e7.msun.resolution.to.avoid.in.typical.fire.sims}. 
And as already noted, convergence in the DF at any given $r$ in MC/PIC/NB methods is extremely slow: one must simultaneously increase the size of some averaging kernel while also increasing total $N$. 
Even a very simple characterization of the DF requires measuring the 10 3D first and second moments, i.e.\ to characterize the mean velocity, dispersion/stress, and anisotropy in each direction -- in NB methods, a meaningful measurement of those moments typically requires an ensemble of thousands of particles \citep[as demonstrated explicitly in Fig.~\ref{fig:test.hernquist.vsample}, where good measurement of even two lowest-order moments requires hundreds of particles;][]{necib:2019.dm.velocity.distribution.fire.vs.obs,zhang:2026.modeling.dm.velocity.distrib.from.obs.mw.analogs.fire2}. 
This is related to cusp/core behavior requiring $\gtrsim 2000$ particles for a robust result \citep{power:2003.nfw.models.convergence,hopkins:fire2.methods,2004MNRAS.353..624D,2019MNRAS.488.3663L}. 
But those 10 moments are the same information contained within a \textit{single} basis function, for a \textit{single} resolution element, in our method. 

Direct/Eulerian/lattice-Boltzmann Vlasov methods would likely struggle with this problem in two respects: first, at the resolution we consider, it would simply be too expensive to run, and second, because the dissipation which is the most severe issue in our method should be much larger from large motion across the fixed phase-space grid. 
Moments or wavefunction methods could approximately capture the ``hot start'' version so long as it was started from the true equilibrium DF and is effectively static. Their ability to capture the true DF would depend on the closures imposed. But more importantly they would fail catastrophically with the ``cold start'' setup.

\section{Potential Applications \&\ Variations}
\label{sec:application}

We have outlined a very general method designed to represent, in principle, any weakly-collisional flow. Before concluding, we briefly describe potential applications, together with the modifications each would need. Several of these will be the subject of future studies using the implementation in \GIZMO, but this is only a partial ``motivational'' list, and our hope is that the method will prove useful, and be extended and improved, well beyond it.

In general, it is useful to compare the resolution scale $\Delta x$ to three characteristic physical scales. First, the inter-particle separation and/or hard-scattering or Debye-like scale $\Delta x_{\rm discrete}$: a method like ours, representing a continuous DF, only makes sense if $\Delta x \gg \Delta x_{\rm discrete}$. Second, the collisional/deflection/isotropization mean-free-path $\Delta x_{\rm MFP}$: if this is everywhere small ($\Delta x \gg \Delta x_{\rm MFP}$, the strongly-collisional limit), our method formally still works but reduces to the underlying fluid scheme (with extra overhead for unused bases), so is not practically useful there. Third, some physics -- most notably magnetization of charged particles (Lorentz forces), though strong self-gravitating clumping can behave similarly -- introduces a scale $\Delta x_{\rm coll} > \Delta x_{\rm discrete}$ (e.g.\ the gyro-scale) where these forces strongly modify the dynamics. If $\Delta x_{\rm coll} \gg \Delta x$ this can be treated as just another resolved external force; if $\Delta x_{\rm coll} \ll \Delta x$ (with $\Delta x_{\rm discrete} \ll \Delta x \ll \Delta x_{\rm MFP}$), it requires the modifications discussed below.

\subsection{Dark Matter} 
\label{sec:apply:dm}

An obvious application we have already referred to throughout is dark matter (DM) dynamics \&\ cosmology, for which the implementation here requires essentially no modification. Even in the purely collisionless limit there are many advantages over standard N-body methods (\S~\ref{sec:compare:nbody}) in accuracy, smoothness, convergence, and information content. These advantages grow, and the cost-multiplier shrinks, for models with non-trivial dark-sector interactions (self-interacting [SIDM], excited-state, atomic, mirror, light-mediator, baryon-scattering DM): the method can correctly integrate between weakly- and strongly-collisional limits, capture the ``effective conductivity'' in the strongly-self-interacting limit, resolve tail effects in strongly velocity-dependent SIDM, and represent the thermal state of the DM needed for DM-baryon exchange -- none of which N-body can do directly. Models like warm DM, historically challenging for N-body without artificial fragmentation, become trivial.

\subsection{Stellar Dynamics} 
\label{sec:apply:star}

Stellar dynamics in the large-$N$ limit shares all the same advantages as DM. Minimizing N-body heating and allowing an explicit DF representation is especially valuable for narrow/cold resonant dynamics, where traditional N-body methods converge painfully slowly. Two caveats apply, both minor. First, since our basis functions exchange mass, ``micro-particle state labels'' that matter dynamically -- stellar abundances, ages, masses, e.g.\ in simulations with feedback/stellar evolution -- must either be advected with the mass flux (as gas-phase metals follow gas) or tracked via independent, separately-exchanged basis sets; this is simply bookkeeping. Second, one must respect the small-$N$/hard-N-body limit: at sufficiently high resolution a DF representation of stars breaks down, just as for any application here, but this limit is reached more easily for stars, particularly when (as in some galactic star-formation models) the ``micro-particle'' represents an unresolved star cluster rather than a single star -- there, N-body is the correct dynamics.

\subsection{Weakly-Collisional Neutral (Atomic/Molecular) Gas Dynamics} 
\label{sec:apply:neutral}

Neutral fluid dynamics in the weakly-collisional regime is similar to the dark matter case, and the implementation here can be applied directly. Like for stars, ``tags'' for different quantities (e.g.\ abundances, different species) can be included in fluxes, or tracked in a true multi-fluid fashion via different bases. Collisionality can easily be incorporated according to the methods described in \S~\ref{sec:collisions.howcollisional} (indeed neutral collisions are often specifically approximated as hard-sphere scattering in this limit), with a typical $\sigma_{nn} \sim 10^{-16}-10^{-15}\,{\rm cm^{2}}$ -- so there are interesting weakly-collisional dynamics on a wide range of scales from $\Delta x_{\rm discrete} \sim 1\,{\rm cm}\,(n_{n}/{\rm cm^{-3}})^{-1/3}$ to $\Delta x_{\rm MFP} \sim 10^{16}\,{\rm cm}\,(n_{n}/{\rm cm^{-3}})^{-1}$. 

\subsection{Particulate Dynamics (Dust, Pebbles, etc.)} 
\label{sec:apply:dust}

The methods here are especially useful for dust and particulate dynamics, weakly-collisional in almost all regimes of astrophysical, terrestrial, and atmospheric interest. Our method captures the key advantages of the two most popular approaches without their compromises: Monte-Carlo/MHD-PIC methods (``dust as particles'') can represent arbitrarily complicated DFs, self-crossing streams, etc., but at the cost of substantial numerical noise, slow convergence, and ill-defined force-softening and collision operators; fluid methods (``dust as advected scalar'') are smooth but are essentially a 1-basis, 1-moment approximation, unable to capture non-linear dust dynamics, instabilities, or grain-grain collisions. Adding physics like dust-dust collisions or velocity-dependent effects (sputtering, condensation, coagulation, bouncing, shattering) only strengthens this advantage.

Background forces from the gas -- drag and Lorentz forces -- are straightforward external forces. Drag is especially trivial, a simple operator-split damping of the DF: for the velocity-independent drag ${\rm d}_{t} {\bf v}_{\rm p} = t_{\rm s}^{-1}\,({\bf v}_{\rm gas}-{\bf v}_{\rm p})$ appropriate for subsonic Epstein/Stokes/Coulomb drag, one simply has $\langle {\bf v}_{\alpha} \rangle \to \psi \langle {\bf v}_{\alpha} \rangle + (1-\psi)\,{\bf v}_{\rm gas}$, ${\bf S}_{\alpha} \to {\bf S}_{\alpha} \psi^{2}$ with $\psi \equiv \exp{(-\Delta t/t_{\rm s})}$, and our well-defined volume partition makes assigning the equal-and-opposite back-reaction to the gas no different than in any other two-fluid method. Lorentz forces fall into the resolved/unresolved-gyro-radius dichotomy discussed above: resolved (often true, given particulates' high mass-to-charge ratios) is again a trivial external force; unresolved requires the anisotropy treatment described for ionized plasmas below. The ability to smoothly follow evolution from weakly- to strongly-collisional regimes is uniquely valuable for e.g.\ planet(esimal) formation via streaming instabilities or pebble-pile collapse/accretion.

Since particulate dynamics often involve a background ``gas'' (e.g.\ MHD-PIC), the volume partition offers a choice: our default is ``dust/particulate cells'' that move with respect to the ``gas cells,'' but one could instead anchor the partition to the gas mesh exactly. This only changes the face definition and velocity, and has advantages -- no separate particulate grid, perfect co-location for drag/back-reaction, and exact gas-following in the perfect-coupling limit $t_{\rm s}\to0$ -- at the cost of no longer being Lagrangian with the particulate, so clumping \textit{relative to} gas (the standard outcome of resonant drag instabilities; \citealt{squire.hopkins:2018.rdi.streaming.instability}) is always under-resolved, as in ``dust as fluid'' methods.

\subsection{Multi-Phase/Polydisperse Flows} 
\label{sec:apply:disperse}

Many numerical methods have been proposed which track multi-phase and/or polydisperse flows as essentially akin to either weakly-collisional neutral or particulate dynamics. For example, if there are some condensates/precipitates/droplets/bubbles/clouds (ranging from aerosolized sprays through ISM molecular clouds, e.g.\ \citealt{huang:2022.subgrid.wind.model.with.subgrid.clouds.ala.iryna.in.gizmo,butsky:2024.twophase.cold.gas.subgrid.model.methods.paper}), these can be evolved as a set of ``particles'' (MHD-PIC-like) or a ``weakly collisional fluid scalar set'' (tracking their abundance in each gas cell but flowing through cells), subject to some ambient interactions (e.g.\ drag and mass exchange from hydrodynamic interactions, collisions/coagulation/etc.). Indeed, the EQMOM-like methods we discuss in \S~\ref{sec:compare:moments}, which are closely related to the methods developed here, were developed for these cases. Obviously, the method here is just as advantageous for these multiphase flows as it is for weakly-collisional neutral and particulate dynamics.

\subsection{Radiation Transport \&\ RHD} 
\label{sec:apply:rhd}

The methods here cross-apply very directly to radiation transport (RT) and RHD -- mostly a simple change of variables, and in some ways simpler in the ultra-relativistic limit. The DF $f({\bf x},\,{\bf v},\,t) = f({\bf x},\,{\bf p},\,t)$ maps directly to the intensity $I_{\nu}({\bf x},\,\nu,\,{\bf n},\,t) = I_{\nu}({\bf x},\,{\bf p},\,t)$ (${\bf p} = {\bf p}_{\gamma} = h \nu {\bf n}/c$), with a trivial map from our basis functions in ${\bf v}$ or ${\bf p}$ to ${\bf p}_{\gamma}$. The 10 moments defined here for each basis map directly to the radiation energy density $e_{\gamma}$, flux ${\bf F}_{\gamma}$, and all components of the radiation pressure/Eddington tensor $\mathbb{P}_{\gamma}$, $\mathbb{D}_{\gamma}$. Because all speeds are $c$, the upwind flux computation actually simplifies (trivial versions of ours with modified $\tilde{w}^{\prime}$, $\zeta^{\prime}$, $\omega^{\prime}$), and coupling to reduced-speed-of-light (RSOL) methods is trivial: these simply augment the evolution equations for all radiation moments (${\rm d}_{t} X \rightarrow (c/\tilde{c})\, {\rm d}_{t} X$; \citealt{gnedin.abel.2001:otvet,hopkins.most:2025.time.dilation.methods.extreme.timestepping}). All non-hyperbolic source/sink terms in the intensity equation are simply operator-split.

This allows construction of truly general RHD methods that avoid the pitfalls of classical moment (flux-limited-diffusion or M1-type) methods -- e.g.\ mishandling the optically-thin, crossing-beam, or anisotropic-scattering/source limits -- while keeping their advantages (smooth, hyperbolic evolution equations, trivial RSOL coupling): we are simply evolving the full intensity equation for an arbitrary DF. Two choices lead to interesting variations. For the basis functions:
\begin{itemize}
\item{General/Spectral Bases:} Bases with finite width in $|{\bf p}|$ and evolved $\langle |{\bf p}|\rangle$, as here, represent a smooth continuum spectrum across all frequencies $\nu$ (integrating a ${\bf p}$-basis over ${\bf n}$ gives the spectrum). One could adapt the bases to a form more natural for radiation, e.g.\ $\tilde{g}_{\alpha}({\bf p}) \rightarrow \tilde{g}_{\alpha}(|{\bf p}| =h \nu/c,\, \hat{\bf p} = {\bf n}) = B_{\nu}^{\alpha}(\nu,\,T_{\alpha})\,\zeta({\bf n},\,\langle {\bf n} \rangle_{\alpha})$ -- a Planck function (basis ``radiation temperature'' $T_{\alpha}$ encoding $\langle |{\bf p}| \rangle$) times an arbitrarily isotropic or beamed angular function. This resembles adaptive-grey-body M1 methods \citep{levermore:1984.FLD.M1,grudic:starforge.methods}, but extended with full anisotropic directionality and collisionless behavior.

\item{Narrow or Fixed-Bin Bases:} Alternatively, force bins to fixed $|{\bf p}|$ (bases $\delta(|{\bf p}|-|{\bf p}|_{0})$), free only in direction ${\bf n}$. This reduces dimensionality and the moments needed, but discards spectral evolution (absorption/re-emission, blue/redshifts); many existing RHD implementations effectively do this via fixed, narrow $\nu$ bins.
\end{itemize}
And the choice of volume partition: our default is effectively Lagrangian with the radiation, so for a pure beam the cells move with $c$ (or RSOL $\tilde{c}$).\footnote{Mesh motion here is a purely numerical construct with no physical meaning: the radiation energy flux between two parallel same-direction-$c$ beam bases in neighboring cells is identically zero (${\bf v}-{\bf v}_{F}=\mathbf{0}$), since we do not transform relativistically into the frame when computing fluxes.} This is analogous to on-the-fly Monte-Carlo RT or ray-tracing, but with all our advantages over pure Monte Carlo. Alternatively, anchoring the partition to the gas mesh (as for particulates above) is more analogous to typical moments-RHD implementations: the direct-ray method of \citet{jiang:2014.rhd.solver.local} is the special case with the partition anchored to a fixed Eulerian gas grid and basis functions fixed in both $\nu$ and ${\bf n}$ ($f \propto I_{\nu} \propto \delta(|{\bf p}|-|{\bf p}|_{0},\,{\bf n}-{\bf n}_{0,\,i})$), i.e.\ a multi-basis, 1-moment scheme.

\subsection{Cosmic Rays \&\ Relativistic Particles} 
\label{sec:apply:cr}

Numerically, cosmic ray (CR) transport and fully-coupled CR-MHD dynamics are basically identical to particulates \citep{bai:2015.mhd.pic,bai:2019.cr.pic.streaming,ji:2021.cr.mhd.pic.dust.sims}, with a few added terms involving the Lorentz factor $\gamma$; Lorentz forces are always important here, so again there are two regimes. If CR gyro-radii are \textit{resolved} ($\Delta x \lesssim \Delta x_{\rm coll}$), our method is an improvement on MHD-PIC methods \citep{bai:2015.mhd.pic,bai:2019.cr.pic.streaming,ji:2021.cr.mhd.pic.dust.sims} evolving the entire DF, in the same way it improves N-body or particulate dynamics: Lorentz forces are simply an external force, and CR scattering and streaming instabilities are explicitly resolved. As with dust, this improves dramatically on the smoothness and information content of PIC/Monte-Carlo methods, whose noise can destroy slow-growing instabilities -- the original motivation for the implicit MHD-PIC ``$\delta f$'' formalism \citep{bai:2019.cr.pic.streaming}, which is unnecessary here since $f$ is evolved directly and is trivially smooth (a $\delta f$-like scheme is still trivial to construct, if desired, as simply a different basis choice). Collisions/annihilation/spallation/decay and other reactions are easily added as operator-split DF operations, and an RSOL is as trivial to add as for RHD above, in the same way one does for MHD-PIC \citep{ji:2021.mhd.pic.rsol}.

In the limit of \textit{unresolved} gyro radii, the method is better thought of as an extension of existing CR-MHD fluid moment methods \citep{hopkins:m1.cr.closure,hopkins:cr.spectra.accurate.integration,thomas.pfrommer.18:alfven.reg.cr.transport}, in which gyro-averaging projects the transport along field lines, so fluxes become truly 1D (along $\hat{\bf b}$) and correspondingly simpler (also simpler in the ultra-relativistic limit, as \S~\ref{sec:apply:rhd}). Those prior methods effectively adopt a 1-basis, 2-moment model at each $|{\bf p}|$, discarding the pitch-angle dimension of the gyro-averaged DF ($\bar{f}(|{\bf p}|,\,\mu) \to \tilde{f}(|{\bf p}|,\,\langle \mu\rangle)$) in favor of an M1-like closure \citep{hopkins:m1.cr.closure,thomas:2021.compare.cr.closures.from.prev.papers}; our method trivially restores the full pitch-angle DF here. In this limit one cannot explicitly resolve collisionless CR scattering off unresolved field-line fluctuations, so some sub-grid scattering/collision operator is still required -- not just for the effective transport parameters (diffusion coefficient, streaming speed), but to regulate the DF itself against kinetically-unstable configurations (e.g.\ antiparallel streams) that a well-posed effective collision operator should isotropize rapidly.

As with RHD, it may help in both regimes to use different basis functions -- e.g.\ gyro/polar coordinates ${\bf p}\rightarrow(|{\bf p}|,\,\hat{\bf p} \cdot \hat{\bf b},\,\phi)$, or bases binned in $|{\bf p}|$ -- and there is the same choice of whether to tie the volume partition to the CRs or to the gas decomposition (the default in the gyro-unresolved case above).

\subsection{Ionized Plasmas} 
\label{sec:apply:plasma}

Akin to charged-particulate (\S~\ref{sec:apply:dust}) and cosmic-ray (\S~\ref{sec:apply:cr}) dynamics, applying the methods here to \textit{ionized} weakly-collisional plasmas (cf.\ neutral fluids, \S~\ref{sec:apply:neutral}) depends on whether the gyro radii are resolved or unresolved -- immediately raising the question of \textit{which} gyro radii (electrons or ions), given their large disparity, a classic practical difficulty in PIC-type simulations that our approach does not resolve.

With that caveat, in the resolved/kinetic limit ($\Delta x \lesssim \Delta x_{\rm coll} \sim r_{\rm gyro}$) the methods here are a straightforward extension of PIC, as throughout: removing Monte-Carlo/discreteness noise, evolving the full DF, resolving kinetic instabilities, with no special tricks needed. Collisionless scattering is resolved, Lorentz forces trivial to add, arbitrary collision operators straightforward, and computing/re-projecting smooth ${\bf E}$ and ${\bf B}$ fields (often a technical subtlety in PIC) becomes computationally cleaner. Per \S~\ref{sec:moment.order}, a 13- rather than 10-moment basis may converge faster here for speed, but since the instabilities regulating the DF shape are \textit{resolved} in this regime, the 13-moment closure adds no representational power over 10-moment-with-more-bases (only a convergence/cost tradeoff), and the 10-moment form remains robust arbitrarily far from equilibrium while 13-moment realizability requires care if not near-Maxwellian. So 13-moment methods converge faster and suppress spurious instabilities (e.g.\ bump-on-tail) closer to equilibrium, while 10-moment-with-more-bases is more robust far from it. And any existing 13-moment code generalizes almost trivially to our method by adding bases (a super-set of $N_{\rm basis}$ independent 13-moment sets), the only qualitatively new machinery needed being our basis-matching/pairing operation.

The small-gyro-radius but weakly-collisional (gyrokinetic) limit, as for particulates and CRs, is more conceptually challenging. As usual it helps to decompose the DF into $p=|{\bf p}|$, pitch angle $\mu \equiv \hat{\bf p} \cdot \hat{\bf b}$, and gyro phase $\phi$; averaging over $\phi$ reduces dimensionality, projects fluxes onto $\hat{\bf b}$, and reduces the moments needed, at the cost of (1) simple, trivially-incorporated source terms, and (2) requiring a sub-grid model for the unresolved (gyro-scale) collisionless scattering and instabilities that would otherwise regulate the DF -- the latter being the genuinely non-trivial part of the problem.

\section{Discussion}
\label{sec:discussion}

We have presented a new numerical method to directly integrate the Vlasov equation in a Lagrangian, finite-element manner, combining ideas from recent meshless multi-fluid hydrodynamic,  radiation-transport, and plasma physics schemes. Where standard MC/PIC/NB methods discretize the underlying DF into $\delta$-function ``super-particles'' -- forcing the force softening, continuum field estimators (density, electric/magnetic fields, back-reaction, etc.), and any collisional operators to be {ad hoc} -- the present method directly evolves a continuous DF locally represented in each mesh element by a small set of basis-function components. We present an implementation retaining finite-mass mesh elements that couple to standard MC/PIC/NB machinery (tree or fast multipole or particle-mesh methods, fully-adaptive force-softening, domain decomposition) without modification. The result is a method that solves the continuum collisionless Vlasov/Boltzmann equation but inherits and extends upon the favorable cost scalings and flexibility of MC/PIC/NB, with a modest fixed overhead per element for the per-basis Riemann solve and pairing operations, bringing improved smoothness and accuracy and convergence properties.

We have shown that the method is numerically stable and accurate and manifestly conserves mass, momentum, and energy on a number of controlled test problems in 1D and 3D, and have outlined a set of applications in the 3D \GIZMO\ implementation we make public, including but not limited to: cosmology and dark matter; stellar dynamics; weakly-collisional neutral hydrodynamics; particulate/dust/grain/pebble dynamics; multiphase and polydisperse flows; radiation transport/hydrodynamics; cosmic ray transport and magnetohydrodynamics; and ionized plasma particle physics. 
Compared to traditional MC/PIC/NB methods (\S~\ref{sec:compare:nbody}), the method here has well-defined formal convergence properties; carries full local phase-space information, which allows arbitrarily-complex local phase-space operators (anisotropic, velocity-dependent, or direction-dependent scattering, annihilation/decay/spallation/destruction/collisions) to be applied trivially and consistently; uses a continuum density reconstruction, which makes force-softening a well-defined operation tied to the same volume partition that produces the inter-cell faces; and eliminates spurious discreteness, two-body, and relaxation effects. The cost is a modest, but not prohibitive, increase in per-element complexity, together with some numerical diffusivity that converges with $N_{\rm basis}$ and the spatial resolution in the usual finite-volume sense. Compared to Eulerian or semi-Lagrangian direct-Vlasov methods (\S~\ref{sec:compare.eulerian}), the present approach is many orders of magnitude less expensive at a given effective velocity resolution because the basis functions are Lagrangian in velocity space and adapt to the local flow, rather than being fixed on a global $6$D phase-space grid. Compared to ``moments''-based methods or Schr{\"o}dinger-Poisson reformulations (\S~\ref{sec:compare:moments}), it correctly represents multiple streams, phase-wrappings, and arbitrary DFs by construction. And compared to tessellation or waterbag methods (\S~\ref{sec:compare.waterbag}), the per-step cost does not grow exponentially with the number of phase-wrappings, so the method remains tractable in the late, highly-mixed, multiply-streaming regimes that arise generically in weakly-collisional dynamics.

We discuss and explore a number of subtle aspects of the scheme, including different implementation choices (like choice of volume partition, basis function definitions, reconstruction and slope-limiters, basis-function matching, number of bases, approximate Riemann solvers, etc.). We also discuss how to smoothly and stably make this method fully-continuous into the strongly-collisional regime (requiring only a small modification to the Riemann problem to ``add back'' the collisional effects between faces), as well as how to add other types of collision operators to the phase-space DF. But many further natural improvements and generalizations remain, particularly to the simple implementation we describe and test here. The piecewise-linear reconstruction in \S~\ref{sec:reconstruction} is straightforwardly extended to quadratic or WENO-type reconstructions, which would reduce the diffusivity further at the cost of additional per-pair operations. The basis functions we consider are just a couple of many possibilities; Hermite-type expansions about the bulk velocity, or mixtures of skew-Gaussians, might represent the same information more compactly in some regimes, at the cost of a more complex closure, and for some problems (e.g.\ radiation or cosmic-ray dynamics) other spectral functions (e.g.\ Planck functions or piecewise-power-laws in $|{\bf p}|$) may be more natural. The realizability-repair operator of \S~\ref{sec:repair} preserves $(m,\,{\bf p},\,{\bf T})$ exactly but discards higher moments; a CQMOM- or EQMOM-style construction \citep{fox:2008.eqmom.gaussian.quadrature.moment.method,marchisio.fox:2013.qmom.book.multiphase.moments} which preserves a finer moment hierarchy is straightforward to implement. The single-basis pairing rules of \S~\ref{sec:pairing} could be replaced by a fractional / soft assignment based on optimal-transport-style weights \citep{sinkhorn:1964.relationship.positive.matrices.doubly.stochastic,cuturi:2013.sinkhorn.distances.optimal.transport}, although we have not yet found a test problem on which this gives a clear improvement. And the fluxes (already supporting an arbitrarily anisotropic DF) could be made to project that anisotropy along special directions, like the magnetic field direction for convenient representation of cases with unresolved gyro-radii. The core of the method is conceptually quite simple, and as such, while we intend to study some specific applications in \GIZMO, our hope is that this class of method can be implemented in many different codes with different choices improving beyond the simple examples here.

\acknowledgments{Support for PFH was provided by a Simons Investigator Grant \&\ NSF-AAG grant \#2509270. Numerical calculations were run on Frontera (LRAC AST20016) and the resources of the Texas Advanced Computing Center (TACC), including the Vista GPU system (NAIRR Startup NAIRR260139), and on the Caltech Resnick HPC cluster, under allocations AST20016 and AST21010. Generative AI (Claude Opus 4.6-4.8, Fable 5, and GPT Codex 5.5) was used in testing and debugging the code implementation in \GIZMO\ and in the writing and editing of this manuscript.}

\bibliography{ms_extracted}

\end{document}